\documentclass[11pt,oneside]{book}%
\usepackage[dvips]{graphics}%

\usepackage[centertags]{amsmath}
\usepackage{amsfonts}
\usepackage{amssymb}
\usepackage{amsthm}
\usepackage{newlfont}

\begin{document}%

\setlength{\unitlength}{1mm} \baselineskip .85cm
\large \baselineskip .85cm
\begin{titlepage}
\title{\Huge\vspace{-2cm} {\bf Quantum Cloning and Deletion in Quantum Information Theory} \vspace*{5cm} \\
\Large {\bf THESIS SUBMITTED FOR THE DEGREE OF DOCTOR OF
PHILOSOPHY (SCIENCE) OF  BENGAL ENGINEERING AND SCIENCE
UNIVERSITY}\\\vspace{1cm}
    }

\author{ {\bf BY} \vspace*{2cm} \\
\LARGE{\bf \textsl{SATYABRATA~~ADHIKARI}}  \vspace*{2cm}\\
\large {\bf DEPARTMENT OF MATHEMATICS} \\
\large {\bf  BENGAL ENGINEERING AND SCIENCE UNIVERSITY UNIVERSITY}\\
\small  {\bf HOWRAH- 711103, West Bengal} \\
\large {\bf INDIA}\\\\ {\bf 2006}}
\date{}
\maketitle
\end{titlepage}

\pagenumbering{roman}
\newpage
\vspace*{2.5cm}
\begin{center}
 {\large {\bf\Large DECLARATION} }
\end{center}
\vspace*{1.5cm} I declare that the thesis entitled ``{\it
\textit{\textbf{Quantum Cloning and Deletion in Quantum
Information Theory}}}" is composed by me and that no part of this
thesis has formed the basis for the award of any Degree, Diploma,
Associateship, Fellowship or any other similar title to me.\\

\vspace*{3cm} \noindent(Satyabrata Adhikari)~~~~~~~~~~~~~~~~~~~~~~~~~~~~~~~~~~~~~~~~~~ Date:\\
Department of Mathematics,\\
Bengal Engineering and Science University, Shibpur\\
Howrah- 711103, West Bengal\\
India

\newpage
\vspace*{2.5cm}
\begin{center}
 {\large {\bf\Large CERTIFICATE} }
\end{center}
\vspace*{1.5cm} This is to certify that the thesis entitled ``{\it
\textit{\textbf{Quantum Cloning and Deletion in Quantum
Information Theory}}}" submitted by Satyabrata Adhikari who got
his name registered on 21.12.2002 for the award of Ph.D.(Science)
degree of Bengal Engineering and Science University, is absolutely
based upon his own work under my supervision in the Department of
Mathematics, Bengal Engineering and Science University, Shibpur,
Howrah-711103, and that neither
this thesis nor any part of it has been submitted for any degree / diploma or any other academic award anywhere before.\\

\vspace*{3cm} \noindent(Binayak Samaddar Choudhury)~~~~~~~~~~~~~~~~~~~~~~~~~~~~~~~~~~~ Date:\\
 Professor\\
Department of Mathematics,\\
Bengal Engineering and Science University, Shibpur\\
Howrah- 711103, West Bengal\\
India

\newpage
\vspace*{1cm}
\begin{center}
 {\huge Acknowledgements}
\end{center}
\vspace*{1cm} A journey is easier when you travel together.
Interdependence is certainly more valuable than independence. This
thesis is the result of five years of work whereby I have been
accompanied and supported by many people. It is a pleasant
circumstance that I have now the opportunity to express my
gratitude for all of them.\\
The first person I would like to thank is my respected supervisor
Dr. Binayak S.Choudhury. I have been with him since 2001. During
these years I have known him as a sympathetic and
principle-centered person. His integral view on research and his
mission for providing 'only high quality work and not less', has
made a deep impression on me. He could not even realize how much I
have learned from him. This thesis would not have been possible
without his help, kind support and encouragement.\\
I am deeply indebted to Dr. A.K.Pati from the Institute of
Physics, Bhubaneswar, India and Dr. G.P.Kar from the Indian
Statistical Institute, Kolkata, India for their help, stimulating
suggestions and encouragements which helped me in the time of
research and writing of this thesis. Again I like to thank Dr.
A.K.Pati and his Institute IOP (Bhubaneswar,India) for giving worm hospitality and facilities for some works.\\
I would like to thank all people who have taught me mathematics:
my school math teachers, my undergraduate teachers and my postgraduate teachers.\\
I wish to extend my warmest thanks to all the academic and support
staffs who have helped me with my work in the Department of
Mathematics, Bengal Engineering and Science University, Shibpur, West Bengal, India.\\
I am very grateful to Samit De who is working in the library of
Saha Institute of Nuclear Physics, Kolkata, India for helping me
in many different ways.\\
I wish to thank my friends for helping me get through the
difficult times, and for all the emotional supports,
entertainments, academic discussions and care they provided.\\
This research has been supported and funded by Council of
Scientific and Industrial Research, government of India, under
project F.No.8/3(38)/2003-EMR-1, New Delhi. I would like to thank
them for their confidence in me.\\
Lastly, and most importantly, I feel a deep sense of gratitude to
my Father and Mother who formed part of my vision and taught me
the good things that really matter in life. They bore me, raised
me, supported me, taught me and loved me. To them I dedicate this
thesis.

\vspace*{2cm}
\noindent Satyabrata Adhikari\\
Department of Mathematics\\
Bengal Engineering and Science University, Shibpur \\
Howrah - 711103, West Bengal\\
India

\newpage
\vspace*{1cm}
\begin{center}
 {\large {\bf\Large \textit{Preface}} }
\end{center}
\vspace*{1cm}

Quantum ideas when incorporated into the domains of computation
and information science have revolutionized the subject. In terms
of potential performance it is recognized that quantum computation
and information theories are far richer than their classical
counterparts. There are two important results of quantum
mechanics, namely impossibility of creating exact copies of
quantum states, commonly known as 'no-cloning theorem' and the
impossibility of deleting one of the two given identical quantum
states commonly known as 'no-deletion principle' have
simultaneously enhanced the scope of such theories in certain
respects and restricted the theories in other directions. This
thesis is concerned with some aspects of quantum cloning and
deletion and their implications in quantum computation and
information problems.
The thesis is organized into five chapters. In the following we state the chapter wise summaries of the thesis:\\
In chapter-1, general introduction is given. In this introduction
we discuss about various existing quantum cloning machines and quantum deletion machines.\\
In chapter-2, we discuss the combination of independently existing
quantum cloning machines known as hybrid quantum cloning machine.
In this chapter we study the state dependent and state independent hybrid quantum cloning machines.\\
Chapter-3 deals with the cloning of entanglement using local
quantum cloners. This concept is popularly known as broadcasting
of entanglement. We discuss the broadcasting of two-qubit
entanglement using state dependent quantum cloners. Further we
find that here that the broadcasting of three-qubit entanglement is also possible.\\
In chapter-4, we investigate about the universal quantum deletion
machine. We show that universal quantum deletion machine exists
with better fidelity of deletion only if an additional
unitary operator called transformer is added with the unitary operator called deleter.\\
In chapter-5, we study the effect on an arbitrary qubit as a
result of concatenation of two quantum operations viz. unitary
quantum cloning and deleting transformations.\\
Lastly, the list of publications are given and the list of
references are given under the broad heading 'Bibliography'. We
also have attached copies of our five published and
one communicated works on which the thesis is based.\\

\baselineskip .7cm
\tableofcontents

\newpage

\pagenumbering{arabic}

\large \baselineskip .85cm

\chapter{General Introduction}
\setcounter{page}{1} \markright{\it CHAPTER~\ref{chap1}. General
Introduction}
\label{chap1}%
All of modern physics is governed by that magnificent and
thoroughly confusing discipline called quantum mechanics ... It
has survived all tests and there is no reason to believe that
there is any flaw in it.... We all know how to use it and how to
apply it to problems; and so we have learned to live with the fact
that nobody can understand it - Murray Gell-Mann

\section{\emph{Entanglement: A Non-Local Resource}}
\subsection{\emph{What is entanglement?}}
In 1935, Einstein, Podolsky and Rosen (EPR) \cite{einstein1}
presented a paradox that still surprises us today. Consider two
physical systems that once interacted but are remote from each
other now and do not interact. The two systems are still entangled
if their quantum state does not factor into a product of states of
each system. Entangled particles have correlated properties, and
these correlations are at the heart of the paradox. Entanglement
between quantum systems is a purely quantum mechanical phenomenon.
It is closely related to the superposition principle and describes
correlation between quantum systems that are much stronger and
richer than any classical correlation could be.
Mathematically entanglement can be defined in a following way:\\
Let us consider a system consisting of two subsystems where each
subsystem is associated with a Hilbert space. Let $H_{A}$ and
$H_{B}$ denote these two Hilbert spaces. Let $|i\rangle_{A}$ and
$|j\rangle_{B}$ (where i,j=1,2,3,.....) represent two complete
orthonormal basis for $H_{A}$ and $H_{B}$ respectively. The two
subsystems taken together is associated with the Hilbert space
$H_{A}\otimes H_{B}$, spanned by the states
$|i\rangle_{A}\otimes|j\rangle_{B}$. Any linear combination of the
basis states $|i\rangle_{A}\otimes|j\rangle_{B}$ is a state of the
composite system AB and any pure state $|\psi\rangle_{AB}$ of the
system can be written as
\begin{eqnarray}
|\psi\rangle_{AB}=
\sum_{i,j}c_{ij}|i\rangle_{A}\otimes|j\rangle_{B}
\end{eqnarray}
where the $c_{ij}s$ are complex coefficients and $\sum_{i,j}|c_{ij}|^{2}=1$.\\
If $|\psi\rangle_{AB}$ factors into a normalized state
$|\psi\rangle_{A}=\sum_{i}^{k=dim(H_{A})}c_{i}|i\rangle_{A}$ in
$H_{A}$ and a normalized state
$|\psi\rangle_{B}=\sum_{i}^{k=dim(H_{B})}c_{j}|j\rangle_{B}$ in
$H_{B}$, i.e. $|\psi\rangle_{AB}=|\psi\rangle_{A}\otimes
|\psi\rangle_{B}$, then the state $|\psi\rangle_{AB}$ is called
separable state or product state.\\
If a state belonging to the Hilbert space $H_{A}\otimes H_{B}$ is
not a product state, then such a state is called entangled state.
\subsection{\emph{Pure state entanglement}}
If $|\psi\rangle_{AB}$ represents a pure state of a composite
system consisting of two Hilbert spaces $H_{A}$ and $H_{B}$, then
$|\psi\rangle_{AB}$ can always be written in Schmidt form (Schmidt
decomposition) \cite{nielsen2} as
\begin{eqnarray}
|\psi\rangle_{AB}= \sum_{i}^{k\leq min\{dim H_{A},dim
H_{B}\}}\sqrt{\lambda_{i}}|i\rangle_{A}\otimes|i\rangle_{B}
\end{eqnarray}
where $|i\rangle_{A}$, $|i\rangle_{B}$ are two orthonormal bases
of systems A and B respectively and $\lambda_{i}\geq0$,
$\sum\lambda_{i}=1$. The non-negative real numbers $\lambda_{i}$
are known as Schmidt coefficients. If two or more Schmidt
coefficients are non-zero, then the state $|\psi\rangle$ is
referred to as Pure entangled state. If only one Schmidt
coefficient is non-zero and all others are zero, then the state
$|\psi\rangle$ is called product state. In particular, two qubit
pure state can be written in the schmidt form as
\begin{eqnarray}
|\psi\rangle_{AB}=
\sum_{i=1}^{2}\sqrt{\lambda_{i}}|i\rangle_{A}\otimes|i\rangle_{B}
\end{eqnarray}
where the Schmidt coefficients $\lambda_{1},\lambda_{2}$ satisfy
the normalization condition i.e. $\lambda_{1}+\lambda_{2}=1$.\\
It has been shown that every pure entangled state violates some
Bell-type inequality \cite{gisin5}, while no product state does.
Entangled states cannot be prepared from unentangled states by any
sequence of local actions of two distant partners, even with the
help of classical communication. The most familiar example of pure
entangled state is the singlet state of two spin-$\frac{1}{2}$
particles
\begin{eqnarray}
|\psi^{-}\rangle=
\frac{1}{\sqrt{2}}(|\uparrow\downarrow\rangle-|\downarrow\uparrow\rangle)
\end{eqnarray}
which cannot be reduced to direct product by any transformation of
the bases pertaining to each one of the particles.
\subsection{\emph{Measure of Pure state entanglement}}
\textbf{1. Entropy of entanglement \cite{bennett5}:} Let Alice(A)
and Bob(B) share a pure entangled state $|\psi\rangle_{AB}$.
Quantitatively, a pure state's entanglement is conveniently
measured by its entropy of entanglement,
\begin{eqnarray}
E|\psi_{AB}\rangle= S(\rho_{A})= S(\rho_{B})
\end{eqnarray}
Here $S(\rho)=-Tr(\rho log_{2}\rho)$ is the von-Neumann entropy
and $\rho_{A}= Tr_{B}(|\psi\rangle_{AB}\langle\psi|),
\rho_{B}=Tr_{A}(|\psi\rangle_{AB}\langle\psi|)$ denote the reduced
density matrices obtained by tracing the whole system's pure-state
density matrix $|\psi\rangle_{AB}\langle\psi|$ over Bob's and
Alice's degrees of freedom respectively.
\subsection{\emph{Mixed state entanglement}}
Due to decoherence effect we usually deal with mixed states. A
mixed state of quantum system consisting of two subsystems is
supposed to represent entanglement if it is inseparable
\cite{bennett5,horo4,ishizaka1,kent1,osborne1,werner2} i.e. cannot
be written in the form
\begin{eqnarray}
\rho=\sum_{i}p_{i}(\rho_{i}^{A}\otimes\rho_{i}^{B}),~~~~~~p_{i}\geq0,~~~\sum_{i}p_{i=1}
\end{eqnarray}
where $\rho_{i}^{A}$ and $\rho_{i}^{B}$ are states for the two
subsystems A and B respectively.  A test for separability of
$(2\times2)$ systems is the Peres-Horodecki criterion
\cite{horo2,peres1}, which states that a necessary and sufficient
condition for the state $\hat{\rho}$ of two spins ½ to be
inseparable is that at least one of the eigen values of the
partially transposed operator defined as
$\rho_{m\mu,n\nu}^{T_{2}}= \rho_{m\nu,n\mu}$ is negative. This is
equivalent to the condition that at least one
of the two determinants\\
$W_3$ = \begin{tabular}{|c c c|}
 $\rho_{00,00}$ & $\rho_{01,00}$ & $\rho_{00,10}$ \\
 $\rho_{00,01}$ & $\rho_{01,01}$ & $\rho_{00,11}$ \\
 $\rho_{10,00}$ & $\rho_{11,00}$ & $\rho_{10,10}$ \\
\end{tabular}~~~~  and $W_4$= \begin{tabular}{|c c c c|}
  $\rho_{00,00}$ & $\rho_{01,00}$ & $\rho_{00,10}$ & $\rho_{01,10}$ \\
  $\rho_{00,01}$ & $\rho_{01,01}$ & $\rho_{00,11}$ & $\rho_{01,11}$ \\
  $\rho_{10,00}$ & $\rho_{11,00}$ & $\rho_{10,10}$ & $\rho_{11,10}$ \\
  $\rho_{10,01}$ & $\rho_{11,01}$ & $\rho_{10,11}$ & $\rho_{11,11}$ \\
\end{tabular}~~ is negative \\\\ and
$W_2$ = \begin{tabular}{|c c|}
 $\rho_{00,00}$ & $\rho_{01,00}$ \\
 $\rho_{00,01}$ & $\rho_{01,01}$ \\
 \end{tabular} ~~~is non-negative.\\
 We will use these conditions for inseparability in the subsequent
 chapter.
\subsection{\emph{Measure of Mixed state entanglement}}
The fundamental law of quantum information processing says that
the mean entanglement cannot be increased under local operation
and classical communication (LOCC). The law actually says that
there is some probability with which the two distant partners can
obtain more entangled state. Then, however, with some other
probability they will obtain less entangled states so that on
average the mean entanglement will not increase. Under LOCC
operations, one can only change the form of entanglement. It is
known that concentration of entanglement is possible using local
operation and classical communication. Therefore, to measure the
efficiency with which one can perform this concentration, some
measures of entanglement is introduced. Entanglement measures
answer the following question 'how much entanglement is needed to
create a given quantum state by local operation and classical
communication alone?' or inversely 'how many singlets one can
prepare from a supply of non-maximally entangled states?'.\\
Now we have listed below the conditions which every 'decent' measure of entanglement should satisfy.\\
C1: The measure of entanglement for any separable state should be zero, i.e. $E(\rho)=0$.\\
C2: The amount of entanglement in any state $\rho$ should be
unaffected for any local unitary transformation of the form $U_{A}
\otimes U_{B}$, i.e. $E(\rho)=
E(U_{A} \otimes U_{B}\rho U_{A}^{\dagger} \otimes U_{B}^{\dagger})$.\\
C3: Local operations, classical communication and sub-selection
cannot increase the expected entanglement i.e.
$E(\rho)\geq\sum_{i}p_{i}E(\rho_{i})$, where $p_{i}$ denotes
the probability with which the state $\rho_{i}$ occurs.\\
C4: For any two given pairs of entangled particles in the total
state $\rho=\rho_{1}\otimes\rho_{2}$, we should have
$E(\rho)=E(\rho_{1})+E(\rho_{2})$.\\
There are many measures of mixed state entanglement but in this thesis, we mention briefly just four of them.\\
\textbf{1. Entanglement of formation
\cite{chen1,fei1,plenio1,wootters2}:} The entanglement of
formation $E_{F}$ of a bipartite mixed state $\rho_{AB}$ is
defined by
\begin{eqnarray}
E_{F}(\rho_{AB})=
min_{\rho_{AB}=\sum_{i}p_{i}|\psi_{i}\rangle_{AB}\langle\psi_{i}|}\sum_{i}p_{i}E(|\psi_{i}\rangle\langle\psi_{i}|)
\end{eqnarray}
The minimization in equation (1.7) is taken over all possible
decompositions of the density operator $\rho_{AB}$ into pure
states $|\psi\rangle$. Entanglement of formation gives an upper
bound on the efficiency of purification procedures. In addition it
also gives the amount of entanglement that has to be
used to create a given quantum state.\\
\textbf{2. Relative entropy of Entanglement
\cite{audenaert1,miranowicz1,plenio1}:} The relative entropy of
entanglement are based on distinguishability and geometrical
distance. The idea is to compare a given quantum state $\sigma$ of
a pair of particles with separable states. The relative entropy of
entanglement of a given state $\sigma$ is defined by
\begin{eqnarray}
E_{RE}(\sigma)= min_{\rho\epsilon M}D(\sigma||\rho)
\end{eqnarray}
Here 'M' denotes the set of all separable states and D can be any
function that describes a measure of separation between two
density operators. A particular form of the function D is the
relative entropy which is defined as $S(\sigma||\rho)=Tr\{\sigma
ln \sigma- \sigma ln\rho\}$.\\
\textbf{3. Concurrence
\cite{audenaert2,miranowicz1,verstraete1,wootters2}:} Wootters
gave out, for the mixed state $\hat{\rho}$ of two qubits, the
concurrence is
\begin{eqnarray}
C = max ( \lambda_{1}-\lambda_{2}-\lambda_{3}-\lambda_{4}, 0)
\end{eqnarray}
where the $\lambda_{i}$, in decreasing order, are the square roots
of the eigen values of the matrix
$\rho^{\frac{1}{2}}(\sigma_{y}\otimes\sigma_{y})\rho^{*}
(\sigma_{y}\otimes\sigma_{y})\rho^{\frac{1}{2}}$ with $\rho^{*}$
denoting the complex conjugation of $\rho$ in the computational
basis $\{|00\rangle,|01\rangle,|10\rangle,|11\rangle\}$ and
$\sigma_{y}$ denoting the Pauli operator.\\
The entanglement of formation $E_{F}$ can then be expressed as a
function of C given in (1.9), namely
\begin{eqnarray}
E_{F}=
-\frac{1+\sqrt{1-C^{2}}}{2}~~log_{2}\frac{1+\sqrt{1-C^{2}}}{2}-
\frac{1-\sqrt{1-C^{2}}}{2}~~log_{2}\frac{1-\sqrt{1-C^{2}}}{2}
\end{eqnarray}
The concurrence and entanglement of formation satisfy convexity.
For the two qubit state given in equation (1.3), the concurrence
is given by
\begin{eqnarray}
C = 2\sqrt{\lambda_{1}\lambda_{2}}
\end{eqnarray}
The concurrence in equation (1.11) can also be written as
\begin{eqnarray}
C = \sqrt{2(1-Tr\rho_{A}^{2})}
\end{eqnarray}
\textbf{4. Negativity \cite{audenaert2,miranowicz1,verstraete1}:}
The negativity $E_{N}$ of a mixed state $\rho$ is defined by
\begin{eqnarray}
E_{N}(\rho)= 2 \sum_{j}max(0,-\mu_{j})
\end{eqnarray}
where $\mu_{j}$ are the eigenvalues of the partial transpose
$\rho^{\Gamma}$ of the density matrix $\rho$ of the system.\\
The negativity \cite{vidal2} of a bipartite system described by
the density matrix $\rho$ can also be expressed in the form as
\begin{eqnarray}
E_{N}(\rho)= \frac{\parallel\rho^{T_{A}}\parallel_{1}-1}{2}
\end{eqnarray}
where $\rho^{T_{A}}$ is the partial transpose with respect to the subsystem A, and $||..||$ denotes the
trace norm.\\
Negativity does not change under LOCC. It measures how negative
the eigenvalues of the density matrix are after the partial
transpose is taken. For pure states it has been proven that the
negativity is exactly equal to the concurrence \cite{vidal1}. For
mixed states, Eisert and Plenio \cite{eisert1} conjectured that
negativity never exceeds concurrence. K.Audenaert, F.Verstraete,
T.D.Bie and B.D.Moor \cite{audenaert2} proved the conjecture of
Eisert and Plenio that concurrence can never be smaller than
negativity. For higher dimension, the negativity can be
generalized as \cite{lee1}
\begin{eqnarray}
E_{N}(\rho)= \frac{\parallel\rho^{T_{A}}\parallel_{1}-1}{d-1}
\end{eqnarray}
where d is the smaller of the dimensions of the bipartite system.
\subsection{\emph{Entanglement swapping}}
Entanglement swapping \cite{bose1,zukowski1}  is a method that
enables one to entangle two quantum systems that do not have
direct interaction with one another. In order to illustrate
entanglement swapping, we first define Bell states as
$\phi^\pm\equiv\frac{(|00\rangle\pm|11\rangle)}{\sqrt{2}}$ and
$\psi^\pm\equiv\frac{(|01\rangle\pm|10\rangle)}{\sqrt{2}}$.
Suppose two distant parties, Alice and Bob, share $\phi_{12}^{+}$
and $\phi_{34}^{+}$ where Alice has qubits '1' and '4', and Bob
possesses '2' and '3'. A measurement is performed on qubits '2'
and '3' with the Bell basis $\{\phi^\pm,\psi^\pm\}$, then the
total state $\phi_{12}^{+}\otimes\phi_{34}^{+}$ is projected onto
$|\eta_1\rangle=\phi_{23}^{+}\otimes\phi_{14}^{+},
|\eta_2\rangle=\phi_{23}^{-}\otimes\phi_{14}^{-},
|\eta_3\rangle=\psi_{23}^{+}\otimes\psi_{14}^{+},
|\eta_4\rangle=\psi_{23}^{-}\otimes\psi_{14}^{-}$ with equal
probability of $\frac{1}{4}$ for each. Previous entanglement
between qubits '1' and '2', and '3' and '4' are now swapped into
entanglement between qubits '2' and '3', and '1' and '4'. Although
we considered entanglement swapping with the initial state
$\phi_{12}^{+}\otimes\phi_{34}^{+}$, similar results can be
achieved with other Bell states.\\
S.Bose et.al. \cite{bose1} generalized the procedure of
entanglement swapping and obtained a scheme for manipulating
entanglement in multiparticle systems. They showed that this
scheme can be regarded as a method of generating entangled states
of many particles. An explicit scheme that generalizes
entanglement swapping to the case of generating a 3-particle GHZ
state from three Bell pairs has been presented by Zukowski et.al.
\cite{zukowski1} The standard entanglement swapping helps to save
a significant amount of time when one wants to supply two distant
users with a pair of atoms or electrons (or any particle
possessing mass) in a Bell state from some central source. The
entanglement swapping can be used, with some probability which we
quantify, to correct amplitude errors that might develop in
maximally entangled states during propagation.
\section{\emph{Distance measure between quantum states}}
\subsection{ \emph{Fidelity}}
A measure of distance between quantum states is the fidelity
\cite{nielsen2}. The fidelity of states $\rho$ and $\sigma$ is
defined to be
\begin{eqnarray}
F(\rho,\sigma)\equiv Tr
\sqrt{\rho^{\frac{1}{2}}\sigma\rho^{\frac{1}{2}}}
\end{eqnarray}
When $\rho$ and $\sigma$ commute the quantum fidelity
$F(\rho,\sigma)$ reduces to the classical fidelity.\\
The fidelity between a pure state $|\psi\rangle$ and an arbitrary
state $\rho$ is defined by
\begin{eqnarray}
F(|\psi\rangle,\rho) = \sqrt{\langle\psi|\rho|\psi\rangle}
\end{eqnarray}
\textbf{Properties of Fidelity:}\\
1. The fidelity is symmetric in
its inputs, i.e. $F(\rho,\sigma)=F(\sigma,\rho)$.\\
2. The fidelity $F(\rho,\sigma)$ lies between 0 and 1 (including
0 and 1), i.e. $0\leq F(\rho,\sigma)\leq 1$.\\
(i)$F(\rho,\sigma)=0$, iff $\rho$ and $\sigma$ have support on
orthogonal subspaces.\\
(ii) $F(\rho,\sigma)=1$, iff $\rho=\sigma$.\\
3. The fidelity is not a metric but it can be converted into a
metric by suitably defining it. If we define the angle between
states $\rho$ and $\sigma$ by $A(\rho,\sigma)\equiv arccos
F(\rho,\sigma)$, then fidelity turn out to be a metric
because the angle between two points on the sphere is a metric.\\
4. Fidelity is monotonic under quantum operation.\\
5. Fidelity has the strong concavity property.
\subsection{\emph{Hilbert-Schmidt (H-S) distance}}
The Hilbert-Schmidt distance \cite{filip2,ozawa1} is defined by
\begin{eqnarray}
D_{HS}(\sigma,\rho)=\parallel\sigma-\rho\parallel^{2}=Tr[(\sigma-\rho)^{2})]
\end{eqnarray}
The Hilbert-Schmidt distance defined in (1.18) satisfies all the criterion of the distance function D
which is defined below:\\
Let S be set of density operators on the Hilbert space H.\\
Let $D:S\times S\rightarrow R$ be a function satisfying the
following conditions:\\
D1. $D(\sigma,\rho)\geq 0$ for any $\sigma,\rho~\epsilon~S$ and
the equality holds when $\sigma=\rho$.\\
D2. $D(\Theta \sigma,\Theta\rho)\leq  D(\sigma,\rho)$ for any
$\sigma,\rho~\epsilon~S$ and for any completely positive trace
preserving map $\Theta$ on the space of operators on H.\\
H-S distance is easier to calculate and also it serves as a good
measure of quantifying the distance between the pure states. It is
conjectured that the Hilbert-Schmidt distance is a reasonable
candidate of a distance to generate an entanglement measure
\cite{vedral2}. Also it is shown that the quantum relative
entropy and the Bures metric satisfy (D1) and (D2) \cite{vedral1}.\\
Later in this thesis we will use frequently the H-S distance
measure.
\section{\emph{No-Cloning Theorem: A Brief History}}
\subsection{\emph{Wigner's  prediction}}
In 1961, E.P.Wigner \cite{wigner1}, assumed that there be a
'living state' $|\psi\rangle$ which is given in a quantum
mechanical sense in a finite dimensional Hilbert space $H^{N}$. He
then noticed that, among all the possible unitary transformations,
those that transform $|\Psi_{i}\rangle =|\psi\rangle |w\rangle$ to
$|\Psi_{f}\rangle=|\psi\rangle |\psi\rangle|r\rangle$, where
$|r\rangle$ is the rejected part of the nutrient state
$|w\rangle$, are a negligible set but he failed to notice that no
transformation realizes that task for arbitrary $|\psi\rangle$,
which would have been the no-cloning theorem. From his observation
Wigner concluded that biological reproduction "appears to be a
miracle from the point of view of the physicist". But nowadays we
know that his description of the living state as a quantum
mechanical state is not correct because quantum mechanics would
not permit the accurate replication of biological information. On
the other hand genetic information encoded in a living state can
be safely copied because it is never encoded in superposition
states - they would be instantly destroyed by decoherence
\cite{zurek1}. Pati \cite{pati11} had shown that Wigner's
replicating machine for a species can be ruled out simply based on
the linearity of the quantum theory. Thus there does not exist any
replicating machine for a living organism in the quantum
mechanical sense.
\subsection{\emph{Herbert's  Argument}}
In 1982, Nick Herbert \cite{herbert1} put forward an
unconventional proposal "FLASH", where he used quantum
correlations \cite{einstein1} to communicate faster than light.
The word FLASH is an acronym for "First Light Amplification
Superluminal Hookup". His apparatus consisted of idealized laser
gain tube which would have macroscopically distinguishable outputs
when the input was a single arbitrarily polarized photon. The
output of the apparatus contains noise which is due to the
combined effect of 'stimulated emission' and 'spontaneous
emission'. In spite of the production of noise his claim was that
at least statistically, the noise would not prevent identifying
the polarization of the
incoming photon. Herbert's argument can also be given in the following way:\\
Let us consider two distant parties, Alice and Bob, sharing two qubits in the singlet
state
\begin{eqnarray}
|\psi^{-}\rangle_{AB}=\frac{1}{\sqrt{2}}(|01\rangle_{AB}-|10\rangle_{AB})
\end{eqnarray}
The first qubit A belongs to Alice and the second qubit B belongs
to Bob. Now, Alice measures $\sigma_{x}$ or $\sigma_{z}$ in the
basis $\{|+\rangle,|-\rangle\}$ or $\{|0\rangle,|1\rangle\}$ on
her qubit, where $|\pm\rangle=
\frac{1}{\sqrt{2}}(|0\rangle\pm|1\rangle$.\\
If Alice measures $\sigma_{x}$ on her qubit, then
\begin{eqnarray}
|\psi^{-}\rangle_{AB}~~~ \textrm{transforms to}~~~
\frac{-1}{\sqrt{2}}[(\sigma_{x}|-\rangle)_{A}\otimes|+\rangle_{B}+
(\sigma_{x}|+\rangle)_{A}\otimes|-\rangle_{B}]
\end{eqnarray}
If Alice measures $\sigma_{z}$ on her qubit, then
\begin{eqnarray}
|\psi^{-}\rangle_{AB}~~~ \textrm{transforms
to}~~~\frac{1}{\sqrt{2}}[(\sigma_{z}|0\rangle)_{A}\otimes|1\rangle_{B}+
(\sigma_{z}|1\rangle)_{A}\otimes|0\rangle_{B}]
\end{eqnarray}
After Alice's measurement and without any communication with her,
Bob finds his qubit in the random mixture. For the first case
(1.20), Bob finds his qubit in the mixed state described by the
density operator
$\frac{1}{2}(|+\rangle\langle+|+|-\rangle\langle-|)=\frac{1}{2}I$
and for the second case (1.21), he finds his qubit in the random
mixture
$\frac{1}{2}(|0\rangle\langle0|+|1\rangle\langle1|)=\frac{1}{2}I$.
Now Bob cannot identify the basis in which Alice perform her
measurement for the following two reasons: (i) The mixed state
appeared in Bob's place due to Alice's measurement is random in
both cases and (ii) We are taking into account the fact that there
is no classical communication in between them.\\
But if Bob uses his perfect cloner to copy his qubit, then according to the
measurement performed by Alice, Bob finds his qubit in two different mixed states.\\
If Alice measures $\sigma_{x}$, then the mixture at Bob's place is given by
\begin{eqnarray}
\rho_{x}&=& \frac{1}{2}(|++\rangle\langle++|+|--\rangle\langle--|){}\nonumber\\&=&
\frac{1}{4}(|00\rangle\langle00|+
|00\rangle\langle11|+|01\rangle\langle01|+|01\rangle\langle10|+|10\rangle\langle01|
+|10\rangle\langle10|+|11\rangle\langle00|+{}\nonumber\\& & |11\rangle\langle11|)
\end{eqnarray}
If Alice measures $\sigma_{z}$, then the mixed state at Bob's place is given by the
reduced density operator
\begin{eqnarray}
\rho_{z}= \frac{1}{2}(|00\rangle\langle00|+|11\rangle\langle11|)
\end{eqnarray}
From equations (1.22) and (1.23), it is clear that
$\rho_{x}\neq\rho_{z}$. Therefore, by applying the perfect cloner,
Bob can distinguish the measurement performed by Alice without any classical communication with her.\\
Putting the argument in this way, Herbert thought that the two
distant partners can communicate with each other faster than the
speed of light. But his thought experiment was proved to be wrong
by Wootters and Zurek.
\subsection{\emph{Wootters and Zurek No-cloning Theorem}}
In 1982, Wootters and Zurek \cite{wootters1}, in their pioneering
work proved that "an arbitrary quantum state cannot be cloned".
This theorem is popularly known as 'no-cloning theorem'.
Therefore, no-cloning theorem ruled out the Herbert's argument on
superluminal communication between the two distant partners. The
impossibility of cloning of an unknown qubit makes quantum
information different from classical information. In quantum
regime, no-cloning theorem only prohibits the construction of an
apparatus which will amplify arbitrary non-orthogonal states but
it does not rule out the possibility of a device which amplifies
the orthogonal states. Since orthogonal states can be thought of
as a different states of classical information so there is no
question of contradiction between the no-cloning theorem and the
cloning of classical information. Next we will give the proof of
no-cloning theorem using linearity and unitarity of the
quantum mechanics.\\
\textbf{1. Proof of No-cloning theorem by Linearity}\\
If possible, let there exist a perfect amplifying device which would have the
following effect on an incoming arbitrary quantum state $|s\rangle$:
\begin{eqnarray}
|s\rangle|\Sigma\rangle|Q_{i}\rangle\rightarrow|s\rangle|s\rangle|Q_{f}\rangle
\end{eqnarray}
where $|Q_{i}\rangle$ and $|Q_{f}\rangle$ are the initial and
final state of the device respectively. $|\Sigma\rangle$
represents a blank state on which the input state is copied.\\
Let an arbitrary pure quantum state $|s\rangle$ be given by
\begin{eqnarray}
|s\rangle = \alpha|0\rangle+\beta|1\rangle
\end{eqnarray}
with $\alpha^{2}+|\beta|^{2}=1$.\\
The action of the amplifier on the two orthogonal states
$|0\rangle$ and $|1\rangle$ separately is given by
\begin{eqnarray}
|0\rangle|\Sigma\rangle|Q_{i}\rangle\rightarrow|0\rangle|0\rangle|Q_{0}\rangle
\end{eqnarray}
\begin{eqnarray}
|1\rangle|\Sigma\rangle|Q_{i}\rangle\rightarrow|1\rangle|1\rangle|Q_{1}\rangle
\end{eqnarray}
Now due to the linear structure of the quantum mechanics, the
interaction between the amplifying device and the state (1.25) is
given by
\begin{eqnarray}
|s\rangle|\Sigma\rangle|Q_{i}\rangle \rightarrow
\alpha|00\rangle|Q_{0}\rangle+\beta|11\rangle|Q_{1}\rangle
\end{eqnarray}
Since we assume that the amplifier is perfect, so
\begin{eqnarray}
|s\rangle|\Sigma\rangle|Q_{i}\rangle\rightarrow|s\rangle|s\rangle|Q_{f}\rangle=
(\alpha^{2}|00\rangle+\alpha\beta|01\rangle+\beta\alpha|10\rangle+\beta^{2}|11\rangle)|Q_{f}\rangle
\end{eqnarray}
Therefore, in general, the equations (1.28) and (1.29) are not
identical. Thus, we arrive at a contradiction. Hence, there does
not exist any perfect amplifier which could copy an arbitrary quantum state.\\
The equations (1.28) and (1.29) are identical only when $\alpha=0$
or $\beta=0$. This observation tells us that the information
stored in the state $|0\rangle$ or $|1\rangle$ can be perfectly
copied while the information stored
in the arbitrary superposition of the states $|0\rangle$ and $|1\rangle$ cannot be perfectly copied.\\\\
\textbf{2. Proof of No-cloning theorem by Unitarity}\\
Let us assume that we have a perfect quantum cloning machine which can copy an unknown
quantum state $|\psi\rangle$. Suppose $|\Sigma\rangle$ represents a blank state onto
which the unknown quantum state is to be copied. Therefore, before interaction with
the copying machine, the joint state is given by
\begin{eqnarray}
|\psi\rangle\otimes|\Sigma\rangle
\end{eqnarray}
Let U be the unitary evolution which governs the copying procedure
of the perfect quantum cloning machine and its effect on the
combined state (1.30) is given by
\begin{eqnarray}
U(|\psi\rangle\otimes|\Sigma\rangle)=|\psi\rangle\otimes|\psi\rangle
\end{eqnarray}
Suppose equation (1.31) is valid for two pure unknown quantum
states, $|\xi\rangle$ and $|\eta\rangle$. Then we have
\begin{eqnarray}
U(|\xi\rangle\otimes|\Sigma\rangle)=|\xi\rangle\otimes|\xi\rangle
\end{eqnarray}
\begin{eqnarray}
U(|\eta\rangle\otimes|\Sigma\rangle)=|\eta\rangle\otimes|\eta\rangle
\end{eqnarray}
The inner product of (1.32) and (1.33) gives
\begin{eqnarray}
\langle\eta|\xi\rangle=(\langle\eta|\xi\rangle)^{2}{}\nonumber\\
\Rightarrow\langle\eta|\xi\rangle(1-\langle\eta|\xi\rangle)=0{}\nonumber\\
\Rightarrow\textrm{ either }\langle\eta|\xi\rangle=0\textrm{ or
}\langle\eta|\xi\rangle=1
\end{eqnarray}
Equation (1.34) implies that the quantum states $|\xi\rangle$ and
$|\eta\rangle$ are either orthogonal or identical. Therefore,
perfect quantum cloner can be constructed for only orthogonal
states $|0\rangle$ and $|1\rangle$.
\section{\emph{Description of quantum cloning machines}}
\subsection{\emph{State dependent quantum cloning machines}}
In this section, we study the state dependent quantum cloning
machines. When the quality of the copies at the output of the
quantum cloner depends on the input state, the machine is said to
be state dependent quantum cloning machine
\cite{bruss2,chefles1,han1,wootters1}.\\
\textbf{\textrm{1. Wootters and Zurek quantum cloning machine}}\\
In the computational basis states $|0\rangle$ and $|1\rangle$,
Wootters and Zurek quantum cloning transformation is given by,
\begin{eqnarray}
|0\rangle_{a}|\Sigma\rangle_{b}|Q_{i}\rangle_{x}\longrightarrow|0\rangle_{a}|0\rangle_{b}|Q_0\rangle_{x}\\
|1\rangle_{a}|\Sigma
\rangle_{b}|Q_{i}\rangle_{x}\longrightarrow|1\rangle_{a}|1\rangle_{b}|Q_1\rangle_{x}
\end{eqnarray}
The system lebeled by 'a' is the input mode, while the other
system 'b' represents the qubit onto which information is copied
and is analogous to "blank paper" in a copier. $|Q_{i}\rangle_{x}$
is the initial machine state vector and $|Q_0\rangle_{x}$ and
$|Q_1\rangle_{x}$ are the final machine state vectors.\\
Unitarity of the transformation gives,
\begin{eqnarray}
\langle Q_{i}|Q_{i}\rangle=\langle Q_0|Q_0\rangle=\langle Q_1|Q_1\rangle=1
\end{eqnarray}
Let us now consider pure superposition state given by,
\begin{eqnarray}
|\chi\rangle_{a}=\alpha|0\rangle_{a}+\beta|1\rangle_{a}
\end{eqnarray}
For simplicity we will assume the probability amplitudes to be
real and $\alpha^2+\beta^2=1$.\\
The density matrix of the state $|\chi\rangle$ in the input mode 'a' is given by,
\begin{eqnarray}
\rho_{a}^{id}=|\chi\rangle\langle
\chi|=\alpha^2|0\rangle\langle0|+\alpha\beta|0\rangle\langle1|+\alpha\beta
|1\rangle\langle0|+\beta^2|1\rangle\langle1|
\end{eqnarray}
After applying the cloning transformation (1.35-1.36) the
arbitrary quantum state $|\chi\rangle$ given in equation (1.38)
takes the form
\begin{eqnarray}
|\chi^{out}\rangle_{abx}=\alpha|0\rangle_{a}|0\rangle_{b}|Q_0\rangle_{x}+\beta|1\rangle_{a}|1\rangle_{b}
|Q_1\rangle_{x}
\end{eqnarray}
If it is assumed that two copying machine states $|Q_0\rangle$ and
$|Q_1\rangle$ are orthonormal then the reduced density operator
$\rho_{ab}^{(out)}$ is given by
\begin{eqnarray}
\rho_{ab}^{(out)}=Tr_x[\rho_{abx}^{(out)}]=\alpha^2|00\rangle\langle00|+\beta^2|11\rangle\langle11|
\end{eqnarray}
where $\rho_{abx}^{(out)}=|\chi^{out}\rangle_{abx}\langle\chi^{out}|$.\\
The reduced density operators describing the original and the copy
mode are given by,
\begin{eqnarray}
\rho_{a}^{(out)}=Tr_b[\rho_{ab}^{(out)}]=\alpha^2|0\rangle\langle0|+\beta^2|1\rangle\langle1|\\
\rho_{b}^{(out)}=Tr_a[\rho_{ab}^{(out)}]=\alpha^2|0\rangle\langle0|+\beta^2|1\rangle\langle1|
\end{eqnarray}
The copying quality i.e. the distance between the density matrix
of the input state $\rho_{a}^{(id)}$ and the reduced density
matrix $\rho_{a}^{(out)}$ ($\rho_{b}^{(out)}$) of the output state
can be measured by Hilbert-Schmidt norm given by
\begin{eqnarray}
D_a=Tr[\rho_{a}^{(id)}-\rho_{a}^{(out)}]^2
\end{eqnarray}
In this case,
\begin{eqnarray}
D_a=2\alpha^2\beta^2=2\alpha^2(1-\alpha^2)
\end{eqnarray}
$D_{a}$ is called copy quality index.\\
Wootters and Zurek (W-Z) quantum cloning machine can be regarded
as a state- dependent quantum cloning machine because the distance
between the pure states depends on the input state parameter
$\alpha$. Thus, for some values of $\alpha$, the distance is
minuscule while for some values of $\alpha$, the distance is very
large. Hence Wootters and Zurek (W-Z) quantum cloning machine
works perfectly for some inputs and badly for some others. Since
$D_a$ depends on $\alpha^2$, the average distortion is calculated
over all input states, i.e., over all $\alpha^2$ lying between 0
and 1, which is
\begin{eqnarray}
\overline{D_a}=\int^1_0 D_a(\alpha^2)d\alpha^2=\frac{1}{3}
\end{eqnarray}\\\\
\textbf{(a) The copy quality indices and the entanglement
indices:}\\
Buzek and Hillery \cite{buzek1} expresses the entanglement indices
$D_{ab}^{1}$,~$D_{ab}^{2}$,~$D_{ab}^{3}$ in terms of the copy
quality indices $D_{a}$ and $D_{b}$. $D_{ab}^{1}$ expresses the
"H-S distance" between the actual two mode density operator
$\rho_{ab}^{out}$ and a tensor product of density operators
$\rho_{a}^{out}$ and $\rho_{b}^{out}$. $D_{ab}^{2}$ measures the
"H-S distance" between density operator $\rho_{ab}^{out}$ and a
tensor product of $\rho_{a}^{id}$ and $\rho_{b}^{id}$.
$D_{ab}^{3}$ represents the "H-S distance" between the tensor
product of density operators $\rho_{a}^{id}$ and $\rho_{b}^{id}$
and a tensor product of $\rho_{a}^{out}$ and
$\rho_{b}^{out}$.\\
Also we note that the two outputs produced by the Wootters and
Zurek quantum cloning machine are same and thus $D_{a}$= $D_{b}$ =
$2\alpha^{2}\beta^{2}$ (equ. 1.45). The entanglement indices
can be expressed in terms of $D_{a}$ or $D_{b}$ or both in the following way:\\
\begin{eqnarray}
D_{ab}^{1}&=& Tr[\rho_{ab}^{(out)}-\rho_{a}^{(out)}\otimes\rho_{b}^{(out)}]^2
{}\nonumber\\&=& (2\alpha^{2}\beta^{2}).(2\alpha^{2}\beta^{2}){}\nonumber\\&=&
D_{a}.D_{b}{}\nonumber\\&=& D_{a}^{2}= D_{b}^{2}
\end{eqnarray}
\begin{eqnarray}
D_{ab}^{2}&=&Tr[\rho_{ab}^{(out)}-\rho_{a}^{(id)}\otimes\rho_{b}^{(id)}]^2
{}\nonumber\\&=&8\alpha^{4}\beta^{4}+
4\alpha^{2}\beta^{2}(\alpha^{4}+\beta^{4}){}\nonumber\\&=&2\alpha^{2}\beta^{2}
+2\alpha^{2}\beta^{2}{}\nonumber\\&=&D_{a}+ D_{b}{}\nonumber\\&=&2D_{a}=2D_{b}
\end{eqnarray}
\begin{eqnarray}
D_{ab}^{3}&=&Tr[\rho_{a}^{(id)}\otimes\rho_{a}^{(id)}-\rho_{a}^{(out)}\otimes\rho_{b}^{(out)}]^2
{}\nonumber\\&=&4\alpha^{4}\beta^{4}+
4\alpha^{2}\beta^{2}(\alpha^{4}+\beta^{4}){}\nonumber\\&=&2\alpha^{2}\beta^{2}
+2\alpha^{2}\beta^{2}-(2\alpha^{2}\beta^{2}).(2\alpha^{2}\beta^{2}) {}\nonumber\\ &=&
D_{a}+ D_{b}-D_{a}.D_{b}{}\nonumber\\&=&D_{a}(2-D_{a})=D_{b}(2-D_{b})
\end{eqnarray}
\emph{\textbf{(b) Wootters and Zurek quantum cloning machine in higher dimension:}}\\
In two dimensional Hilbert space the relationships between the
copying quality indices and the entanglement indices are
established by Buzek and Hillery. So it is natural to ask a
question whether those relationships between the copying quality
indices and the entanglement indices depend on the dimension of
the state space? The answer is given in the affirmative by M.Ying
\cite{ying1}. He studied the Wootter's and Zurek quantum copying
machine on a higher dimensional state space and established the
inequalities which described the relationship among the copying
quality indices $D_{a}$ and $D_{b}$ and the entanglement indices
$D_{ab}^{1}$,~$D_{ab}^{2}$,~$D_{ab}^{3}$.\\
The transformation rule for Wootters and Zurek quantum copying
machine in 'n' dimension can be defined by
\begin{eqnarray}
|k\rangle_{a}|\Sigma\rangle_{b}|Q_{i}\rangle_{x}\rightarrow|k\rangle_{a}|k\rangle_{b}|Q_{k}\rangle_{x}&
& (k=0,1,............,n-1)
\end{eqnarray}
The unitarity of the transformation gives
\begin{eqnarray}
\langle Q_{i}|Q_{i}\rangle=\langle Q_k|Q_k\rangle=1 & & (k=0,1,............,n-1)
\end{eqnarray}
Further, it is assumed that the machine state vectors are mutually
orthogonal, i.e. $\langle Q_{k}|Q_{l}\rangle=0$ for $0\leq\textrm{
k,j }\leq n-1$
and $\textrm{ k }\neq\textrm{ j }.$\\
The n-dimensional pure state is given by
\begin{eqnarray}
|\chi\rangle_{a}= \sum_{k=0}^{n-1}\alpha_{k}|k\rangle_{a}
\end{eqnarray}
with $\sum\alpha_{k}^{2}=1$.\\
For simplicity it is assumed that the probability amplitudes
$\alpha_{k}$ (k=0,1,.............,n-1) are all real numbers.\\
The density operator for the generalised n-dimensional input state is
\begin{eqnarray}
\rho_{a}^{(id)}=\sum_{k=0}^{n-1}\sum_{j=0}^{n-1}\alpha_{k}\alpha_{j}|k\rangle\langle
j|
\end{eqnarray}
 The action of the cloning machine on the n-dimensional input
state is given by
\begin{eqnarray}
|\chi\rangle_{a}|\Sigma\rangle_{b}|Q_{i}\rangle\rightarrow|\psi\rangle_{abx}^{out}
\equiv\sum_{k=0}^{n-1}\alpha_{k}|k\rangle_{a}|k\rangle_{b}|Q_{k}\rangle_{x}
\end{eqnarray}
After tracing out the machine state vector in mode 'x', the reduced density operator
describing the output state is given by
\begin{eqnarray}
\rho_{ab}^{(out)}=Tr_x[\rho_{abx}^{(out)}]=\sum_{k=0}^{n-1}\alpha_{k}^2|kk\rangle\langle
kk|
\end{eqnarray}
Furthermore, the reduced density operator describing the state in mode 'a' and 'b' is
given by
\begin{eqnarray}
\rho_{a}^{(out)}=Tr_b[\rho_{ab}^{(out)}]=\sum_{k=0}^{n-1}\alpha_{k}^2|k\rangle\langle
k|
\end{eqnarray}
and
\begin{eqnarray}
\rho_{b}^{(out)}=Tr_a[\rho_{ab}^{(out)}]=\sum_{k=0}^{n-1}\alpha_{k}^2|k\rangle\langle
k|
\end{eqnarray}
The copying quality indices are given by\\
$D_a(n)=Tr[\rho_{a}^{(id)}-\rho_{a}^{(out)}]^2$,
$D_b(n)=Tr[\rho_{b}^{(id)}-\rho_{b}^{(out)}]^2$ \\
and the entanglement indices are given by\\
$D_{ab}^{1}(n)=Tr[\rho_{ab}^{(out)}-\rho_{a}^{(out)}\otimes\rho_{b}^{(out)}]^2$,
$D_{ab}^{2}(n)=Tr[\rho_{ab}^{(out)}-\rho_{ab}^{(id)}]^2$,\\
$D_{ab}^{3}(n)=Tr[\rho_{ab}^{(id)}-\rho_{a}^{(out)}\otimes\rho_{b}^{(out)}]^2$,\\
where $\rho_{ab}^{(id)}= \rho_{a}^{(id)}\otimes\rho_{b}^{(id)}$,
$\rho_{a}^{(id)}=\rho_{b}^{(id)},\rho_{a}^{(out)}=\rho_{b}^{(out)}$.\\
The relations between the entanglement indices and the copying quality indices are the following \cite{ying1}:\\
\textbf{First Inequality:}
\begin{eqnarray}
D_{a}(n).D_{b}(n)-\frac{(n-1)(n-2)}{n^{2}}\leq D_{ab}^{1}(n)\leq D_{a}(n).D_{b}(n)
\end{eqnarray}
The minimum value of $D_{ab}^{1}(n)-D_{a}(n).D_{b}(n)$ is attained
when
$\alpha_{0}^{2}=\alpha_{1}^{2}=.....=\alpha_{n-1}^{2}=\frac{1}{n}.$
The maximum value of $D_{ab}^{1}(n)-D_{a}(n).D_{b}(n)$ is attained
at each of the points
$(\alpha_{0}^{2},\alpha_{1}^{2},.....,\alpha_{n-1}^{2})
\equiv(1,0,....,0),(0,1,0,....,0),.....,(0,0,....,1).$\\\\
\textbf{Second Inequality:}
\begin{eqnarray}
[D_{a}(n)+ D_{b}(n)]-\frac{(n-1)(n-2)}{n^{2}}\leq D_{ab}^{2}(n)\leq D_{a}(n)+ D_{b}(n)
\end{eqnarray}
The minimum value of $D_{ab}^{2}(n)-[D_{a}(n)+ D_{b}(n)]$ is
attained when
$\alpha_{0}^{2}=\alpha_{1}^{2}=.....=\alpha_{n-1}^{2}=\frac{1}{n}.$
The maximum value of $D_{ab}^{2}(n)-[D_{a}(n)+ D_{b}(n)]$ is
attained at each of the points
$(\alpha_{0}^{2},\alpha_{1}^{2},.....,\alpha_{n-1}^{2})
\equiv(1,0,....,0),(0,1,0,....,0),.....,(0,0,....,1).$\\\\
\textbf{Third Inequality:}
\begin{eqnarray}
D_{a}(n)+ D_{b}(n)-D_{ab}^{1}(n)-\frac{(n-1)(n-2)}{n^{2}}\leq
D_{ab}^{3}(n)\leq D_{a}(n)+ D_{b}(n)-D_{ab}^{1}(n)
\end{eqnarray}
The minimum value of $D_{ab}^{3}(n)-[D_{a}(n)+
D_{b}(n)-D_{ab}^{1}(n)]$ is attained when
$\alpha_{0}^{2}=\alpha_{1}^{2}=.....=\alpha_{n-1}^{2}=\frac{1}{n}.$
The maximum value of $D_{ab}^{2}(n)-[D_{a}(n)+
D_{b}(n)-D_{ab}^{1}(n)]$ is attained at each of the points
$(\alpha_{0}^{2},\alpha_{1}^{2},.....,\alpha_{n-1}^{2})
\equiv(1,0,....,0),(0,1,0,....,0),.....,(0,0,....,1).$\\\\
\textbf{\textrm{2. Bruss, DiVincenzo, Ekert, Fuchs, Macchiavello
and Smolin's quantum cloning machine}}\\
The deterministic state dependent quantum cloner was first studied
by Bruss, DiVincenzo, Ekert, Fuchs, Macchiavello and Smolin
\cite{bruss2}. They designed a optimal state dependent cloner
which copy the qubit selected from an ensemble containing only two
equiprobable non-orthogonal quantum states $|a\rangle$ and
$|b\rangle$. They have also shown that a priori knowledge about
the input state makes the performance of the quantum cloning
machine better than the universal quantum cloning machine in the sense of higher fidelity.\\
The optimal fidelity as obtained by Bruss et.al. \cite{bruss2} for
state dependent cloner is  given by the formula:
\begin{eqnarray}
F&=&
\frac{1}{2}+\frac{\sqrt{2}}{32S}(1+S)(3-3S+\sqrt{9S^{2}-2S+1})
{}\nonumber\\& & \times\sqrt{3S^{2}+2S-1+(1-S)\sqrt{9S^{2}-2S+1}}
\end{eqnarray}
where $S=|\langle a|b\rangle|$.\\
The minimum value of F is approximately equal to 0.987 and it is
achieved for $S=\frac{1}{2}$.\\
In quantum cryptography \cite{bennett1,bennett6,ekert1,shor1}, the
eavesdropper's main concern is not in copying the quantum
information encoded in the two non-orthogonal quantum states but
rather in optimizing the tradeoff between the classical
information made available to her versus the disturbance inflicted
upon the original qubit. In this respect, optimal state dependent
quantum cloner played a crucial role. Eavesdropper can use the
state dependent quantum cloning machine to copy the original qubit
in transit between a sender and a receiver. Then she resent the
original qubit to the receiver. Thereafter, she can obtain some
information from the cloned qubit by measuring it. Also since the
disturbance upon the original qubit is very low due to the
application of state dependent cloner, eavesdropper can steal the
information in the midway without giving any clue to
sender and receiver.\\
If one could get a success to construct a nearly perfect cloner
then it will certainly solve some problem like state estimation
problem \cite{bruss3,maruyama1,zhang2} but sidewise it will create
a problem in quantum cryptography. In quantum cryptography, the
third party Eve could steal information using quantum cloning
machine which is nearly perfect. So the optimal quantum cloning
machine not only helps us but also it has some negative effects
like it also helps the information hacker 'Eve'. We observe that
for this job state dependent cloner would be the more efficient
candidate than state independent cloner.
\subsection{\emph{State independent (Universal) quantum cloning machine}}
In this section, we study the state independent quantum cloning machines. When the quality of
the two identical copies at the output are independent of the input state, the machine is said
to be state independent or universal \cite{bruss2,buzek1,buzek4,fan2,fiurasek1,gisin1}.\\
\emph{\textbf{1. Buzek and Hillery quantum cloning machine $(1\rightarrow 2~ type)$ }}\\
Wootters and Zurek considered a quantum copying machine
(1.35-1.36) which demonstrated that if it copies two basis vectors
perfectly then it cannot copy the superpositions of these vectors
perfectly, i.e. there does not exist a quantum copier which can
copy arbitrary qubit without introducing errors. Even though ideal
copying is prohibited by the laws of quantum mechanics for an
arbitrary state, it is still possible to design quantum copiers
that operate reasonably well. The first universal quantum cloning
machine was constructed by Buzek and Hillery (in 1996) \cite{buzek1}.\\
In particular, the universal quantum cloning machine is specified
by the following conditions \cite{buzek4}:\\
(i) The state of the original system and its quantum copy at the output of the quantum
cloner, described by the density operators $\rho_{a}^{out}$ and $\rho_{b}^{out}$,
respectively are identical, i.e.,
\begin{eqnarray}
\rho_{a}^{out}= \rho_{b}^{out}
\end{eqnarray}
(ii) If no a priori information about the input state of the
original system is available, then to fulfil the requirement of
equal copy quality of all pure states, one should design a quantum
copier in such a way that the distances between the density
operators of each system at the output $\rho_{j}^{out}$ where
$j=a,b$ and the ideal density operator $\rho_{j}^{id}$ which
describes the input state of the original mode are input state
independent. Quantitatively this means that if we employ the
square of the Hilbert-Schmidt norm
\begin{eqnarray}
d(\hat{\rho}_{1};\hat{\rho}_{2})= Tr[( \hat{\rho}_{1}-\hat{\rho}_{2})^{2}]
\end{eqnarray}
as a measure of distance between two operators, then the quantum copier should be such
that
\begin{eqnarray}
d(\hat{\rho}_{j}^{out};\hat{\rho}_{j}^{id})= constant,~~~j=a,b
\end{eqnarray}
Here we note that other measures of the distance between two
density operators such as Bures distance and trace norm can also
be used to specify the universal cloning transformation
\cite{kwek1}. The final form of the transformation does not depend on the choice of the measure.\\
(iii) Finally, to make the quality of the copies better, it is
required that the copies are as close as possible to the input
state. \\
To construct the universal quantum cloning machine which
obeys the above three conditions, Buzek and Hillery proposed a
cloning transformation given below:
\begin{eqnarray}
|0\rangle_{a}|\Sigma\rangle_{b}|Q\rangle_{x}\longrightarrow
|0\rangle_{a}|0\rangle_{b}|Q_0\rangle_{x}+[|0\rangle_{a}|1\rangle_{b}+|1\rangle_{a}|0\rangle_{b}]|Y_0\rangle_{x}\\
|1\rangle_{a}|\Sigma\rangle_{b}|Q\rangle_{x}\longrightarrow
|1\rangle_{a}|1\rangle_{b}|Q_1\rangle_{x}+[|0\rangle_{a}|1\rangle_{b}+|1\rangle_{a}|0\rangle_{b}]|Y_1\rangle_{x}
\end{eqnarray}
The unitarity of the transformation gives
\begin{eqnarray}
\langle Q_i|Q_i\rangle + 2\langle Y_i|Y_i\rangle =1,~~~i=0,1 \\
\langle Y_0|Y_1\rangle=\langle Y_1|Y_0\rangle=0
\end{eqnarray}
Further, it is assumed that
\begin{eqnarray}
\langle Q_i|Y_i\rangle=0,~~i=0,1\\
\langle Q_0|Q_1\rangle=0
\end{eqnarray}
The quantum cloning machine (1.65-1.66) copies the input state
(1.38) and produces two-qubit output described by the density
operator
\begin{eqnarray}
\rho_{ab}^{(out)}&=&\alpha^2|00\rangle\langle00|\langle
Q_0|Q_0\rangle+\sqrt{2}\alpha\beta|00\rangle\langle+|\langle
Y_1|Q_0\rangle + \sqrt{2}\alpha\beta|+\rangle\langle00|\langle
Q_0|Y_1\rangle{}\nonumber\\&&+[2\alpha^2\langle
Y_0|Y_0\rangle+2\beta^2\langle Y_1|Y_1\rangle]|+\rangle\langle+|
+\sqrt{2}\alpha\beta|+\rangle\langle11|\langle
Q_1|Y_0\rangle{}\nonumber\\&&+\sqrt{2}\alpha\beta|11\rangle\langle+|\langle
Y_0|Q_1\rangle+\beta^2|11\rangle\langle11|\langle Q_1|Q_1\rangle
\end{eqnarray}
where $|+\rangle=\frac{1}{\sqrt{2}}(|10\rangle+|01\rangle)$.\\
The reduced density operator describing the original mode can be
obtained by taking partial trace over the copy mode and it reads
as
\begin{eqnarray}
\rho_a^{(out)}= [\alpha^2+\xi(\beta^2-\alpha^2)] |0\rangle\langle0| +
\alpha\beta\eta(|0\rangle\langle1|+ |1\rangle\langle0|)+
[\beta^2+\xi(\beta^2-\alpha^2)] |1\rangle\langle1|
\end{eqnarray}
where $\langle Y_0|Y_0\rangle=\langle Y_1|Y_1\rangle=\xi$,~~~ $\langle
Y_0|Q_1\rangle=\langle Q_0|Y_1\rangle=\langle
Q_1|Y_0\rangle=\langle Y_1|Q_0\rangle=\frac{\eta}{2}$,\\
$\xi$ and $\eta$ are two free parameters and they can be
determined from condition (ii) mentioned above and the criteria
that the distance between the two-qubit
output density matrix $ \rho_{ab}^{(out)}$  and the ideal two-qubit output $ \rho_{ab}^{(id)}$  be input state independent.\\
The density operator $ \rho_b^{(out)}$ describing the copy mode is
exactly same as the density operator $\rho_a^{(out)}$ describing the original mode.\\
Now the Hilbert Schmidt norm for the density operators $\rho_a^{(id)}$ and
$\rho_a^{(out)}$ is given by,
\begin{eqnarray}
D_a=2\xi^2(4\alpha^4-4\alpha^2+1)+2\alpha^2\beta^2(\eta-1)^2
\end{eqnarray}
with $0\leq\xi\leq\frac{1}{2}$ and
$0\leq\eta\leq2\sqrt{\xi(1-2\xi)}\leq\frac{1}{\sqrt{2}}$ which
follows from Schwarz inequality.\\
Now to satisfy the condition (ii) of universal quantum cloning machine described
above, the Hilbert Schmidt norm $D_a$ must be independent of the parameter $\alpha^2$.
\begin{eqnarray}
\frac{\partial D_a}{\partial\alpha^2} = 0 \Longrightarrow
\eta=1-2\xi
\end{eqnarray}
Using the relation $\eta=1-2\xi$, the equation (1.73) reduces to
\begin{eqnarray}
D_a=2\xi^2
\end{eqnarray}
The value of the parameter $\xi$ can be determined from the
condition that the distance between two-qubit density operators
$\rho_{ab}^{(id)}$ and $\rho_{ab}^{(out)}$ be input state
independent, i.e.
\begin{eqnarray}
\frac{\partial D_{ab}^2}{\partial\alpha^2} = 0
\end{eqnarray}
where $D_{ab}^2=Tr[\rho_{ab}^{(out)}-\rho_{ab}^{(id)}]^2= 8\xi^{2}+2\alpha^{2}(1-\alpha^{2})(1-6\xi)$.\\
Solving equation (1.76), we find $\xi=\frac{1}{6}$. For this value
of $\xi$, the norm $D_{ab}^2$ is independent of $\alpha^2$ and its
value is equal to $\frac{2}{9}$.\\
Also for $\xi=\frac{1}{6}$, the distance between the single qubit state at the output
of the copying machine and the input state is given by
\begin{eqnarray}
D_a=\frac{1}{18}
\end{eqnarray}
Next we will show that the two-qubit density operator
$\rho_{ab}^{out}$ given by the equation (1.71) is inseparable.\\
In the case of two qubits, we can utilize the necessary and
sufficient condition \cite{horo2,peres1} which states that the
partial transpose of a $4\times4$ density
matrix\\\\
$W_4$=\begin{tabular}{|c c c c|}
  $\rho_{00,00}$ & $\rho_{01,00}$ & $\rho_{00,10}$ & $\rho_{01,10}$ \\
  $\rho_{00,01}$ & $\rho_{01,01}$ & $\rho_{00,11}$ & $\rho_{01,11}$ \\
  $\rho_{10,00}$ & $\rho_{11,00}$ & $\rho_{10,10}$ & $\rho_{11,10}$ \\
  $\rho_{10,01}$ & $\rho_{11,01}$ & $\rho_{10,11}$ & $\rho_{11,11}$ \\
\end{tabular}~~~~ be negative for non-separability to hold.\\\\
Using the value of the parameter $\xi=\frac{1}{6}$, the equation
(1.71) can be rewritten as
\begin{eqnarray}
\rho_{ab}^{out}&=&
\frac{2\alpha^{2}}{3}|00\rangle\langle00|+\frac{\alpha\beta}{3}(|00\rangle\langle01|+
|00\rangle\langle10|+|01\rangle\langle00|+|10\rangle\langle00|)+
\frac{\alpha\beta}{3}(|01\rangle\langle11|{}\nonumber\\&&+
|10\rangle\langle11|+|11\rangle\langle01|+|11\rangle\langle10|)+\frac{1}{6}(|01\rangle\langle01|
+|01\rangle\langle10|+|10\rangle\langle01|{}\nonumber\\&&+|10\rangle\langle10|)+
\frac{2\beta^{2}}{3}|11\rangle\langle11|
\end{eqnarray}
The partial transpose of a $4\times4$ density matrix is given by
\begin{eqnarray}
W_4=\begin{tabular}{|c c c c|}
  $\frac{2\alpha^{2}}{3}$ & $\frac{\alpha\beta}{3}$ & $\frac{\alpha\beta}{3}$ & $\frac{1}{6}$ \\
  $\frac{\alpha\beta}{3}$ & $\frac{1}{6}$ & $0$ & $\frac{\alpha\beta}{3}$ \\
  $\frac{\alpha\beta}{3}$ & $0$ & $\frac{1}{6}$ & $\frac{\alpha\beta}{3}$ \\
  $\frac{1}{6}$ & $\frac{\alpha\beta}{3}$ & $\frac{\alpha\beta}{3}$ & $\frac{2\beta^{2}}{3}$
  \end{tabular}=-\frac{1}{6^{4}}
\end{eqnarray}
Equation (1.79) implies that $W_{4}$ is negative for all values of
$\alpha^{2}$, therefore the two qubit
at the output of the quantum copier is inseparable for any arbitrary input state.\\
Furthermore, Gisin and Massar \cite{gisin1} showed that the
universal quantum cloning machine satisfies criteria (iii) i.e.
the universal quantum cloning transformation defined
by equations (1.65-1.66) is optimal in the sense that it produces best quality copies at the output.\\
Therefore the optimal unitary transformation which implements the
universal quantum cloning machine is given by\\
\begin{eqnarray}
|0\rangle_{a}|\rangle_{b}|Q\rangle_{x}\longrightarrow\sqrt{\frac{2}{3}}|00\rangle_{ab}|\uparrow\rangle_{x}
+\sqrt{\frac{1}{3}}|+\rangle_{ab}|\downarrow\rangle_{x}
\end{eqnarray}
\begin{eqnarray}
|1\rangle_{a}|\rangle_{b}|Q\rangle_{x}\longrightarrow\sqrt{\frac{2}{3}}|11\rangle_{ab}|\downarrow\rangle_{x}
+\sqrt{\frac{1}{3}}|+\rangle_{ab}|\uparrow\rangle_{x}
\end{eqnarray}
The state space of the cloning machine is two dimensional and it
is spanned by the orthogonal vectors $|\uparrow\rangle_{x}$ and
$|\downarrow\rangle_{x}$.\\
The reduced density operators describing the state of both copies
at the output of the universal quantum cloner (1.80-1.81) are
equal and they can be expressed as
\begin{eqnarray}
\hat{\rho_{a}^{out}}=\hat{\rho_{b}^{out}}=\frac{5}{6}|\psi\rangle_{a}\langle\psi|+\frac{1}{6}|\psi_{\perp}\rangle_{a}\langle\psi_{\perp}|
\end{eqnarray}
where
\begin{eqnarray}
|\psi\rangle_{a}= \alpha|0\rangle_{a}+\beta|1\rangle_{a},~~
|\psi_{\perp}\rangle_{a}=
\beta^{*}|0\rangle_{a}-\alpha^{*}|1\rangle_{a}
\end{eqnarray}
From equation (1.79), it is clear that the mixed state at the
output of the cloner is entangled. Bruss and Macchiavello
\cite{bruss8} studied the entanglement properties of the output
state of a universal quantum cloning machine. Without any loss of
generality, they studied the entanglement structure of the cloning
output for an input basis state $|0\rangle$ given in equation
(1.80). Their investigations are about the amount of entanglement
between the two clones and between an ancilla and a clone.\\
The reduced density matrices $\rho_{ab}$ for two clones and $\rho_{ax}$ for one clone
and ancilla are given by
\begin{eqnarray}
\rho_{ab}=\begin{tabular}{|c c c c|}
$\frac{2}{3}$ & $0$ & $0$ & $0$ \\
$0$ & $\frac{1}{6}$ & $\frac{1}{6}$ & $0$ \\
$0$ & $\frac{1}{6}$ & $\frac{1}{6}$ & $0$ \\
$0$ & $0$ & $0$ & $0$
\end{tabular} ~~~ and ~~~
\rho_{ax}=\begin{tabular}{|c c c c|}
$\frac{2}{3}$ & $0$ & $0$ & $\frac{1}{3}$ \\
$0$ & $\frac{1}{6}$ & $0$ & $0$ \\
$0$ & $0$ & $0$ & $0$ \\
$\frac{1}{3}$ & $0$ & $0$ & $\frac{1}{6}$
\end{tabular}
\end{eqnarray}
In the case of the cloning output, the concurrences defined in
equation (1.9), for the mixed states described by density matrices
$\rho_{ab}$ and $\rho_{ax}$, are calculated to be
$C_{ab}=\frac{1}{3}$ and $C_{ax}=\frac{2}{3}$. Further using the
relation (1.10), the entanglement of formation for the density
matrices $\rho_{ab}$ and $\rho_{ax}$ are found out to be
$E_{F,ab}\simeq 0.1873$ and $E_{F,ax}\simeq 0.55$ respectively.
Therefore, we can observe that the entanglement between clone and
ancilla is higher than between the two copies. Hence, after the
cloning procedure, the information is distributed in such a way
that some of it are in the copies, some are in the entanglement
between the copies, some are in the copy machine, and some are in
the entanglement between the copies and the copy machine. The
information in the entanglement and in the copy machine is effectively lost.\\
Almost simultaneously with Gisin and Massar \cite{gisin1} but
independently Bruss, DiVincenzo,Ekert,Fuchs,Machiavello and Smolin
\cite{bruss2} have constructed the class of unitary
transformations for the optimal universal symmetric quantum cloner.\\
Here the class of unitary cloning transformation is given by
\begin{eqnarray}
U|0\rangle_{a}|\Sigma\rangle_{b}|Q\rangle_{x}=\sqrt{\frac{2}{3}}
~e^{i\delta_{a}}|00\rangle_{ab}|Q_{0}\rangle_{x}
+\sqrt{\frac{1}{6}}~e^{i\delta_{\bar{a}}}(|01\rangle_{ab}+|10\rangle_{ab})|Q_{1}\rangle_{x}
\end{eqnarray}
\begin{eqnarray}
U|1\rangle_{a}|\Sigma\rangle_{b}|Q\rangle_{x}=\sqrt{\frac{2}{3}}~e^{i\delta_{\bar{a}}}|11\rangle_{ab}
|Q_{1}\rangle_{x}+\sqrt{\frac{1}{6}}~e^{i\delta_{a}}(|01\rangle_{ab}+|10\rangle_{ab}|Q_{0}\rangle_{x}
\end{eqnarray}
where $\langle Q_{0}|Q_{1}\rangle=0$ and $\delta_{a}$, $\delta_{\bar{a}}$ denotes the phase factors.\\
Each output state of the quantum cloning machine is input state
independent if and only if the reduced density operator takes the
form
$\rho^{(out)}=\eta|\psi\rangle^{(in)}\langle\psi|+\frac{1}{2}(1-\eta)I$.
Therefore, the quality of the clones is defined by the fidelity
$F= ~^{(in)}\langle\psi|\rho^{(out)}|\psi\rangle^{(in)}=
\frac{1}{2}(1+\eta)$, where $\eta$ denotes the reduction factor.\\
Bruss et.al. \cite{bruss2} found that the quantum cloning transformation (1.85-1.86) would be optimal if $\eta=\frac{2}{3}$.\\
The corresponding optimal cloning fidelity is
\begin{eqnarray}
F_{opt}= \frac{5}{6}
\end{eqnarray}
Also we note that if $\delta_{a}=\delta_{\bar{a}}=0$ then it
reduces to Buzek-Hillery
universal quantum cloning transformation.\\\\
\emph{\textbf{2. Universal symmetric cloner $(1\rightarrow M~type~ and~ N\rightarrow M~type)$}}\\
In 1997, Gisin and Massar \cite{gisin1} introduced the idea of
generating M identical copies from one qubit input. They also
extended their
idea for an arbitrary number $N (< M)$ of input qubits.\\
The $1\rightarrow M$ quantum cloning machine, when acting on an arbitrary input state
$|\psi\rangle$, is described by the following unitary operator:
\begin{eqnarray}
U_{1\rightarrow M}|\psi\rangle\otimes R = \sum_{j=0}^{M-1}\alpha_{j}|(M-j)\psi,
j\psi^{\perp}\rangle\otimes R_{j}(\psi)
\end{eqnarray}
where $\alpha_{j}=\sqrt{\frac{2(M-j)}{M(M+1)}}$ and $R_{j}(\psi)$ represents the
internal state of the quantum cloning machine with
$\langle R_{j}(\psi)|R_{k}(\psi)\rangle = 0$ for $j\neq k$.\\
The density matrix $\rho^{(out)}= F_{1,M}|\psi\rangle\langle\psi|+
(1-F_{1,M})|\psi^{\perp}\rangle\langle\psi^{\perp}|$ describing
the output qubits is the same for all copies, where the fidelity
$F_{1,M}$ is given by
\begin{eqnarray}
F_{1,M}&=& \sum_{j=0}^{M-1}Prob (j~errors~in~the~(M-1)~last~qubits) {}\nonumber\\&=&
\frac{2M+1}{3M}.
\end{eqnarray}
A more general quantum cloning machine that takes N identical qubits all prepared in
the state $|\psi\rangle$ into $M (> N)$ identical copies is described by
\begin{eqnarray}
U_{N\rightarrow M}|\psi\rangle^{\otimes N}\otimes R =
\sum_{j=0}^{M-N}\alpha_{j}|(M-j)\psi, j\psi^{\perp}\rangle\otimes R_{j}(\psi)
\end{eqnarray}
where $\alpha_{j}=\sqrt{\frac{N+1}{M+1}}\sqrt{\frac {(M-N)!(M-j)!}
{(M-N-j)!M!}}$.\\
The fidelity of each output qubit of the more generalized quantum
cloning machine is given by
\begin{eqnarray}
F_{N,M}= \frac{M(N+1)+N}{M(N+2)}.
\end{eqnarray}
For pure input states the fidelity given in equation (1.91) which
is achieved by the cloning transformation (1.90) was shown to be
optimal in \cite{bruss1}. The fidelity $F_{N,M}$ tends to
$\frac{N+1}{N+2}$ as $M\rightarrow\mathcal{1}$, which is the
optimal fidelity
achievable by carrying out a measurement on N identical input qubits.\\
Moreover, Gisin and Massar discussed some special cases of the
generalized quantum cloning machine defined in equation (1.90).\\
\textbf{Case-1:}~~If N=1 and M=2, then $N\rightarrow M$ quantum
cloning machine reduces to $1\rightarrow 2$ Buzek-Hillery
universal quantum cloning machine. The fidelity of each output
qubit is $\frac{5}{6}$.\\
\textbf{Case-2:}~~If N=1 and M=$M_{1}(>1)$ but finite, then
$N\rightarrow M$ quantum cloning machine reduces to $1\rightarrow
M_{1}$ quantum cloning machine. The fidelity of each output qubit
is $F_{1,M_{1}}=\frac{2M_{1}+1}{3M_{1}}$. Now if we assume that
the quantum cloning machine could produce infinite number of
copies i.e. if $M_{1}\rightarrow\mathcal{1}$ then
$F_{1,\;\mathcal{1}}\rightarrow\frac{2}{3}$. In this case the
cloning fidelity is equal to the fidelity of measurement i.e. the
overlapping between the states before and after measurement for a
given single unknown quantum state. Further, we
note that when the number of clones $M_{1}$ grows for fixed N, the cloning fidelity decreases. \\
\textbf{Case-3:}~~If N=$N_{1}$ and M=$N_{1}+1$ both are finite,
then $N\rightarrow M$ quantum cloning machine reduces to
$N_{1}\rightarrow N_{1}+1$ quantum cloning machine. In this case,
The fidelity of each output qubit reduces to
$F_{N_{1},N_{1}+1}=\frac{N_{1}^{2}+3N_{1}+1}{N_{1}^{2}+3N_{1}+2}$
which tends to 1 as $N_{1}\rightarrow\mathcal{1}$.\\\\
\emph{\textbf{3. Universal symmetric quantum cloner in d-Dimension \cite{buzek4,zanardi1}}}\\
In 1998, Buzek and Hillery \cite{buzek4} proposed a universal cloning transformation of states in a d-dimensional Hilbert space.\\
The cloning transformation in a d-dimensional Hilbert space is given by
\begin{eqnarray}
|\Psi_{i}\rangle_{a}|\Sigma\rangle_{b}|Q\rangle_{x}&\rightarrow&\sqrt{\frac{2}{d+1}}
|\Psi_{i}\rangle_{a}|\Psi_{i}\rangle_{b}|Q_{i}\rangle_{x}+\sqrt{\frac{1}{2(d+1)}}
\sum_{j\neq i}^{d}(|\Psi_{i}\rangle_{a}
|\Psi_{j}\rangle_{b}{}\nonumber\\&&+|\Psi_{j}\rangle_{a}
|\Psi_{i}\rangle_{b})|Q_{j}\rangle_{x}
\end{eqnarray}
The density operator describing each copy at the output can be written in the scaled
form as
\begin{eqnarray}
\rho_{j}^{(out)}= \eta \rho_{j}^{(id)}+ \frac{1-\eta}{d}\hat{I}
\end{eqnarray}
where $\rho_{j}^{(id)}=|\psi\rangle\langle\psi|$ is the density
operator describing the input state which is going to be cloned
and $\eta$ is called reduction factor or scaling factor which is
given by
\begin{eqnarray}
\eta=\frac{d+2}{2(d+1)}
\end{eqnarray}
As $d\rightarrow\mathcal{1}$, the scaling factor $\eta\rightarrow\frac{1}{2}$.\\
The fidelity of the copies in terms of the reduction factor is given by
\begin{eqnarray}
F(d)= \frac{\eta(d-1)+1}{d}
\end{eqnarray}
For 2-dimensional case, the scaling factor and the fidelity takes the value
$\eta=\frac{2}{3}$ and $F(2)= \frac{5}{6}$ which was shown to be optimal value by
Gisin and Massar and also by Bruss et.al. Further we note that if we consider the
cloning transformation of states in an infinite dimensional Hilbert space then one can
copy the quantum information with at most fidelity $\frac{1}{2}$. That means in an
infinite dimensional Hilbert space, we cannot extract much information about an
arbitrary quantum state using quantum cloning machine. The reason behind the poor
copying of quantum information contained in a higher dimensional state space can be
explained by the von Neumann entropy. The von Neumann entropy measures the degree of
entanglement between the copies and the copier and is given by
\begin{eqnarray}
S= ln(d+1)-\frac{2~ln2}{d+1}
\end{eqnarray}
It is clear from equation (1.96) that the entropy does not depend
on the input state to be copied but it depends on the dimension of
the state space. Also the entropic function is an increasing
function of the dimension d. Therefore, with the increase of the
dimension d, the entanglement between the copies and the copier
also increases. Probably this is the reason why cloning machine
fails to produce better quality copies in higher dimension.\\
Fan, Matsumoto and Wadati \cite{fan2} constructed an optimal N to
M $(N<M)$ quantum cloning transformation for d-dimensional quantum
system. Their proposed cloning transformation is given by
\begin{eqnarray}
U_{N\rightarrow M}|\textbf{n}\rangle\otimes
R=\sum_{j=0}^{M-N}\alpha_{\textbf{nj}}|\textbf{n+j}\rangle\otimes R_{\textbf{j}}
\end{eqnarray}
where $|\textbf{n}\rangle=|n_{1},.....,n_{d}\rangle$ is a
completely symmetric and normalized state, $R_{\textbf{j}}$
denotes the orthogonal normalized internal states of the universal
quantum cloning machine,\\
$\sum_{k=1}^{d}j_{k}= M-N$ and
\begin{eqnarray}
\alpha_{\textbf{n,j}}=\sqrt{\frac{(M-N)!(N+d-1)!}{(M+d-1)!}}
\sqrt{\prod_{k=1}^{d}\frac{(n_{k}+j_{k})!}{n_{k}!j_{k}!}}
\end{eqnarray}
The state of each d-level clone is described by the reduced density operator
\begin{eqnarray}
\rho^{out}=\eta_{N,M}(d)|\psi\rangle\langle\psi|+ \frac{(1-\eta_{N,M}(d))}{d}\hat{I}
\end{eqnarray}
where $\eta_{N,M}(d)=\frac{N(M+d)}{M(N+d)}$ denotes the scaling
factor for qudits.\\
The fidelity of the copy qudit produced from the quantum cloning
machine (1.97) is given by
\begin{eqnarray}
F_{N,M}(d)=\frac{N(d-1)+M(N+1)}{(d+N)M}
\end{eqnarray}
The fidelity $F_{N,M}(d)$ was shown to be optimal by Werner
\cite{werner1} and Keyl and Werner \cite{keyl1}.\\
Note: (i) For d=2, the fidelity $F_{N,M}(2)$ reduces to the
fidelity $F_{N,M}$ of the generalised $N\rightarrow M$ quantum cloning machine for qubit.\\
(ii) When N=1 and M=2, the performance of the quantum cloning
machine for qudit is given in terms of the fidelity $F_{1,2}(d)=\frac{d+3}{2(d+1)}$.\\
(iii) In the limit $d\rightarrow\mathcal{1}$, the fidelity
$F_{N,M}\mathcal{(1)}$ tends to $\frac{N}{M}$. Further, if we want
to produce large number of copies of finite number of input in
higher dimensional systems i.e. if $N\ll M$, then the existing
quantum cloning machine produces poor quality copies.\\
(iv) For sufficiently large M and finite N and d, the fidelity
$F_{N,\mathcal{1}}(d)$ tends to $\frac{N+1}{N+d}$ which is the
optimal fidelity for state estimation of N copies of a
d-dimensional quantum system. For d=2, the cloning fidelity tends
to the optimal measurement fidelity $\frac{N+1}{N+2}$ for N
identical qubits. This expression for measurement fidelity
originally derived by Massar and Popescu \cite{massar1}.
\subsection{\emph{Probabilistic cloning}}
One could design a quantum cloning machine which will copy only
states from a particular set of allowed input states. This type of
cloning machine produces better quality copies than the universal
quantum cloning machine. In fact, it is possible for the copier to
be perfect for certain  small size of the input sets. If the input
set contains any two non-orthogonal states, then it is impossible
to build a perfect quantum copier \cite{hillery1}. Also it is not
possible to build a perfect copier for input sets of more than two
states. Therefore, the quantum copier which produces perfect
copies and works every time does not exist in nature. But if we
relax the latter condition i.e. if we allow the quantum cloning
machine to fail to produce perfect copies for sometime, then such
type of quantum cloning machine exists in nature and they are
called probabilistic quantum cloning machines
\cite{azuma1,duan1,fiurasek2,qiu3,ting1,zhang1,zhang3,zhang4}.
Probabilistic quantum cloning machine performs measurements and
unitary operations, with a post selection of the measurement
results and hence the desired copies are produced only with
certain probabilities. Also we cannot exclude the fact that there
are some probability for which the cloning machine fails to
produce the
perfect copies and in those cases the copies would be discarded.\\
In 1997, Duan and Guo \cite{duan1,duan3} first proposed such type
of quantum cloning machine which produces with some probability
perfect copies of two non-orthogonal states. They showed that two
non-orthogonal quantum states secretly chosen from a certain set
$S=\{|\Psi_{0}\rangle,|\Psi_{1}\rangle\}$ can be perfectly cloned
with some probability less than unity. Few months later, they
generalised their result and showed that non orthogonal quantum
states secretly chosen from a certain set $S=
\{|\Psi_{1}\rangle,\Psi_{2}\rangle,..........,|\Psi_{n}\rangle\}$
can be cloned probabilistically by a unitary evolution together with a reduction process.\\
\textbf{Theorem 1.1:} The n non-orthogonal states
$|\Psi_{1}\rangle,|\Psi_{2}\rangle,..........,|\Psi_{n}\rangle$
can be probabilistically cloned with unit fidelity by the same cloning machine if and only if they are linearly-independent.\\
C-W Zhang, Z-Y Wang, C-F Li and G.C.Guo \cite{zhang1} have
considered the realizations of quantum probabilistic identifying
and cloning machines by physical means. They showed that the
unitary representation and the Hamiltonian of probabilistic
cloning and identifying machines are determined by the
probabilities of successes. The logic networks are obtained by
decomposing the unitary representation into universal quantum
logic operations. They also discussed the robustness of the
networks and found that if error occurs in the input target
system, it can be detected and the to-be-cloned states can be recycled.\\
C-W Zhang, C-F Li, Z-Y Wang and G.C.Guo \cite{zhang4} proposed a
probabilistic quantum cloning scheme using
Greenberger-Horne-Zeilinger (GHZ) states \cite{greenberger1}, Bell
basis measurements, single-qubit unitary operations and
generalized measurements. The single-qubit generalized measurement
is performed by the unitary transformation on the composite system
of that qubit and the auxiliary probe with reduction measurement
of the probe. They showed that their scheme may be used in
experiment to clone the states of one particle to those of two
different particles with higher probability and less GHZ resources.\\
J.Fiurasek \cite{fiurasek2} have investigated the optimal
probabilistic realizations of several important quantum
information processing tasks such as the optimal cloning of
quantum states and purification of mixed quantum states. The
performance of these probabilistic operations is quantified by the
average fidelity between the ideal (generally mixed) states
$\rho^{in}$ and actual output pure states $|\psi^{out}\rangle$. He
derived a simple formula for the maximum achievable average
fidelity and provided an explicit prescription how to construct a
trace-decreasing completely positive map that reaches the maximum
average fidelity $F_{max}$ given by
\begin{eqnarray}
F_{max}= max [eig(A^{-1}R)]
\end{eqnarray}
Where, $A=\int_{S_{in}}(\rho_{in}^{T}\otimes I_{out})~d\rho_{in}$
and $R=\int_{S_{in}}(\rho_{in}^{T}\otimes \psi_{out})~d\rho_{in}$.
Further, it was shown that the fidelity of probabilistic cloning
can be strictly higher than the maximal fidelity of deterministic
cloning even if the set of the cloned states is linearly dependent
and continuous. This improvement in fidelity is achieved at the
expense of a certain fraction of unsuccessful events when the
probabilistic transformation fails and does not produce any output
state.\\
K.Azuma, J.Shimamura, M.Koashi and N.Imoto \cite{azuma1} studied
the probabilistic cloning of a mutually non-orthogonal set of pure
states $\{|\psi_{1}\rangle,|\psi_{2}\rangle\}$, with the help of
supplementary information in the form of pure states
$\{|\phi_{1}\rangle,|\phi_{2}\rangle\}$. They showed that the best
efficiency of producing m copies is always achieved by a two-step
protocol in which the helping party first attempts to produce
$m-1$ copies from the supplementary state, and if it fails, then
the original state is used to produce m copies. To perform the
two-step protocol, two types of probabilistic cloning machines are
used:\\
(i) Original state $\{|\psi_{i}\rangle\}_{i=1,2}$ is copied by the
machine
$\{|\psi_{i}\rangle\longrightarrow^{\gamma_{i}^{A}}|\psi_{i}\rangle^{\otimes
m}\}_{i=1,2}$ with probability $\gamma_{1}^{A}=\gamma_{2}^{A}=
\frac{1-\langle\psi_{1}|\psi_{2}\rangle}{1-|\langle\psi_{1}|\psi_{2}\rangle|^{m}}$\\
(ii) Supplementary information in terms of pure state
$\{|\phi_{i}\rangle\}_{i=1,2}$ is copied by the machine
$\{|\phi_{i}\rangle\longrightarrow^{\gamma_{i}^{B}}|\psi_{i}\rangle^{\otimes
m-1}\}_{i=1,2}$ with probability $\gamma_{1}^{B}=\gamma_{2}^{B}=
\frac{1-|\langle\phi_{1}|\phi_{2}\rangle|}{1-|\langle\psi_{1}|\psi_{2}\rangle|^{m-1}}$.\\
Therefore, using these machines in the two-step protocol, an
overall success probability $\gamma_{totmax}$ is given by
\begin{eqnarray}
\gamma_{totmax}=\gamma_{1}^{B}+(1-\gamma_{1}^{B})\gamma_{1}^{A}=
\frac{1-|\langle\psi_{1}|\psi_{2}\rangle\langle\phi_{1}|\phi_{2}\rangle|}{1-|\langle\psi_{1}|\psi_{2}\rangle|^{m}}
\end{eqnarray}
It was further shown that when the number of states exceeds two,
the best efficiency is not always achieved by such a protocol.
\subsection{\emph{Phase covariant quantum cloning}}
Bruss, Cinchetti, Ariano, and Macchiavello \cite{bruss6} were the
first who studied the cloning transformations for a restricted set
of pure input states of the form
\begin{eqnarray}
|\psi_{\phi}\rangle=\frac{1}{\sqrt{2}}[\;|0\rangle+e^{i\phi}|1\rangle\;]
\end{eqnarray}
where the parameter $\phi$ represents the angle between the Bloch
vector and the x-axis and $\phi~\mathcal{2}~[0,2\pi)$. The qubits
of this form are called equatorial qubits because the z-component
of their Bloch vector is zero. The importance of equatorial qubits
lies in the fact that the quantum cryptographic experiments
require the equatorial states rather than the states that span the
whole Bloch sphere. The cloning transformations that can clone
arbitrary equatorial qubits are called phase covariant quantum
cloners \cite{bruss6,chiara1,chiara2,fan1,fan4,karimipour1}
because they keep the quality of the copies same for all
equatorial qubits or in other words the
fidelity does not depend on the parameter $\phi$.\\
Bruss, Cinchetti, Ariano, and Macchiavello \cite{bruss6} proposed
the following 1 to 2 cloning transformation for the input (1.103),
\begin{eqnarray}
|0\rangle_{a}|\Sigma\rangle_{b}|Q\rangle_{x}\rightarrow
[(\frac{1}{2}+\sqrt{\frac{1}{8}})|00\rangle_{ab}+
(\frac{1}{2}-\sqrt{\frac{1}{8}})|11\rangle_{ab}]|\uparrow\rangle_{x}
+\frac{1}{2}|+\rangle_{ab}|\downarrow\rangle_{x}
\end{eqnarray}
\begin{eqnarray}
|1\rangle_{a}|\Sigma\rangle_{b}|Q\rangle_{x}\rightarrow
[(\frac{1}{2}+\sqrt{\frac{1}{8}})|11\rangle_{ab}+
(\frac{1}{2}-\sqrt{\frac{1}{8}})|00\rangle_{ab}]|\downarrow\rangle_{x}
+\frac{1}{2}|+\rangle_{ab}|\uparrow\rangle_{x}
\end{eqnarray}
where $|\uparrow\rangle_{x}$ and $|\downarrow\rangle_{x}$ denote
the orthogonal machine state vectors and \\
$|+\rangle=\frac{1}{\sqrt{2}}(|01\rangle+|10\rangle)$.\\
The reduced density operator of both copies at the output can be expressed as
\begin{eqnarray}
\rho^{out}=(\frac{1}{2}+\sqrt{\frac{1}{8}})|\psi_{\phi}\rangle\langle\psi_{\phi}|+
(\frac{1}{2}-\sqrt{\frac{1}{8}})|\psi_{\phi,\perp}\rangle\langle\psi_{\phi,\perp}|
\end{eqnarray}
where the state $|\psi_{\phi,\perp}\rangle$ is orthogonal to the state $|\psi_{\phi}\rangle$.\\
The optimal fidelity of 1 to 2 phase covariant cloning transformation is given by
\begin{eqnarray}
F^{phase}_{1,2}= \frac{1}{2}+ \sqrt{\frac{1}{8}}.
\end{eqnarray}
We note that the fidelity of the phase covariant quantum cloning
machine is greater than the fidelity of the universal quantum
cloning machine. This is due to the fact that more information
about the input qubit is given to the phase covariant quantum
cloning machine. For \textit{x-y} equatorial qubits, Fan,
Matsumoto, Wang, Wadati's \cite{fan3} conjecture was that the
general N to M $(M>N)$ quantum cloning transformation would be\\
(i) when $M-N$ is even,
\begin{eqnarray}
U_{N\rightarrow M}|(N-j)\uparrow,j\downarrow\rangle\otimes
R=|(\frac{M+N-2j}{2})\uparrow,(\frac{M-N+2j}{2})\downarrow\rangle\otimes R_{L}
\end{eqnarray}
(ii) when $M-N$ is odd,
\begin{eqnarray}
{}\nonumber\\&&U_{N\rightarrow
M}|(N-j)\uparrow,j\downarrow\rangle\otimes
R=\frac{1}{\sqrt{2}}|(\frac{M+N-2j+1}{2})\uparrow,(\frac{M-N+2j-1}{2})\downarrow\rangle\otimes
R_{L}{}\nonumber\\&&+\frac{1}{\sqrt{2}}|(\frac{M+N-j-1}{2})\uparrow,(\frac{M-N+2j+1}{2})\downarrow\rangle\otimes
R_{L+1}
\end{eqnarray}
The corresponding fidelities for the above two cases are given
by\\
(i) when $M-N$ is even,
\begin{eqnarray}
F^{phase}_{N,M}=\frac{1}{2}+\frac{1}{2^{N}}\sum_{j=0}^{N-1}\frac{N!}{j!(N-j-1)!}
\times\sqrt{\frac{(M-N+2j+2)(M+N-2j)}{4M^{2}(j+1)(N-j)}}
\end{eqnarray}
(ii) when $M-N$ is odd,
\begin{eqnarray}
F^{phase}_{N,M}=\frac{1}{2}+\frac{1}{2^{N+1}}\sum_{j=0}^{N-1}\frac{N!}{j!(N-j-1)!}
\times\frac{1}{\sqrt{4M^{2}(j+1)(N-j)}}\times
{}\nonumber\\(\sqrt{(M-N+2j+1)(M+N-2j+1)}{}\nonumber\\+\sqrt{(M-N+2j+3)(M+N-2j-1)})
\end{eqnarray}
\textbf{Note: }\\
1. When N=1 and $M>1$,
\begin{eqnarray}
F^{phase}_{1,M}&=& \frac{1}{2}+\frac{\sqrt{M(M+2)}}{4M}\textrm{ ~~~~when M is even
}{}\nonumber\\&=& \frac{1}{2}+\frac{(M+1)}{4M}\textrm{~~~~~~~~~~~~when M is odd }
\end{eqnarray}
2. In particular, when N=1 and M=2, the fidelity of each output
qubit is given by
$F^{phase}_{1,2}=\frac{1}{2}+\sqrt{\frac{1}{8}}$. Moreover, Fan,
Matsumoto, Wang, Wadati \cite{fan3} showed that the copied qubits
are separable for the case of optimal phase-covariant quantum
cloning. This observation makes the equatorial states unique in
the sense that they are the only states which give rise to
separable density matrix for the output copies.\\
3. When N=1 and M=3, the fidelity of 1 to 3 phase covariant quantum cloning machine is
$\frac{5}{6}$ which is equal to the fidelity of 1 to 2 universal
Buzek-Hillery quantum cloning machine.\\
4. If we make $M\rightarrow\infty$ in equations (1.110) and
(1.111), then the fidelity in the limiting sense is given by
\begin{eqnarray}
F^{phase}_{N,\infty}=\frac{1}{2}+\frac{1}{2^{N+1}}\sum_{j=0}^{N-1}\frac{N!}{j!(N-j-1)!}
\times\sqrt{\frac{1}{(j+1)(N-j)}}
\end{eqnarray}
The conjecture \cite{fan3} about N to M phase covariant quantum
cloning transformation seems to be correct because the optimal
fidelity of $N\rightarrow\infty$ quantum cloning equals to the
corresponding fidelity for optimal covariant quantum phase
estimation of equatorial qubits originally derived by Derka, Buzek and Ekert \cite{derka1}.\\
Further, Karimipour and Rezakhani \cite{karimipour1} investigated
the phase covariant quantum cloning of the states on the Bloch
sphere which have a definite z component of spin. They showed that
it is always possible to clone a spin state $|\textbf{n}\rangle$
with a fidelity higher than the universal value and that of
equatorial states, if the third component of its spin
$\langle\textbf{n}|\sigma_{z}|\textbf{n}\rangle$ is known.\\
Now to clone a general two level state
$|\textbf{n}\rangle=\cos\frac{\theta}{2}|0\rangle +
e^{i\phi}\sin\frac{\theta}{2}|1\rangle$, they considered the
following cloning transformation
\begin{eqnarray}
U|0\rangle_{a}|\Sigma\rangle_{b}|Q\rangle_{x}=\nu|00\rangle_{ab}|\uparrow\rangle_{x}
+\mu(|01\rangle_{ab}+|10\rangle_{ab})|\downarrow\rangle_{x}
\end{eqnarray}
\begin{eqnarray}
U|1\rangle_{a}|\Sigma\rangle_{b}|Q\rangle_{x}=\nu|11\rangle_{ab}|\downarrow\rangle_{x}
+\mu(|01\rangle_{ab}+|10\rangle_{ab})|\uparrow\rangle_{x}
\end{eqnarray}
where $|\uparrow\rangle,|\downarrow\rangle$ denotes the orthogonal machine state vectors.\\
When the state $|\textbf{n}\rangle$ is acted upon by the cloning
machine (1.114-1.115), each copy at the output is described by the
reduced density operator
\begin{eqnarray}
\rho^{out}= \mu^{2}I + 2\mu\nu|\textbf{n}\rangle\langle\textbf{n}|
+ (\nu^{2}-2\mu\nu)\cos^{2}\frac{\theta}{2}|0\rangle\langle0|+
\sin^{2}\frac{\theta}{2}|1\rangle\langle1|)
\end{eqnarray}
where I denotes the identity operator in 2-dimensional Hilbert space.\\
The fidelity of the cloning is given by
\begin{eqnarray}
F(\theta)=
\frac{1}{2}+\mu\sqrt{1-2\mu^{2}}+(\frac{1-2\mu^{2}}{2}-\mu\sqrt{1-2\mu^{2}})\langle\sigma_{z}\rangle^{2}
\end{eqnarray}
where
$\langle\sigma_{z}\rangle\equiv\langle\textbf{n}|\sigma_{z}|\textbf{n}\rangle=\cos\theta$\\
Also
\begin{eqnarray}
D_{ab}^{1}(\theta)=
K_{1}\cos^{8}\frac{\theta}{2}+K_{2}\cos^{6}\frac{\theta}{2}+K_{3}\cos^{4}\frac{\theta}{2}
+K_{4}\cos^{2}\frac{\theta}{2}+K_{5}
\end{eqnarray}
Where $K_{1}=576\mu^{8}-768\mu^{6}+ 352\mu^{4}-64\mu^{2}+4$,
$K_{2}=-1152\mu^{8}+1536\mu^{6}-704\mu^{4}+128\mu^{2}-8$,
$K_{3}=672\mu^{8}-928\mu^{6}+424\mu^{4}-72\mu^{2}+4$,
$K_{4}=-96\mu^{8}+160\mu^{6}-72\mu^{4}+8\mu^{2}$,$K_{5}=4\mu^{8}+2\mu^{4}$
\begin{eqnarray}
D_{ab}^{2}(\theta)=8\mu^{4}-(6\mu^{4}+\mu^{2}+2\mu\nu-1)\sin^{2}\theta
\end{eqnarray}
Now for some fixed value of $\theta$, the fidelity $F(\theta)$
attains its optimal value when
\begin{eqnarray}
\mu^{2}=\frac{1}{4}(1-\frac{1}{\sqrt{1+2\tan^{4}\theta}})
\end{eqnarray}
It can be observed that the fidelity of cloning spin states with a
definite component of spin along the z direction is higher than
the fidelity of cloning spin states with zero third component of
its spin. Also the distances $D_{ab}^{1}$ and $D_{ab}^{2}$ is
minimized when $\mu^{2}$ takes the form given in equation (1.120).\\
Now we discuss here some other interesting results which one can
get from equations (1.117-1.119).\\
\textbf{Result-1:} If
$\frac{1-2\mu^{2}}{2}-\mu\sqrt{1-2\mu^{2}}=0$ then
$\mu=\frac{1}{\sqrt{6}}$. Therefore, $\mu$ does not depend on the
parameter $\theta$ and hence the fidelity F and the distances
$D_{ab}^{1}$ and $D_{ab}^{2}$ are also independent of $\theta$.
Thus the cloning machine (1.114-1.115) reduces to Buzek-Hillery
universal quantum cloning machine. The values of the fidelity and
the distances are given by $F=\frac{5}{6}$,
$D_{ab}^{1}=\frac{19}{324}$ and $D_{ab}^{2}=\frac{2}{9}$.\\
\textbf{Result-2:} If $\langle\sigma_{z}\rangle=0$, then
$\theta=\frac{\pi}{2}$.
$F(\frac{\pi}{2})=\frac{1}{2}+\mu\sqrt{1-2\mu^{2}}$.
$F(\frac{\pi}{2})$ attains its maximum value when
$\mu=\frac{1}{2}$. In this case, the cloning machine (1.114-1.115)
reduces to optimal phase covariant quantum cloning machine.
Therefore, the optimal value of the fidelity and the distances are
given by $F^{opt}(\frac{\pi}{2})=\frac{1}{2}+\frac{1}{\sqrt{8}}$,
$D_{ab}^{1}(\frac{\pi}{2})=\frac{9}{64}$,
$D_{ab}^{2}(\frac{\pi}{2})=\frac{7}{8}-\frac{1}{\sqrt{2}}$.\\\\
Fan, Imai, Matsumoto and Wang \cite{fan4} studied the
phase-covariant quantum cloning machine for d-level quantum
systems. The optimal 1 to 2 phase-covariant quantum cloning
transformation for input state
$|\Psi\rangle^{(in)}=\frac{1}{\sqrt{d}}\sum_{j=0}^{d-1}e^{i\phi_{j}}|j\rangle\rangle$
is given by
\begin{eqnarray}
U|j\rangle_{a}|\Sigma\rangle_{b}|Q\rangle_{x}=\alpha|j\rangle_{a}|j\rangle_{b}|Q_{j}\rangle_{x}
+\frac{\beta}{\sqrt{2(d-1)}}\sum_{k\neq
j}^{d-1}(|j\rangle_{a}|k\rangle_{b}+|k\rangle_{a}|j\rangle_{b})|Q_{k}\rangle_{x}
\end{eqnarray}
where $\alpha^{2}=\frac{1}{2}-\frac{d-2}{2\sqrt{d^{2}+4d-4}}$,
$\beta^{2}=\frac{1}{2}+\frac{d-2}{2\sqrt{d^{2}+4d-4}}$ and
$|Q_{j}\rangle_{x}$ denote the orthonormal machine states.\\
The reduced density matrix for a single qudit at the output of the cloning machine is
given by
\begin{eqnarray}
\rho^{(out)}=\frac{1}{d}\sum_{j=0}^{d-1}|j\rangle\langle j|+(
\frac{\alpha\beta}{d}\sqrt{\frac{2}{d-1}}+\frac{\beta^{2}(d-2)}{2d(d-1)})\sum_{j\neq
k}e^{i(\phi_{j}-\phi_{k}})|j\rangle\langle k|
\end{eqnarray}
The optimal fidelity of 1 to 2 phase-covariant quantum cloning machine is given by
\begin{eqnarray}
F_{1,2~ opt}^{phase}(d)=\frac{1}{d}+\frac{1}{4d}(d-2+\sqrt{d^{2}+4d-4})
\end{eqnarray}
\textbf{Note:}\\
1. If we consider the two-level quantum system, then the
phase-covariant quantum cloning transformation (1.121) produces
two copies of the equatorial qubit with optimal fidelity
$\frac{1}{2}+\sqrt{\frac{1}{8}}$ which agree with the previous
result given in equation (1.107).\\
2. In case of qutrits i.e. for 3-dimensional quantum system, the
optimal fidelity of phase-covariant quantum cloning machine is
found to be $F_{1,2~ opt}^{phase}(3)=\frac{5+\sqrt{17}}{12}$. The
same result was also obtained by D'Ariano, Presti
\cite{d'ariano1}; Cerf, Durt, Gisin \cite{cerf4}; and
Karimipour,Rezakhani \cite{karimipour1}. Karimipour and Rezakhani
\cite{karimipour1} also studied the phase covariant quantum
cloning of d-level system that lies on the Bloch sphere with a
definite z component of spin.
\subsection{\emph{Economical quantum cloning}}
Until now we have discussed those quantum cloning transformations
(cloning machines) in which an additional system called ancilla or
machine state is present. But the presence of ancilla
significantly affects the recent NMR experiments that were
realized for the implementation of cloning operations. The
negative effects occurs due to sensitiveness of the ancilla
towards decoherence and as a result the achieved cloning fidelity
reduces. It was difficult to avoid such effects because the
cloning network contained at least ten single-qubit gates and five
two-qubit gates. To overcome this problem, it is necessary to
construct a cloning network in which less number of quantum gates
are required and also it should keep the fidelity of cloning at
its optimal level. Fortunately, Niu and Griffiths \cite{niu2}
designed a 2-qubit cloning network in which no external ancilla is
required and that it requires only two single-qubit gates and one
two-qubit gate and it requires to control the entanglement of a
pair of qubits only. It is thus likely to be quite less noisy than
its 3 qubit counter part. The cloning procedure which does not
require an extra ancilla or machine state is termed as economical quantum cloning
\cite{buscemi1,cerf3,durt2,niu2} . \\
Durt and Du  \cite{durt1} analyzed the possibility to reduce 3
qubit one-to-two phase-covariant quantum cloning \cite{fuchs1} to
2 qubit economic quantum cloning \cite{niu2}. They derived a
necessary and sufficient condition to characterize the
reducibility of 3 qubit cloners to 2 qubit cloners.
They showed that when this condition is fulfilled, economic cloning is possible.\\
Durt, Fiurasek and Cerf \cite{durt2} proved that universal 1 to 2
cloning transformation for any dimension $d\geq2$ is not possible
to implement in an economic way i.e. without an ancilla, just by
applying a two-qudit unitary transformation to the original state
and a blank copy. They also showed that the 2-dimensional optimal
phase-covariant cloner can be realized economically while an
economical phase-covariant cloner cannot be constructed for any
dimension $d>2$. The impossibility of the construction of
phase-covariant quantum cloner without an ancilla led them to
think about the approximate economical phase-covariant quantum
cloning machine. The optimal economical phase-covariant cloning
transformation which is invariant with respect to the swapping of
the two clones and is also phase covariant is given by
\begin{eqnarray}
U|k\rangle_{a}|l\rangle_{b}&=& |k\rangle_{a}|k\rangle_{b},~~~k=l {}\nonumber\\
&=&\frac{1}{\sqrt{2}}(|k\rangle_{a}|l\rangle_{b}+
|l\rangle_{a}|k\rangle_{b}),~~~k\neq l
\end{eqnarray}
where $l,k~~\mathcal{2}~~\{1,2,........,d-1,d\}$ and initially the
blank state is prepared in the state $|l\rangle_{b}$. \\
The corresponding fidelity of economical cloning is given by
\begin{eqnarray}
F_{1,2~opt}^{econ. phase}(d)= \frac{1}{2d^{2}}[d-1+(d-1+\sqrt{2})^{2}]
\end{eqnarray}
\textbf{Remark:}\\
1.~~~$F_{1,2~opt}^{econ.
phase}~=~~F_{1,2~opt}^{phase}=\frac{1}{2}+\sqrt{\frac{1}{8}},~~~when~~~~
d=2$. In case of qubit, cheaper (or economical) phase-covariant
quantum cloning machine can be constructed with the same fidelity
as optimal phase-covariant quantum cloning machine.\\
2.~~~$F_{1,2~opt}^{econ. phase}~<~~F_{1,2~opt}^{phase},
~~~when~~~~d>2$. In higher dimensional case, approximate
phase-covariant quantum cloning machine without ancilla can be
designed at the cost of lower fidelity of cloning than the optimal
phase-covariant quantum cloning machine with ancilla.\\
3.~~~$F_{1,2~opt}^{econ.
phase},~~F_{1,2~opt}^{phase}\rightarrow\frac{1}{2}~~~as~~~~
d\rightarrow\infty$. That means in case of infinite dimensional
Hilbert space neither approximate economical phase-covariant
quantum cloning machine nor optimal phase-covariant quantum
cloning machine with ancilla play a significant role in cloning
procedure because the fidelity is very low in this case.
\subsection{\emph{Asymmetric quantum cloning}}
N.J.Cerf \cite{cerf5} brought in a new concept of quantum cloning
machine which copies the information of a quantum system into two
non-identical (approximate) clones. To implement his idea, he
introduced a family of Pauli quantum cloning machines that produce
two approximate non-identical copies and later he generalized the
Pauli quantum cloning machine to any arbitrary dimension and
constructed a family of asymmetric Heisenberg quantum cloning
machine. J.Fiurasek, R.Filip, and N.J.Cerf \cite{fiurasek3}
investigated asymmetric universal cloning in arbitrary dimension.
They proved the optimality of the universal asymmetric
$1\rightarrow2$ cloning machines and then extended the idea of
asymmetric cloning to quantum triplicators, which produce three
clones with different fidelity. S.Iblisdir, A.Acin, N.J.Cerf,
R.Filip, J.Fiurasek and N.Gisin \cite{iblisdir1} ; S.Iblisdir,
A.Acin and N.Gisin \cite{iblisdir2} investigated the optimal
distribution of quantum information over multipartite systems by
introducing the optimal asymmetric $N\rightarrow M_{A}+M_{B}$
cloning machine. The cloning machine takes N identical pure input
states and produces two sets of clones $M_{A}$ and $M_{B}$ with
fidelities $F_{A}$ and $F_{B}$ respectively. They also analyzed
the trade-off between these fidelities. The latter group also
generalized the above asymmetric quantum cloning transformation to
more than two sets of clones. The experimental implementations of
an optimal asymmetric $1\rightarrow2$ quantum cloning  of a
polarization state of photon is recently proposed by R.Filip
\cite{filip1}. The cloning transformation $N\rightarrow
M_{A}+M_{B}$ has been proven to be a useful tool when (i) studying
the security of some quantum key distribution scheme and (ii)
studying the estimation of state by keeping finite number of
clones in one set and infinite number (in the limiting sense) of
clones in another set.\\ L.-P.Lamoureux, and N.J.Cerf
\cite{lamoureux1} constructed the class of optimal $1\rightarrow2$
phase-covariant quantum cloning machines in any dimension and then
extended the concept to the class of asymmetric quantum cloning
machines. They studied the balance between the fidelity of two
clones and concluded that the relative
fidelity between two clones decreases with the dimension.\\
The following notions are used in the discussion: \\
\textbf{Definition-1.1:} The cloning machines which produce two
approximate non-identical clones are known as asymmetric cloning
machines. \\
\textbf{Definition-1.2:} A Pauli channel is defined using a group
of four error operators, (the three Pauli matrices
$\sigma_{x},\sigma_{y},\sigma_{z}$ and I) which act on state
$|\psi\rangle$ by either rotating it by one of the Pauli matrices
or leaving it unchanged.\\
\textbf{Definition-1.3:} Asymmetric cloning machines that produce
two output qubits, each emerging from a Pauli channel, are called
asymmetric Pauli cloning machines.\\\\
\textbf{1. Asymmetric Pauli cloning machines }\\
If the input qubit X of the Pauli channel is initially in a fully
entangled state with a reference qubit R that is unchanged while X
is processed by the channel, i.e. if the joint state of input
qubit X and the reference qubit is
$|\psi\rangle_{RX}=|\phi^{+}\rangle$, then the joint state of the
reference qubit R and the output Y is a mixture of the four Bell
states $|\Phi^{\pm}\rangle=\frac{1}{\sqrt{2}}(|00\rangle\pm
|11\rangle)$ and
$|\Psi^{\pm}\rangle=\frac{1}{\sqrt{2}}(|01\rangle\pm |10\rangle)$
and can be written in the form
\begin{eqnarray}
\rho_{RY}=(1-p)|\Phi^{+}\rangle\langle\Phi^{+}|+p_{z}|\Phi^{-}\rangle\langle\Phi^{-}|+
p_{x}|\Psi^{+}\rangle\langle\Psi^{+}|+p_{y}|\Psi^{-}\rangle\langle\Psi^{-}|
\end{eqnarray}
where $p_{x},p_{y},p_{z}$ denote the probabilities with which a
qubit undergoes a phase-flip $(\sigma_{z})$, a bit-flip
$(\sigma_{x})$, or their combination
$(\sigma_{x}\sigma_{z}=-i\sigma_{y})$ in a Pauli channel and
$p=p_{x}+p_{y}+p_{z}$.\\
Instead of defining a Pauli cloning machine by a particular
unitary transformation, Cerf \cite{cerf1,cerf2} characterized a
Pauli cloning machine by the four qubit wave function
$|\psi\rangle_{RABC}$. Thus the family of Pauli cloning machines
relies on a parametrization of 4-qubit wave functions for which
all qubit pairs are in a mixture of Bell states. After cloning,
the four qubits R,A,B,C are in a pure state for which $\rho_{RA}$
and $\rho_{RB}$ are mixtures of Bell states (i.e. A and B emerge
from a Pauli channel). Also it is assumed that $\rho_{RC}$ to be a
Bell mixture.\\
The four qubit wave function $|\psi\rangle_{RABC}$ for the bipartite partition
RA versus BC can be written as a superposition of double Bell states
\begin{eqnarray}
|\psi\rangle_{RA;BC}=\{v|\Phi^{+}\rangle|\Phi^{+}\rangle+z|\Phi^{-}\rangle|\Phi^{-}\rangle
+x|\Psi^{+}\rangle|\Psi^{+}\rangle+y|\Psi^{-}\rangle|\Psi^{-}\rangle\}_{RA;BC}
\end{eqnarray}
where x,y,z,v are complex amplitudes satisfying the condition
$|x|^{2}+|y|^{2}+|z|^{2}+|v|^{2}=1$. For simplicity other possible
permutations of the Bell states are not considered here. The first
output A emerges from a Pauli channel with probabilities
$p_{x}=|x|^{2},p_{y}=|y|^{2}$ and $p_{z}=|z|^{2}$.\\
An interesting property of these double Bell states is that they
transform into superpositions of double Bell states for the two
remaining partitions of the four qubits RABC into two other pairs
(RB versus AC) and (RC versus AB).\\
Therefore, the four qubit wave function $|\psi\rangle_{RABC}$ can
also be written for the partition RB versus AC as
\begin{eqnarray}
|\psi\rangle_{RB;AC}=\{v'|\Phi^{+}\rangle|\Phi^{+}\rangle+z'|\Phi^{-}\rangle|\Phi^{-}\rangle
+x'|\Psi^{+}\rangle|\Psi^{+}\rangle+y'|\Psi^{-}\rangle|\Psi^{-}\rangle\}_{RB;AC}
\end{eqnarray}
with
\begin{eqnarray}
{}\nonumber\\&& v'=\frac{1}{2}(v+z+x+y){}\nonumber\\&&
z'=\frac{1}{2}(v+z-x-y){}\nonumber\\&&
x'=\frac{1}{2}(v-z+x-y){}\nonumber\\&& y'=\frac{1}{2}(v-z-x+y)
\end{eqnarray}
Equation (1.129) implies that when tracing over half of the
system, the second output B emerges from a Pauli channel with
probabilities $q_{x}=|x'|^{2},q_{y}=|y'|^{2}$ and
$q_{z}=|z'|^{2}$.\\
The partition RC versus AB
\begin{eqnarray}
|\psi\rangle_{RC;AB}=\{v''|\Phi^{+}\rangle|\Phi^{+}\rangle+z''|\Phi^{-}\rangle|\Phi^{-}\rangle
+x''|\Psi^{+}\rangle|\Psi^{+}\rangle+y''|\Psi^{-}\rangle|\Psi^{-}\rangle\}_{RC;AB}
\end{eqnarray}
describe the third output C which emerges from a Pauli channel
with probabilities $|x''|^{2}$, $|y''|^{2}$, $|z''|^{2}$ and
$|v''|^{2}$ and these probabilities are related with $|x|^{2}$,
$|y|^{2}$, $|z|^{2}$ and $|v|^{2}$ in the following way:
\begin{eqnarray}
{}\nonumber\\&& v''=\frac{1}{2}(v+z+x-y){}\nonumber\\&&
z''=\frac{1}{2}(v+z-x+y){}\nonumber\\&&
x''=\frac{1}{2}(v-z+x-y){}\nonumber\\&& y''=\frac{1}{2}(v-z-x-y)
\end{eqnarray}
Now the trade-off between the quality of the two copies produced
by an asymmetric Pauli cloning machine can be studied by writing
the wave function of the whole system of four particles RABC in
the following way:
\begin{eqnarray}
|\psi\rangle_{RABC}=
\sum_{m,n=0}^{1}\alpha_{m,n}|\Phi_{m,n}\rangle_{RA}|\Phi_{m,-n}\rangle_{BC}=
\sum_{m,n=0}^{1}\beta_{m,n}|\Phi_{m,n}\rangle_{RB}|\Phi_{m,-n}\rangle_{AC}
\end{eqnarray}
where $|\Phi_{m,n}\rangle$ denotes the Bell basis.\\
The relation between the coefficients $\alpha_{m,n}$ and $\beta_{m,n}$ is given by
\begin{eqnarray}
\beta_{m,n}= (\frac{1}{2})
\sum_{x,y=0}^{1}e^{i\pi(nx-my)}\alpha_{x,y}
\end{eqnarray}
This shows that if one output copy (say, A) is close to perfect,
then second output copy B is close to imperfect (i.e. very noisy)
and vice versa.\\
\textbf{2. Asymmetric Heisenberg cloning machines}\\
Cerf \cite{cerf2} generalized the asymmetric Pauli cloning machine
to systems of arbitrary dimensions d and defined a family of
asymmetric cloning machines that produces two imperfect copies of
the state of an N-dimensional quantum system that emerge from
non-identical
Heisenberg channels.\\
The corresponding family of asymmetric quantum cloning machines
are called asymmetric Heisenberg cloning machines. A Heisenberg
channel is characterized by the $N^{2}$-dimensional probability
distribution $\textbf{p}$ of error operators which a quantum state
undergoes in the channel.\\
The generalized Bell basis is defined by
\begin{eqnarray}
|\Phi_{m,n}\rangle=\frac{1}{\sqrt{d}}\sum_{k=0}^{d-1}exp(\frac{2\pi
i k n}{d})|k\rangle|(k+m)modulo~d\rangle
\end{eqnarray}
Let us suppose that the input state we wish to clone
$|\psi\rangle_{A}$ is prepared in the maximally entangled state
$|\Phi_{0,0}\rangle$, given by (1.134), with a reference state R.
The cloning machine is described by a unitary operation U acting
on a four qubit state, namely the initial state and another two
d-level systems initially prepared in the state $|0\rangle_{B}$
and $|0\rangle_{C}$:
\begin{eqnarray}
U|\Phi_{0,0}\rangle_{RA}|00\rangle_{BC}=|\psi\rangle_{RABC}&=&
\sum_{m,n=0}^{d-1}\alpha_{m,n}|\Phi_{m,n}\rangle_{RA}|\Phi_{m,-n}\rangle_{BC}{}\nonumber\\&=&
\sum_{m,n=0}^{d-1}\beta_{m,n}|\Phi_{m,n}\rangle_{RB}|\Phi_{m,-n}\rangle_{AC}
\end{eqnarray}
where A, B denotes the two clones, C is the ancilla (cloning
machine), $|\Phi_{m,n}\rangle$ is the generalized Bell state and
\begin{eqnarray}
\beta_{m,n}= (\frac{1}{d}) \sum_{x,y=0}^{d-1}exp~(\frac{2\pi
i(nx-my)}{d})~\alpha_{x,y}
\end{eqnarray}
The action of a Heisenberg cloning machine on an arbitrary input
state is given by
\begin{eqnarray}
U|\psi\rangle_{A}|00\rangle_{BC}=|\chi\rangle_{ABC}=\sum_{m,n=0}^{d-1}\alpha_{m,n}U_{m,n}|\psi\rangle_{A}|\Phi_{m,-n}\rangle_{BC}
\end{eqnarray}
where $\sum_{m,n=0}^{d-1}|\alpha_{m,n}|^{2}=1$ and $U_{m,n}$ are
the error operators which define the Heisenberg group:
\begin{eqnarray}
U_{m,n}= \sum_{k=0}^{d-1}exp~(\frac{2\pi i k
n}{d})~|(k+m)~modulo~d\rangle\langle k|
\end{eqnarray}
We note that for two dimensional case, $U_{m,n}$ become the Pauli
operators.\\
After cloning, the two output copies are described by the density
operators
\begin{eqnarray}
\rho_{A}= Tr_{BC}(|\chi\rangle_{ABC}\langle
\chi|)=\sum_{m,n=0}^{d-1}|\alpha_{m,n}|^{2}|\psi_{m,n}\rangle\langle\psi_{m,n}|\\
\rho_{B}= Tr_{AC}(|\chi\rangle_{ABC}\langle
\chi|)=\sum_{m,n=0}^{d-1}|\beta_{m,n}|^{2}|\psi_{m,n}\rangle\langle\psi_{m,n}|
\end{eqnarray}
where
\begin{eqnarray}
|\psi_{m,n}\rangle=U_{m,n}|\psi\rangle
\end{eqnarray}
Furthermore, Ghiu \cite{ghiu1} studied and analyzed the class of
universal asymmetric cloning machines for d-level systems. These
cloning machines are universal in the sense that they generate the
outputs which are independent of the input state. He also showed
that the universal cloning machine is optimal in the sense that a
cloning machine creates one clone with maximal fidelity for a
given fidelity of the other clone.\\
The optimal universal asymmetric Heisenberg cloning machine is
given by
\begin{eqnarray}
{}\nonumber\\&&U|j\rangle|00\rangle=\frac{1}{\sqrt{1+(d-1)(p^{2}+q^{2})}}(|j\rangle|j\rangle|j\rangle
+p\sum_{s=1}^{d-1}|j\rangle|(j+s)~modulo~d\rangle\otimes{}\nonumber\\&&|(j+s)~modulo~d\rangle+
q\sum_{s=1}^{d-1}|(j+s)~modulo~d\rangle|j\rangle|(j+s)~modulo~d\rangle)
\end{eqnarray}
where $|j\rangle$ is the computational basis,
$\alpha_{m,n}=\mu,~\forall~(m,n)\neq(0,0)$, $\alpha_{0,0}=\nu$,
$p=\frac{(\nu-\mu)}{[\nu+(d-1)\mu]}$ and
$q=\frac{d\mu}{[\nu+(d-1)\mu]}~=1-p$.\\
After operating optimal universal asymmetric Heisenberg cloning
transformation on the input state
$|\psi\rangle=\sum_{j=0}^{d-1}\alpha_{j}|j\rangle$, the output
states are described by the density operators
\begin{eqnarray}
\rho_{A}=\frac{1}{\sqrt{1+(d-1)(p^{2}+q^{2})}}\{[1-q^{2}+(d-1)p^{2}]|\psi\rangle\langle\psi|+q^{2}I\}\\
\rho_{B}=\frac{1}{\sqrt{1+(d-1)(p^{2}+q^{2})}}\{[1-p^{2}+(d-1)q^{2}]|\psi\rangle\langle\psi|+p^{2}I\}
\end{eqnarray}
To quantify the quality of the copies produced from universal
asymmetric Heisenberg cloning machine, the fidelities are to be
calculated. The fidelities of the two non-identical clones
described by the density operators $\rho_{A}$ and $\rho_{B}$ are
given by
\begin{eqnarray}
F_{A}=\langle\psi|\rho_{A}|\psi\rangle=\frac{1+(d-1)p^{2}}{1+(d-1)(p^{2}+q^{2})}\\
F_{B}=\langle\psi|\rho_{B}|\psi\rangle=\frac{1+(d-1)q^{2}}{1+(d-1)(p^{2}+q^{2})}
\end{eqnarray}
The universal asymmetric Heisenberg cloning machine produces the
best quality copies when $F_{A}+F_{B}$ takes the maximum value. It
can be shown that the maximum value of $F_{A}+F_{B}$ will be
attained when $p=q=\frac{1}{2}$. Inserting $p=q=\frac{1}{2}$ in
equations (1.145-1.146), we get
\begin{eqnarray}
F_{A}=F_{B}=\frac{d+3}{2(d+1)}
\end{eqnarray}
Therefore, equation (1.142) represents the general expression for
the optimal universal asymmetric Heisenberg cloning machine and it
reduces to optimal universal symmetric cloning machine when
$p=q=\frac{1}{2}$.
\section{\emph{Quantum cloning of mixed state}}
In this section we want to focus on the approximate copying of
mixed state. In quantum cryptography, the sender encodes the
information into two non-orthogonal pure states. Then the
information is communicated through a communication channel.
Actually, in reality a communication channel will inevitably
suffer from noise that will have caused the pure states to evolve
to mixed states. If the third party (Eve) intercepts the message
in a midway and wants to extract the information encoded in a
message without revealing her presence to the sender and receiver,
she has to clone the intercepted mixed state with maximum possible
accuracy. Hence the approximate cloning of mixed state is
interesting and important in the field of quantum cryptography.\\
In 2003, A.E.Rastegin \cite{rastegin2} defined the global fidelity
for mixed states in the same way as for pure states. The global
fidelity for mixed state cloning for the set
$\{\rho_{1},~\rho_{2}\}$ can be defined as
\begin{eqnarray}
F_{G}^{M}=\frac{1}{2}[F(\widetilde{\rho_{1}},\rho_{1}\otimes\rho_{1})
+F(\widetilde{\rho_{2}},\rho_{2}\otimes\rho_{2})]
\end{eqnarray}
where $\widetilde{\rho_{1}},\widetilde{\rho_{2}}$ denotes the
output of the system respectively.\\
He also found the upper bound of the global fidelity $F_{G}^{M}$
of mixed state cloning for the set $S=\{\rho_{1},\rho_{2}\}$:
\begin{eqnarray}
Upper~ bound~of~ F_{G}^{M}= \frac{1}{2}[1+f^{3}+(1-f^{2})\sqrt{1+f^{2}}]
\end{eqnarray}
where $f=\sqrt{F(\rho_{1},\rho_{2}})$.\\
If both states to be cloned are pure, then the definition of
global fidelity of mixed state reduces to the definition of global
fidelity of pure state and also the upper bound of the fidelity
$F_{G}^{M}$ coincides with the upper bound of the fidelity for
pure state cloning. But the state dependent cloner for pure state
cloning cannot be used for mixed state cloning because of the two
reasons: (i) State dependent cloning transformation requires no
auxiliary system and (ii) The initial state of the copy mode is
pure. Further, Rastegin \cite{rastegin3} extended the known
global-fidelity limits of state-dependent cloning to mixed quantum
states. He \cite{rastegin1} also extend the concept of the
relative error to the mixed-state cloning and obtained a lower
bound of it. He had also shown that the lower
bound of the relative error contributes to the stronger no-cloning theorem \cite{jozsa1}. \\
H.Fan \cite{fan5} proposed a quantum cloning machine for arbitrary
mixed states in symmetric subspace and showed that the introduced
quantum cloning machine can be used to copy part of the output
state of another quantum cloning machine (e.g. Gisin-Massar
$1\rightarrow3$ quantum cloning machine) and is useful in quantum
computation and quantum information. His proposed 2 to 3 quantum
quantum cloning machine for mixed state is given by
\begin{eqnarray}
U|2\uparrow\rangle\otimes R=\frac{\sqrt{3}}{2}|3\uparrow\rangle\otimes
R_{\uparrow}+\frac{1}{2}|2\uparrow,\downarrow\rangle\otimes R_{\downarrow}
\end{eqnarray}
\begin{eqnarray}
U|2\downarrow\rangle\otimes R=\frac{1}{2}|\uparrow,2\downarrow\rangle\otimes
R_{\uparrow}+\frac{\sqrt{3}}{2}|\downarrow\rangle\otimes R_{\downarrow}
\end{eqnarray}
\begin{eqnarray}
U|\uparrow,\downarrow\rangle\otimes
R=\frac{1}{\sqrt{2}}|2\uparrow,\downarrow\rangle\otimes
R_{\uparrow}+\frac{1}{\sqrt{2}}|\uparrow,2\downarrow\rangle\otimes R_{\downarrow}
\end{eqnarray}
where state $|2\uparrow,\downarrow\rangle$ is a normalized
symmetric state with 2 spin up and 1 spin down.\\
The quantum cloning machine (1.150-1.152) produced three copies of
the 2-qubit mixed state with fidelity $\frac{79}{108}$.\\
Recently, Fan, Liu and Shi \cite{fan6} studied the quantum cloning
of two identical mixed qubits $\rho\otimes\rho$. They proposed a
general quantum cloning machine which creates M copies from 2
identical mixed qubits. Their quantum cloning transformation is
given by
\begin{eqnarray}
U|\uparrow\uparrow\rangle\otimes
R=\sum_{k=0}^{M-2}\alpha_{0k}|(M-k)\uparrow,k\downarrow\rangle\otimes R_{k}
\end{eqnarray}
\begin{eqnarray}
U|\downarrow\downarrow\rangle\otimes
R=\sum_{k=0}^{M-2}\alpha_{2k}|(M-2-k)\uparrow,(2+k)\downarrow\rangle\otimes R_{k}
\end{eqnarray}
\begin{eqnarray}
\frac{1}{\sqrt{2}}U(|\uparrow\downarrow\rangle+|\downarrow\uparrow\rangle)\otimes
R=\sum_{k=0}^{M-2}\alpha_{1k}|(M-1-k)\uparrow,(1+k)\downarrow\rangle\otimes R_{k}
\end{eqnarray}
\begin{eqnarray}
\frac{1}{\sqrt{2}}U(|\uparrow\downarrow\rangle-|\downarrow\uparrow\rangle)\otimes
R=\sum_{k=0}^{M-2}\alpha_{1k}|(M-1-\widetilde{k)\uparrow},(1+k)\downarrow\rangle\otimes
R_{k}
\end{eqnarray}
where
\begin{eqnarray}
\alpha_{jk}=\sqrt{\frac{6(M-2)!(M-j-k)!(j+k)!}{(2-j)!(M+1)!(M-2-k)!j!k!}},~~~~j=0,1,2
\end{eqnarray}
The state $|i\uparrow,j\downarrow\rangle$ is a completely symmetrical state with i
spins up and j spins down, the state $|i\widetilde{\uparrow,j}\downarrow\rangle$ is
orthogonal to $|i\uparrow,j\downarrow\rangle$. $R_{k}$ represents the final
orthogonal machine states.\\
Each output of the quantum cloning machine is described by the reduced density
operator
\begin{eqnarray}
\rho_{red.}^{(out)}=\frac{M+2}{2M}\rho+\frac{M-2}{4M}I
\end{eqnarray}
Since the reduction factor $\frac{M+2}{2M}$ achieves the optimal
bound so the quantum cloning machine (1.153-1.157) copies the two
identical mixed qubits and two identical pure states optimally.
\section{\emph{Quantum cloning and no-signalling}} The non-local property
of the quantum mechanics cannot be used for superluminal
signalling. This fact has already been vividly discussed in the
past
\cite{ghirardi1,ghirardi2,herbert1,navez1,peres2,scherer1,wootters1}.
If a perfect quantum cloning machine were available, Bob could
generate an infinite number of copies of his state, and therefore
would be able to determine his state with perfect accuracy, thus
knowing what basis Alice decided to use. In this way, transfer of
information between Alice and Bob would be possible. In
particular, if they are space-like separated, information could be
transmitted with superluminal speed, but no-cloning theorem ruled
out this possibility. However, imperfect quantum cloning machines
exist \cite{bruss2,buzek1,gisin1,niu1}. So, naturally a question
arises, whether approximate quantum cloning machine can make
possible the superluminal signalling or not? In 1998, Gisin
\cite{gisin2} first attacked this problem and showed that any
approximate optimal quantum cloning cannot lead to signalling.
Therefore the construction of optimal quantum cloning machine does
not violate the "peaceful coexistence" between quantum mechanics
and relativity. He used the no-signalling constraint to derive a
bound on the fidelity of quantum cloning machine and showed that
this bound coincides with the fidelity of the Buzek-Hillery
universal quantum cloning machine. This result again proves that
Buzek-Hillery universal quantum cloning machine is optimal. After
this work, many work had been done on quantum cloning and
signalling. Ghosh, Kar and Roy \cite{ghosh1} used the
no-signalling constraint to find the optimality of the universal
asymmetric quantum cloning machine of Buzek, Hillery and Bednik \cite{buzek6}.\\
Bruss, Ariano, Macchiavello and Sacchi \cite{bruss4} showed that
any linear trace-preserving map forbids superluminal signalling
but converse is not true. The no-signalling condition implies only
linearity but does not imply the two important properties of
quantum operations namely positivity and trace-preservation.
Hence, there exist some maps that go beyond quantum mechanics, but
still preserve the constraint of no-superluminal signalling. They
also gave an example to explain the fact that the cloning fidelity
is unrelated to the no-signalling condition and hence any bound on
a cloning fidelity cannot be derived from the no-signalling
constraint alone. Quantum mechanics as a complete theory,
guarantees no- superluminal signalling, and gives the correct known upper bounds on
the fidelity of quantum cloning.\\
Duan and Guo \cite{duan1} introduced a probabilistic quantum
cloning machine which can be used to produce the perfect clones of
the quantum states secretly chosen from a certain set of linearly
independent states, with some probability less than unity.
Although probabilistic quantum cloning machine produces perfect
clones but it cannot be used for superluminal signalling
\cite{pati4}. Hardy and Song \cite{hardy1} showed that
no-signalling condition lead to the constraints on probabilistic
quantum cloning machine i.e. if probabilistic quantum cloning
machine produces exact clones of (d+1) number of quantum states,
in which d number of states are linearly independent, then there
will be signalling. Further, Ghosh, Kar, Kunkri and Roy
\cite{ghosh2} used the technique of remote state preparation to
prove the Hardy and Song's result in a more simpler way. They
showed that probabilistic exact cloning of any three different
states of a qubit implies (probabilistic) signalling in the sense,
that one can extract more than 1 classical bit message
probabilistically by communicating 1 classical bit only. They also
generalized this result in d-dimensional Hilbert space.
\section{\emph{Quantum Deletion machine}}
A.K.Pati and S.L.Braunstein \cite{pati5} were the first to observe
the fact that it is not possible to delete the information content
of one or more photons by a physical process. That is, the
linearity of quantum theory forbids deleting one unknown quantum
state against a copy in either a reversible or an irreversible
manner. This phenomenon is called "quantum no-deleting" principle.
This principle is complementary to the "quantum no-cloning
theorem". If quantum deleting could be done, then one would create
a standard blank state onto which one could copy an unknown state
approximately, by deterministic cloning or exactly, by
probabilistic cloning. Therefore, when memory in a quantum
computer is scare, quantum deleting may play an important role,
and one could store new information in an already computed state
by deleting the old information.\\
We can understand the principle behind quantum deletion more
clearly, if we compare quantum deletion with the "Landauer erasure
principle" \cite{land1}. It tells us that a single copy of some
classical information can be erased at some energy cost. It is an
irreversible operation. In quantum information theory, erasure of
a single unknown state may be thought of as swapping it with some
standard state and then dumping it into the environment. Unlike
quantum erasure, quantum deletion is a different concept. Quantum
deletion is more like reversible 'uncopying' of an unknown quantum
state. The essential difference is that irreversible erasure
naturally carries over from the classical to the quantum world,
whereas the analogous uncopying of classical information is
impossible for quantum information. Pati and Braunstein
\cite{pati9} had shown that the violation of no-deletion principle
can lead to superluminal signalling using non-local entangled
states. Therefore, no-deletion principle supports the "peaceful
co-existence" between quantum mechanics and relativity. However,
the (Landauer) erasure of information does not allow for any
signalling. This fact provides another evidence in support of the
statement that quantum deletion is fundamentally a different
operation than
erasure. \\
Although there is not a perfect deleting machine, the
corresponding no-deleting principle does not prohibit us from
constructing the approximate deleting machine. Pati et.al.
\cite{pati6} studied the distribution of the quantum information
among various subsystems during the deletion process. They
introduced a state dependent, approximate
quantum deletion machine and named it as conditional deletion machine.\\
The conditional deletion machine (deletion transformation) for
orthogonal qubits is defined by
\begin{eqnarray}
|0\rangle|0\rangle|A\rangle\rightarrow|0\rangle|\Sigma\rangle|A_{0}\rangle\\
|1\rangle|1\rangle|A\rangle\rightarrow|1\rangle|\Sigma\rangle|A_{1}\rangle\\
|0\rangle|1\rangle|A\rangle\rightarrow|0\rangle|1\rangle|A\rangle\\
|1\rangle|0\rangle|A\rangle\rightarrow|1\rangle|0\rangle|A\rangle
\end{eqnarray}
where $|A\rangle$ is the initial state and
$|A_{0}\rangle$,$|A_{1}\rangle$ are the final states of ancilla
and they are mutually orthogonal to each other. The speciality of
the introduced deletion machine lies in the fact that if the two
input qubits are identical then it deletes a copy but if they are
different then it allows them to pass through without any change.\\
Let
\begin{eqnarray}
|\Psi\rangle=\alpha|0\rangle+\beta|1\rangle
\end{eqnarray}
with $|\alpha|^{2}+|\beta|^{2}=1$ be any arbitrary quantum state.\\
Each of a copy from two copies of an arbitrary quantum state
$|\psi\rangle$ can be approximately deleted by the deletion
transformation (1.159-1.162) and it will create the following
state
\begin{eqnarray}
|\Psi\rangle_{a}|\Psi\rangle_{b}|A\rangle_{c}&\rightarrow&\alpha^{2}|0\rangle_{a}|\Sigma\rangle_{b}|A_{0}\rangle_{c}
+\beta^{2}|1\rangle_{a}|\Sigma\rangle_{b}|A_{1}\rangle_{c}+\alpha\beta(|01\rangle_{ab}+|10\rangle_{ab})|A\rangle_{c}
{}\nonumber\\&&=|\Psi_{out}\rangle_{abc}.
\end{eqnarray}
The reduced density matrix of the two qubits $\textit{ab}$ after
the deletion operation is given by
\begin{eqnarray}
\rho_{ab}=tr_{c}(|\Psi_{out}\rangle_{abc}\langle\Psi_{out}|)&=&|\alpha|^{4}|0\rangle\langle0|\otimes
|\Sigma\rangle\langle\Sigma|+|\beta|^{4}|1\rangle\langle1|\otimes
|\Sigma\rangle\langle\Sigma|+{}\nonumber\\&&2|\alpha|^{2}|\beta|^{2}|\psi^{+}\rangle\langle\psi^{+}|
\end{eqnarray}
where $|\psi^{+}\rangle=\frac{1}{\sqrt{2}}(|01\rangle+|10\rangle)$.\\
The reduced density matrix for the qubit in the mode 'a' and 'b'
respectively are
\begin{eqnarray}
\rho_{a}=tr_{b}(\rho_{ab})=|\alpha|^{4}|0\rangle\langle0|+|\beta|^{4}|1\rangle\langle1|+
|\alpha|^{2}|\beta|^{2}I\\
\rho_{b}=tr_{a}(\rho_{ab})=(1-2|\alpha|^{2}|\beta|^{2})|\Sigma\rangle\langle\Sigma|+
|\alpha|^{2}|\beta|^{2}I
\end{eqnarray}
The fidelity of the qubit in mode 'a' is given by
\begin{eqnarray}
F_{a}=\langle\Psi|\rho_{a}|\Psi\rangle=1-2|\alpha|^{2}|\beta|^{2}
\end{eqnarray}
When $\alpha=\beta=\frac{1}{\sqrt{2}}$, i.e. for an equal
superposition of qubit state, the fidelity of the qubit in mode
'a' reduces to $\frac{1}{2}$. The average fidelity is
$\overline{F_{a}}=\frac{2}{3}\approx0.66$. This shows that the
first mode of the qubit is not faithfully retained during the deletion operation.\\
The fidelity of deletion is given by
\begin{eqnarray}
F_{b}=\langle\Sigma|\rho_{b}|\Sigma\rangle=1-|\alpha|^{2}|\beta|^{2}
\end{eqnarray}
For an equal superposition of qubit state the fidelity of deletion
takes the value $\frac{3}{4}$ which is the maximum limit for
deleting an unknown qubit. Furthermore, we can observe that for a
classical bit, i.e. for either $\alpha=0$ and $\beta=1$ or
$\alpha=1$ and $\beta=0$, the qubit in mode 'b' is perfectly
deleted and
simultaneously the deletion machine faithfully retained the qubit in mode 'a'.\\
Since the fidelity of deletion $F_{b}$ depends on the input state
so it is important to calculate the average fidelity and it is
given by $\overline{F_{b}}=\frac{5}{6}\approx0.83$.\\
As we see that quantum deletion machine introduced by Pati et.al.
is state dependent, so it is natural to ask the question as in the
case of quantum cloning machines, whether there exists any quantum
deletion machine which works in a similar fashion for all
arbitrary input states? D.Qiu \cite{qiu1} was the first who
attempted to answer the above question and got success partially.
He verified that some standard universal quantum deleting machine
does not exist. Not only that he constructed a universal deletion
machine but unfortunately the machine was found to be non optimal
in the sense of low fidelity of deletion.\\
A non-optimal universal quantum deletion machine \cite{qiu1} is
defined by
\begin{eqnarray}
U|0\rangle|0\rangle|Q\rangle=\frac{1}{\sqrt{2}}|0\rangle|A\rangle+\frac{1}{\sqrt{2}}|1\rangle|B\rangle\\
U|1\rangle|1\rangle|Q\rangle=\frac{1}{\sqrt{2}}i|1\rangle|B\rangle-\frac{1}{\sqrt{2}}i|0\rangle|A\rangle\\
U|0\rangle|1\rangle|Q\rangle=|0\rangle|1\rangle\\
U|1\rangle|0\rangle|Q\rangle=|1\rangle|0\rangle
\end{eqnarray}
where for any real numbers $r_{1}$,$r_{2}$ with
$r_{1}^{2}+r_{2}^{2}=1$, if
$|A\rangle=r_{1}|0\rangle+r_{2}|1\rangle$, then $|B\rangle=r_{2}|0\rangle-r_{1}|1\rangle$.\\
Using deletion machine (1.170-1.173), one can delete a copy of a
qubit from two identical copies with fidelity $\frac{1}{2}$.
Although the fidelity of deletion is input state independent but
its value does not give any satisfactory result. Therefore, the
prescribed deletion machine (1.170-1.173) is universal but it is
not an optimal one. Recently, we designed a universal quantum
deletion machine which improves
the fidelity of deletion from 0.5 and takes it to 0.75 in the limiting sense \cite{adhikari2}.\\
Also W.Song, M.Yang and Z-L Cao \cite{song2} constructed a state
dependent quantum deleting machine without considering the
ancilla.\\
It is described by the following unitary transformation
\begin{eqnarray}
U|\psi_{i}^{N}\rangle=|\phi_{i}\rangle|\Sigma\rangle^{\otimes (N-M)}
\end{eqnarray}
where $|\psi_{i}^{N}\rangle$ are the N-fold tensor product states
$|\psi_{i}^{N}\rangle=|\psi_{i}\rangle_{1}\otimes......\otimes|\psi_{i}\rangle_{N}$
which are prepared in the same state, and $|\psi_{i}\rangle$ is
chosen from a set of K non-orthogonal states, $|\phi_{i}\rangle$
is the output state after the machine deleting
$|\psi_{i}\rangle^{\otimes (N-M)}$. The global fidelity which
characterizes the distance between the output state
$|\phi_{i}\rangle$ and the ideal state $|\psi_{i}\rangle^{\otimes
(M)}$ is defined by
\begin{eqnarray}
F= \sum_{i=1}^{K}\eta_{i}|\langle|\psi_{i}^{M}|\phi_{i}\rangle|^{2}
\end{eqnarray}
where $\eta_{i}$ denotes a priori probability of the state $|\psi_{i}\rangle^{\otimes
(N)}$.\\
They found the optimal value of the global fidelity when K=2 and it is given by
\begin{eqnarray}
F^{(opt)}=
\frac{1}{2}\{1+[1-4\eta_{1}\eta_{2}\sin^{2}(2\theta-\varphi_{1}+\varphi_{2})]^{\frac{1}{2}}\}.
\end{eqnarray}
where $\eta_{1}+\eta_{2}=1$ and
$|\psi_{1}^{M}\rangle=\cos\theta|\alpha\rangle+\sin\theta|\beta\rangle$,
$|\psi_{2}^{M}\rangle=\cos\theta|\alpha\rangle-\sin\theta|\beta\rangle$,
$|\phi_{1}^{M}\rangle=\cos\varphi_{1}|\alpha\rangle+\sin\varphi_{1}|\beta\rangle$,
$|\phi_{2}^{M}\rangle=\cos\varphi_{2}|\alpha\rangle+\sin\varphi_{2}|\beta\rangle$. The
states $|\alpha\rangle$ and $|\beta\rangle$ are orthonormal basis  for the subspace
spanned by $|\psi_{1}^{M}\rangle$ and $|\psi_{2}^{M}\rangle$.\\
The optimal global fidelity can attain the value one when one of
the a priori probabilities is zero.\\
The possibility of perfect deletion with some probability less
than one cannot be ruled out \cite{feng1,qiu2}. J. Feng, Y-F Gao,
J-S Wang and M-S Zhan \cite{feng1} designed a probabilistic
quantum deletion machine and showed that each of the two copies of
non-orthogonal and linearly independent quantum states can be
probabilistically deleted by a general unitary-reduction
operation. Their prescribed quantum deletion machine can be
described by the following unitary operation U \cite{feng1}:
\begin{eqnarray}
U(|\psi_{i}\rangle|\psi_{i}\rangle|P_{0}\rangle)=\sqrt{b_{i}}|\psi_{i}\rangle|\Sigma\rangle
|P_{i}\rangle+\sum_{l=1}^{k^{2}}\sqrt{f_{i}^{(l)}}|\mu_{l}\rangle|P_{0}\rangle~~~~~(i=1,2,....,k)
\end{eqnarray}
where $|\psi_{i}\rangle|\psi_{i}\rangle$ (i=1,2,...,k) are the
input normalized states of the system D which belongs to a Hilbert
space of dimension $k^{2}$, and $|\mu_{l}\rangle$
$(l=1,2,....,k^{2})$ are the orthonormal basis states of above
space. $|\Sigma\rangle$ is the normalized standard blank state in
Hilbert space of dimension k and $|P_{i}\rangle$ (i=0,1,2,....,k)
are normalized states of the probe system P with a
$k_{p}$-dimensional Hilbert space $(k_{p}\geq k+1)$.
$|P_{0}\rangle, |P_{1}\rangle,
|P_{2}\rangle,.......,|P_{k}\rangle$ are not generally orthogonal,
but each of
$|P_{i}\rangle$ (i=1,2,.....,k) is orthogonal to $|P_{0}\rangle$.\\
They also generalized the results of $2\rightarrow1$ probabilistic
deleting to the case of $N\rightarrow M$ deleting (N, M are
positive integers and $N>M$).

\large \baselineskip .85cm

\chapter{Hybrid quantum cloning}
\setcounter{page}{55} \markright{\it CHAPTER~\ref{chap2}. Hybrid
quantum cloning}
\label{chap2}%
Insofar as mathematics is about reality, it is not certain, and
insofar as it is certain, it is not about reality - Albert
Einstein\\\\
Quantum mechanics and relativity, taken together, are
extraordinarily restrictive, and they therefore provide us with a
great logical machine. We can explore with our minds any number of
possible universes consisting of all kinds of mythical particles
and interactions, but all except a very few can be rejected on a
priori grounds because they are not simultaneously consistent with
special relativity and quantum mechanics. Hopefully in the end we
will find that only one theory is consistent with both and that
theory will determine the nature of our particular universe -
Steven Weinberg
\section{\emph{Prelude}}
A fundamental restriction in quantum theory is that quantum
information cannot be copied perfectly \cite{wootters1} in
contrast to the information we talk about in classical world.
Similarly, it is known that quantum information cannot be deleted
against a copy. But if we pay some price, then approximate or
probabilistic cloning
\cite{bruss7,cerf3,chiribella1,duan1,niu2,roy1,scarani1,zhang3}
and deletion operations
\cite{adhikari1,adhikari2,adhikari4,pati5,qiu3} are possible. For
example, it does not prohibit the possibility of approximate
cloning of an arbitrary state of a quantum mechanical system. The
existence of Universal Copying Machine' (UCM) created a class of
approximate cloning machines which are independent of the input
state \cite{bruss5,buzek1,buzek3,gisin1,ying2}. The optimality of
such cloning transformations has been verified \cite{gisin1}.
There also exists another class of copying machines which are
state dependent. The original proof of the no-cloning theorem was
based on the linearity of the evolution. Later it was shown that
the unitarity of quantum theory also forbids us from accurate
cloning of non-orthogonal states with certainty
\cite{d'ariano2,yuen1}. But as discussed in section-1.4.3,
non-orthogonal states secretly chosen from a set can be faithfully
cloned with certain probabilities \cite{duan1,duan2} or can evolve
into a linear superposition of multiple-copy states together with
a failure term described by a composite state \cite{pati2} if and
only if the states are linearly independent. In quantum world it
is very important to know various limitations imposed by quantum
theory on quantum information. Recently, some general impossible
operations have been studied  by Pati \cite{pati8} in detail. This
unifies the no-cloning, no-compelementing and no-conjugating
theorems in quantum information theory. Among all the impossible
operations \cite{buzek5,pati5,wootters1,zhou1}, the impossibility
of 'cloning-cum-complementing' quantum machines attracts much
attention here in the sense that it is a combination of cloning
machine and complementing machine where the probabilities of
separately existing cloning machines are $\lambda$ and
$1-\lambda$, respectively. In the same spirit, we can imagine a
hybrid cloning machine which is a superposition of two cloning
machines \cite{pati8}. One can construct hybrid cloning machine by
combining different existing cloning transformations. Hybrid
quantum cloning machines can be divided into two groups: (i) State
dependent and (ii) State independent or Universal. The main
objective of this chapter is to
study the behavior of such types of hybrid cloning machines.\\
Before going into the discussion about the hybrid quantum cloning
machine, we would like to discuss briefly about universal
asymmetric Pauli cloning machine and universal anti-cloning
machine.\\
\textbf{Universal asymmetric Pauli cloning machine:} Asymmetric
cloning transformation \cite{cerf1, cerf2} is given by
\begin{eqnarray}
|0\rangle  |\Sigma\rangle |Q\rangle|\longrightarrow(\frac{1}{\sqrt{1+p^2+q^2}})(|0\rangle|0\rangle|\uparrow\rangle+(p|0\rangle|1\rangle+q|1\rangle|0\rangle)|\downarrow\rangle,\\
|1\rangle |\Sigma\rangle
|Q\rangle|\longrightarrow(\frac{1}{\sqrt{1+p^2+q^2}})(|1\rangle|1\rangle|\downarrow\rangle+(p|1\rangle|0\rangle+q|0\rangle|1\rangle)|\uparrow\rangle.
\end{eqnarray}
where $p + q=1$.\\
Pauli cloning machines (transformations) are nothing but
asymmetric cloning machines that generate two non-identical
approximate  copies of a single quantum bit, each output qubits
emerging from a Pauli channel (discussed in subsection 1.4.6)
\cite{cerf1}. The asymmetric quantum cloning machine play an
important role in the situation in which one of the clones need to
be a bit better than the
other.\\\\
\textsl{Table-2.1: Fidelity of the copies produced from asymmetric
Pauli cloning machine}
\begin{tabular}{| c| c| c| c|}
\hline
  parameter (p) & $(F_1)_{PCM}= \frac{(p^2+1)}{2(p^2-p+1)} $ & $(F_2)_{PCM} =\frac{(p^2-2p+2)}{2(p^2-p+1)}$& Difference between
  qualities \\ &  &  &  of the two copies\\ & & &
  $(F_1)_{PCM}\sim (F_2)_{PCM}$\\
  \hline
  0.0 & 0.50 & 1.00 & 0.50 \\
  \hline
  0.1 & 0.55 & 0.99 & 0.44 \\
  \hline
  0.2 & 0.62 & 0.98 & 0.36 \\
  \hline
  0.3 & 0.69 & 0.94 & 0.25 \\
  \hline
  0.4 & 0.76 & 0.89 & 0.13 \\
  \hline
  0.5 & 0.83 & 0.83 & 0.00 (Symmetric copies) \\
  \hline
  0.6 & 0.89 & 0.76 & 0.13 \\
  \hline
  0.7 & 0.94 & 0.69 & 0.25 \\
  \hline
  0.8 & 0.98 & 0.62 & 0.36 \\
  \hline
  0.9 & 0.99 & 0.55 & 0.44 \\
  \hline
  1.0 & 1.00 & 0.50 & 0.50 \\ \hline
\end{tabular}\\\\
\textbf{\textsl{Illustration of the Table 2.1:\\
\textsl{The above table represents the quality of the two different
outputs from asymmetric Pauli cloning machine in terms of the
fidelity for different values of the parameter p. We find that
when $p=0$ or $p=1$, one of the output is totally undisturbed i.e.
it contains the whole information of the input quantum state while
the overlapping of the other output with the original is found to
be 0.5. For $p=0.5$, the Pauli cloning machine reduces to B-H
symmetric quantum cloning machine. We also observe here that the
Pauli quantum cloning machine gives better quality asymmetric
copies when $p=0.4$ and $p=0.6$ because the difference between the
quality of the copies is small in these cases.}}}\\\\
\textbf{Universal anti- cloning machine:} Few years earlier, Gisin
and Popescu \cite{gisin3} discovered an important fact that
quantum information is better stored in two anti-parallel spins as
compared to two parallel spins. This fact gave birth to a new type
of cloning machine called anti-cloning machine \cite{gisin3,song1}
which generates two outputs, one of the output has the same
direction as the input and the other output has direction opposite
to the input. Song and Hardy \cite{song1} constructed a universal
quantum anti-cloner which takes an unknown quantum state just as
in quantum cloner but its output as one with the same copy while
the second one with opposite spin direction to the input state.
For the Bloch vector, an input \textbf{n}, quantum anti-cloner
would have the input as
$\frac{1}{2}(\textbf{1}+\textbf{n}.\sigma)$, then it generates two
outputs, $\frac{1}{2}(\textbf{1}+\eta \mathbf{n.\sigma})$ and
$\frac{1}{2}(\textbf{1}-\eta\textbf{n}.\mathbf{\sigma})$, where
$0\leq\eta\leq1$ is the shrinking factor and the fidelity is
defined as $F=\langle\textbf{n}|\rho^{out}|\textbf{n}\rangle=
\frac{1}{2} (1+\eta)$. If spin flipping were allowed then
anti-cloner would have the same fidelity as the regular cloner
since one could clone first then flip the spin of the second copy.
However spin flipping of an unknown state is not allowed in
quantum mechanics. They also showed that the quantum state can be
anti-cloned exactly with non-zero probability.\\
The universal anti-cloning transformation is given by
\begin{eqnarray}
|0\rangle  |\Sigma\rangle  |Q\rangle&\longrightarrow&
\frac{1}{\sqrt{6}}|0\rangle|0\rangle|\uparrow\rangle
+((\frac{1}{\sqrt{2}})e^{icos^{-1}(\frac{1}{\sqrt{3}})}|0\rangle|1\rangle-\frac{1}{\sqrt{6}}|1\rangle|0\rangle)|\rightarrow\rangle+{}\nonumber
\\&&\frac{1}{\sqrt{6}}|1\rangle|1\rangle|\leftarrow\rangle,\\
|1\rangle  |\Sigma\rangle
|Q\rangle&\longrightarrow&\frac{1}{\sqrt{6}}|1\rangle|1\rangle|\rightarrow\rangle
+((\frac{1}{\sqrt{2}})e^{icos^{-1}(\frac{1}{\sqrt{3}})}|1\rangle|0\rangle-\frac{1}{\sqrt{6}}|0\rangle|1\rangle)|\uparrow\rangle+
{}\nonumber
\\&& \frac{1}{\sqrt{6}}|0\rangle|0\rangle|\downarrow\rangle,
\end{eqnarray}
where$|\uparrow\rangle$,$|\downarrow\rangle$,$|\rightarrow\rangle$,$|\leftarrow\rangle$
are orthogonal machine states. The fidelity of universal
anti-cloner is same as the fidelity of measurement which is equal
to $\frac{2}{3}$ \cite{massar1}.\\
In the subsequent sections, we will discuss about the state
dependent and state independent hybrid quantum cloning
transformations. This chapter is based on our work "Hybrid quantum
cloning machine".
\section{\emph{State dependent hybrid cloning transformation}}


In this section, we study two state dependent hybrid quantum
cloning machines. The quality of the state dependent cloning
machine depends on the input state so naturally one may ask a
question why this type of cloning machine is important for study?
Here we give a reason for such studies. The importance of the
state dependent cloner lies in the eavesdropping strategy on some
quantum cryptographic system. For example, if the quantum key
distribution protocol is based on two non-orthogonal states
\cite{bennett5}, the optimal state dependent cloner can clone the
qubit in transit between a sender and a receiver. The original
qubit can then be re-sent to the receiver and the clone can stay
with an eavesdropper who by measuring it can obtain some
information about the bit value encoded in the original. The
eavesdropper may consider storing the clone and delaying the
actual measurement until any further public communication between
the sender and the receiver takes place. This eavesdropping
strategy has been discussed in \cite{bruss2,gisin4}.\\
{\bf B-H type cloning transformation:} B-H cloning transformation
generally indicates the optimal universal quantum cloning
transformation but in this paper, we relax one condition of
universality of B-H cloning transformation and hence we rename the
B-H cloning transformation as B-H type cloning transformation.
Therefore, although B-H type cloning transformation is
structurally same as the universal B-H cloning transformation but
it is different in the sense that this type of transformation is
state dependent. State dependent ness of the cloning machine
arises because of the relaxation of the condition $\frac{\partial
D_{ab} }{\partial \alpha^2}=0$.

\subsection{ \emph{Hybridization of two B-H type cloning transformation:}}
Here we investigate a new kind of cloning transformation that can
be obtained by combining two different BH type cloning
transformations. Here we consider two B-H type cloning
transformations which occur separately in the hybrid cloning
transformation with probability $\lambda$ and $1-\lambda$ respectively.\\
The hybrid quantum cloning transformation can be written as
\begin{eqnarray}
|\psi\rangle  |\Sigma\rangle
|Q\rangle\otimes|n\rangle\longrightarrow\sqrt{\lambda}[|\psi\rangle|\psi\rangle|Q_{\psi}\rangle
+(|\psi\rangle|\overline{\psi}\rangle+|\overline{\psi}\rangle|\psi\rangle)|Y_{\psi}\rangle]|i\rangle
\nonumber\\+(\sqrt{1-\lambda})[|\psi\rangle|\psi\rangle|{Q_{\psi}^{\prime}}\rangle
+(|\psi\rangle|\overline{\psi}\rangle+|\overline{\psi}\rangle|\psi\rangle)|{Y_{\psi}^{\prime}}\rangle]|j\rangle.
\end{eqnarray}
Unitarity of the transformation gives
\begin{eqnarray}
\lambda(\langle Q_{\psi}|Q_{\psi}\rangle+2\langle
Y_{\psi}|Y_{\psi}\rangle)
+(1-\lambda)(\langle {Q_{\psi}^{\prime}}|{Q_{\psi}^{\prime}}\rangle+2\langle{Y_{\psi}^{\prime}}|{Y_{\psi}^{\prime}}\rangle)=1, \\
2\lambda(\langle Y_{\psi}|Y_{\bar{\psi}}\rangle)
+2(1-\lambda)(\langle{Y_{\psi}^{\prime}}|{Y_{\bar{\psi}}^{\prime}}\rangle)
= 0.
\end{eqnarray}
Equations (2.6) and (2.7) are satisfied for all values of
$\lambda(0<\lambda<1)$ if
\begin{eqnarray}
\langle Q_{\psi}|Q_{\psi}\rangle+2\langle
Y_{\psi}|Y_{\psi}\rangle =\langle\acute{Q_{\psi}}|\acute{Q_{\psi}}\rangle+2\langle\acute{Y_{\psi}}|\acute{Y_{\psi}}\rangle =1\\
\langle
Y_{\psi}|Y_{\overline{\psi}}\rangle=\langle\acute{Y_{\psi}}|\acute{Y_{\overline{\psi}}}\rangle=0
\end{eqnarray}
Further we assume that
\begin{eqnarray}
\langle Q_{\psi}|Y_{\psi}\rangle=0=\langle
Q_{\psi}|Q_{\overline{\psi}}\rangle.
\end{eqnarray}
Let $|\chi\rangle=\alpha|0\rangle+\beta|1\rangle$ with
$\alpha^2+\beta^2=1$, be the input state. The cloning
transformation (2.5) approximately copies the information
contained in the input state $|\chi\rangle$ into two identical
states described by the density operators $\rho_a^{(out)}$ and
$\rho_b^{(out)}$, respectively. The reduced density operator
$\rho_a^{(out)}$ is given by
\begin{eqnarray}
\rho_a^{(out)}&=&|0\rangle\langle0|[\alpha^2+(\beta^2\langle{Y_{1}^{\prime}}|{Y_{1}^{\prime}}\rangle
-\alpha^2\langle{Y_{0}^{\prime}}|{Y_{0}^{\prime}}\rangle)+\lambda(\beta^2\langle
Y_1|Y_1\rangle-\alpha^2\langle
Y_0|Y_0\rangle-\beta^2\langle{Y_{1}^{\prime}}|{Y_{1}^{\prime}}\rangle+{}\nonumber\\&&\alpha^2\langle{Y_{0}^{\prime}}|{Y_{0}^{\prime}}\rangle)]
+|0\rangle\langle1|[\alpha\beta(\langle
{Q_{1}^{\prime}}|{Y_{0}^{\prime}}\rangle+\langle
{Y_{1}^{\prime}}|{Q_{0}^{\prime}}\rangle)+\lambda\alpha\beta(\langle
Q_{1}|Y_{0}\rangle+\langle
Y_{1}|Q_{0}\rangle-{}\nonumber\\&&\langle
{Q_{1}^{\prime}}|{Y_{0}^{\prime}}\rangle-\langle
{Y_{1}^{\prime}}|{Q_{0}^{\prime}}\rangle)]
+|1\rangle\langle0|[\alpha\beta(\langle
{Q_{1}^{\prime}}|{Y_{0}^{\prime}}\rangle+\langle
{Y_{1}^{\prime}}|{Q_{0}^{\prime}}\rangle)+\lambda\alpha\beta(\langle
Q_{1}|Y_{0}\rangle+{}\nonumber\\&&\langle
Y_{1}|Q_{0}\rangle-\langle
{Q_{1}^{\prime}}|{Y_{0}^{\prime}}\rangle-\langle
{Y_{1}^{\prime}}|{Q_{0}^{\prime}}\rangle)]+
|1\rangle\langle1|[\beta^2-(\beta^2\langle{Y_{1}^{\prime}}|{Y_{1}^{\prime}}\rangle-\alpha^2\langle{Y_{0}^{\prime}}|{Y_{0}^{\prime}}\rangle)
+{}\nonumber\\&&\lambda(\beta^2\langle
Y_1|Y_1\rangle-\alpha^2\langle
Y_0|Y_0\rangle-\beta^2\langle{Y_{1}^{\prime}}|{Y_{1}^{\prime}}\rangle+\alpha^2\langle{Y_{0}^{\prime}}
|{Y_{0}^{\prime}}\rangle)].
\end{eqnarray}
The other output state described by the density operator
$\rho_b^{(out)}$ looks exactly the same as $\rho_a^{(out)}$.\\
Let $\langle Y_0|Y_0\rangle=\langle Y_1|Y_1\rangle=\xi$, $\langle
Q_1|Y_0\rangle=\langle Y_0|Q_1\rangle=\langle
Q_0|Y_1\rangle=\langle Y_1|Q_0\rangle=\frac{\eta}{2}$, \\
$\langle {Y_{0}^{\prime}}|{Y_{0}^{\prime}}\rangle=\langle
{Y_{1}^{\prime}}|{Y_{1}^{\prime}}\rangle={\xi^{\prime}}$ and
$\langle {Q_{1}^{\prime}}|{Y_{0}^{\prime}}\rangle=\langle
{Y_{0}^{\prime}}|{Q_{1}^{\prime}}\rangle=\langle
{Q_{0}^{\prime}}|{Y_{1}^{\prime}}\rangle=\langle
{Y_{1}^{\prime}}|{Q_{0}^{\prime}}\rangle=\frac{{\eta^{\prime}}}{2}$\\
with $0\leq\xi({\xi^{\prime})}\leq1$ and
$0\leq\eta({\eta^{\prime})}\leq2\sqrt{\xi(1-2\xi)}(2\sqrt{{\xi^{\prime}}
(1-2{\xi^{\prime}})})\leq\frac{1}{\sqrt{2}}$.\\
Using above conditions, equation (2.11) can be rewritten as
\begin{eqnarray}
\rho_a^{(out)}=|0\rangle\langle0|[\alpha^2+{\xi^{\prime}}(\beta^2-\alpha^2)+\lambda(\xi-{\xi^{\prime}})(\beta^2-\alpha^2)]+|0\rangle\langle1|[\alpha\beta({\eta^{\prime}}+\lambda(\eta-{\eta^{\prime}}))]
\nonumber\\+|1\rangle\langle0|[\alpha\beta({\eta^{\prime}}+\lambda(\eta-{\eta^{\prime}}))]+|1\rangle\langle1|[\beta^2-{\xi^{\prime}}(\beta^2-\alpha^2)-\lambda(\xi-{\xi^{\prime}})(\beta^2-\alpha^2)].
\end{eqnarray}
To investigate how well our hybrid cloning machine copies the
input state, we have to calculate the fidelity. The fidelity
$F_{HCM}$ is defined by
\begin{eqnarray}
F_{HCM}=\langle\chi|\rho_a^{(out)}|\chi\rangle
&=&\alpha^4[(1-{\xi^{\prime}})-\lambda(\xi-{\xi^{\prime}})]+\beta^4[(1-{\xi^{\prime}})-\lambda(\xi-{\xi^{\prime}})]
{}\nonumber\\&&+2\alpha^2\beta^2[{\xi^{\prime}}+\lambda(\xi-{\xi^{\prime}})+{\eta^{\prime}}+\lambda(\eta-{\eta^{\prime}})].
\end{eqnarray}
The equation $\frac{\partial F_{HCM}}{\partial \alpha^2}=0$ gives
the required relationship between the machine parameters $\xi
,{\xi^{\prime}},\eta, \textrm{and}~{\eta^{\prime}}$ in the form
\begin{eqnarray}
\eta^{\prime}(1-\lambda)+\eta\lambda=1-2\xi^{\prime}-2\lambda(\xi-\xi^{\prime}).
\end{eqnarray}
Using (2.14), equation (2.13) reduces to
\begin{eqnarray}
F_{HCM}=(1-\xi^{\prime})-\lambda(\xi-\xi^{\prime}).
\end{eqnarray}
Now the H-S distance between the two mode density operators
$\rho_{ab}^{(out)}$ and
$\rho_{ab}^{(id)}=\rho_a^{(id)}\otimes\rho_b^{(id)}$ is given by
\begin{eqnarray}
D_{ab}&=&Tr[\rho_{ab}^{(out)}-\rho_{ab}^{(id)}]^2{}\nonumber\\&=&
U_{11}^2+2U_{12}^2+2U_{13}^2+U_{22}^2+2U_{23}^2+U_{33}^2,
\end{eqnarray}
where
\begin{eqnarray}
{}\nonumber\\&&U_{11}=\alpha^4-\alpha^2[\lambda(1-2\xi)+(1-\lambda)(1-2{\xi^{\prime}})],
{}\nonumber\\&&
U_{12}=U_{21}=\sqrt{2}\alpha^3\beta-\sqrt{2}\alpha\beta(\eta\frac{\lambda}{2}+(1-\lambda)\frac{{\eta^{\prime}}}{2}),
{}\nonumber\\&& U_{13}=U_{31}=\alpha^2\beta^2,{}\nonumber\\&&
U_{22}=2\alpha^2\beta^2-(2\xi\lambda+2{\xi^{\prime}}(1-\lambda)),
{}\nonumber\\&&
U_{23}=U_{32}=\sqrt{2}\alpha\beta^3-\sqrt{2}\alpha\beta(\eta\frac{\lambda}{2}+(1-\lambda)\frac{{\eta^{\prime}}}{2}),
{}\nonumber\\&&U_{33}=\beta^4-\beta^2[\lambda(1-2\xi)+(1-\lambda)(1-2{\xi^{\prime}})].
\end{eqnarray}
It is interesting to see that the transformation (2.5) can behave
as a state dependent cloner if we relax the condition
$\frac{\delta D_{ab}}{\delta \alpha^2}=0$. Then, it is natural to
expect that the machine parameters depend on the input state.
Thus, our prime task is to find a relationship between the machine
parameters and the input state that minimizes the distortion
$D_{ab}$. We will get an interesting result if we fix any one of
the machine parameters $\xi$ or ${\xi^{\prime}}$ as $\frac{1}{6}$
. Without any loss of generality we can fix
${\xi^{\prime}}=\frac{1}{6}$ . In doing so, the cloning
transformation (2.5) reduces to the combination of B-H optimal
universal cloning machine and the B-H type cloning machine.\\
Now, substituting ${\xi^{\prime}}=\frac{1}{6}$ in (2.17) and using
(2.14), equation (2.16) can be rewritten as
\begin{eqnarray}
D_{ab}=V_{11}^2+2V_{12}^2+2V_{13}^2+V_{22}^2+2V_{23}^2+V_{33}^2,
\end{eqnarray}
where
\begin{eqnarray}
{}\nonumber\\&&
V_{11}=\alpha^4-\alpha^2[\lambda(1-2\xi)+(1-\lambda)(\frac{2}{3})],
{}\nonumber\\&&
V_{12}=V_{21}=\sqrt{2}\alpha^3\beta-\sqrt{2}\alpha\beta(\frac{1}{3}-\lambda(\xi-\frac{1}{6})),
{}\nonumber\\&& V_{13}=V_{31}=\alpha^2\beta^2,{}\nonumber\\&&
V_{22}=2\alpha^2\beta^2-(2\xi\lambda+(\frac{1}{3})(1-\lambda)),
{}\nonumber\\&&
V_{23}=V_{32}=\sqrt{2}\alpha\beta^3-\sqrt{2}\alpha\beta(\frac{1}{3}-
\lambda(\xi-\frac{1}{6})),{}\nonumber\\&&
V_{33}=\beta^4-\beta^2[\lambda(1-2\xi)+(1-\lambda)(\frac{2}{3})].
\end{eqnarray}
Now we are in a position to determine the relationship between the
machine parameter $(\xi)$ and the input state $(\alpha^{2})$ that
minimizes the distortion $D_{ab}$. To obtain the minimum value of
$D_{ab}$ for given $\alpha$ and $\lambda $, we solve the equation
\begin{eqnarray}
\frac{\delta D_{ab}}{\delta
\xi}=0\Longrightarrow\xi=\frac{(9\alpha^2\beta^2-2(1-\lambda))}{12\lambda},~~provided~~\lambda\neq0.
\end{eqnarray}
Now, the cloning machine is defined by those parameters $\xi$
(specified by equation (2.20)) common to the whole family of
states that one wants to clone i.e. for given $\lambda$, we choose
the values of $\alpha$ and $\beta$ in such a way so that the
machine parameter $\xi$ remains invariant. It is clear from
equation (2.20) that if we want to minimize $D_{ab}$ then the
quantum cloning machine having parameter $\xi$ can be applied on
the family of states such that
$\alpha^{2}\beta^{2}=\alpha^{2}(1-\alpha^{2})=constant$. That
means the cloning machine if applied on just four states
$|\psi^{\pm}\rangle_{1}=\alpha|0\rangle\pm\beta|1\rangle,
|\psi^{\pm}\rangle_{2}=\alpha|1\rangle\pm\beta|0\rangle$ will give minimum $D_{ab}$.\\
Since the value of the machine parameter $\xi$ cannot be negative,
so the parameter $\lambda$ takes values lying in the interval
[$1-\frac{9\alpha^{2}(1-\alpha^{2})}{2}] < \lambda < 1$.\\
Also
\begin{eqnarray}
\frac{\delta^2 D_{ab}}{\delta \xi^2}=16\lambda^2>0.
\end{eqnarray}
Therefore, the equation (2.20) represents the required
relationship between the machine parameter and the input state
which minimizes $D_{ab}$.\\
The minimum value of $D_{ab}$ is given by
\begin{eqnarray}
(D_{ab})_{min}=2\alpha^2\beta^2-\frac{9\alpha^4\beta^4}{2}
\end{eqnarray}
which depends on $\alpha^2$ but not on $\lambda$.\\
Substituting $\xi=\frac{9
\alpha^2(1-\alpha^2)-2(1-\lambda)}{12\lambda}$ and
${\xi^{\prime}}=\frac{1}{6}$ in equation (2.15), we get\\
$F_{HCM}=1-\frac{3\alpha^2\beta^2}{4}$.\\\\\\\\\\
\textsl{Table-2.2: Quality of the copies from hybrid cloning machine (B-H
state independent transformation + B-H state dependent
transformation)}\\
\begin{tabular}{| c| c| c| c| c|}
\hline Input state  & Range of  & Range of machine & $(D_{ab})_{min}$ & $F_{HCM}$\\
parameter $(\alpha^{2})$& parameter $\lambda$&  parameter
$\xi(\lambda)$&&\\
 \hline
0.1~~or~~0.9  & (0.595, 1.0) & (0.0, 0.0675) & 0.14 & 0.93\\
\hline
0.2~~0r~~0.8 & (0.280, 1.0) & (0.0, 0.1200) & 0.21 & 0.88\\
\hline
0.3~~or~~0.7 & (0.055, 1.0) & (0.0, 0.1575) & 0.22 & 0.84\\
\hline
0.4~~or~~0.6 & (0.000, 1.0) & (0.0, 0.1800) & 0.22 & 0.82\\
\hline
0.5 & (0.000, 1.0) & (0.0, 0.1875) & 0.22 & 0.81\\
\hline
\end{tabular}\\\\
\textbf{\textsl{Illustration of the Table-2.2:}\\
\textit{The above table shows that there exists several quantum
cloning machines (for different values of $\xi$) which can clone
the four states $\{|\psi^{\pm}\rangle_{1},
|\psi^{\pm}\rangle_{2}\}$ with the same fidelity. For example, If
one of the input state is chosen from the set $S=
\{\sqrt{0.1}|0\rangle\pm\sqrt{0.9}|1\rangle,
\sqrt{0.9}|0\rangle\pm\sqrt{0.1}|1\rangle \}$ and for a fixed
value of $\lambda$, say $\lambda=0.6$ (chosen from the interval
(0.595,1.0), then there exist a quantum cloning machine with
parameter $\xi=0.0014$ lying in the interval $(0.0,0.0675)$, which
clone the state from the set S with the fidelity 0.93.}}\\
\subsection{ \emph{Hybridization of B-H type cloning transformation and phase-covariant quantum cloning transformation}}
In this subsection, we will show that if B-H type cloning
transformation occurs with probability $\lambda$ and the
phase-covariant quantum cloning transformation occurs with
probability $1-\lambda$ then the resulting hybrid quantum cloning
machine is a state dependent quantum cloning machine.\\
The Hybrid cloning transformation is given by
\begin{eqnarray}
|0\rangle |\Sigma\rangle |Q\rangle |n\rangle
\longrightarrow\sqrt{\lambda}[|0\rangle|0\rangle|Q_{0}\rangle
+(|0\rangle|1\rangle+|1\rangle|0\rangle)|Y_{0}\rangle]|i\rangle
\nonumber\\+(\sqrt{1-\lambda})[((\frac{1}{2}+\frac{1}{\sqrt{8}})|0\rangle|0\rangle
+(\frac{1}{2}-\frac{1}{\sqrt{8}})|1\rangle|1\rangle)|\uparrow\rangle+
\frac{1}{2}|+\rangle|\downarrow\rangle)]|j\rangle,\\
|1\rangle |\Sigma\rangle |Q\rangle
|n\rangle\longrightarrow\sqrt{\lambda}[|1\rangle|1\rangle|Q_{1}\rangle
+(|0\rangle|1\rangle+|1\rangle|0\rangle)|Y_{1}\rangle]|i\rangle
\nonumber\\+(\sqrt{1-\lambda})[((\frac{1}{2}+\frac{1}{\sqrt{8}})|1\rangle|1\rangle
+(\frac{1}{2}-\frac{1}{\sqrt{8}})|0\rangle|0\rangle)|\downarrow\rangle+\frac{1}{2}|+\rangle|\uparrow\rangle)]|j\rangle.
\end{eqnarray}
When $\lambda=1$ cloning transformation reduces to B-H type
cloning transformation and when $\lambda=0$ it takes the form of
phase-covariant quantum cloning transformation.\\The cloning
machine (2.23-2.24) approximately copies the information of the
input state $|\chi\rangle$ given in subsection 2.2.1 into two
identical states described by the same reduced density operator
\begin{eqnarray}
\rho=\lambda[(1-\xi)|\chi\rangle\langle\chi|+\xi|\overline{\chi}\rangle\langle\
\overline{\chi}|]+(1-\lambda)[(\frac{1}{2}+\frac{1}{\sqrt{8}})|\chi\rangle\langle\chi|+(\frac{1}{2}-\frac{1}{\sqrt{8}})|\overline{\chi}\rangle\langle\
\overline{\chi}|]
\end{eqnarray}
where $|\bar{\chi}\rangle$ is an orthogonal state to
$|\chi\rangle$ and $\xi$ is the machine parameter of the B-H type
cloning machine given by $\xi=\langle Y_{0}|Y_{0}\rangle=\langle
Y_{1}|Y_{1}\rangle$.\\
Now, the fidelity is given by
\begin{eqnarray}
F_1=\langle\chi|\rho|\chi\rangle=(\frac{1}{2}+\frac{1}{\sqrt{8}})+\lambda(\frac{1}{2}-\frac{1}{\sqrt{8}}-\xi)
\end{eqnarray}
The hybrid quantum cloning machine constructed by combining the
B-H type cloning transformation and phase-covariant quantum
cloning transformation is state dependent. State dependableness
condition arises from the fact that B-H type cloning
transformation is state dependent. Consequently, the fidelity
$F_1$ depends on the input state as it depends on the machine
parameter $\xi(\alpha^2)=\frac{3\alpha^2(1-\alpha^2)}{4}$. This
relationship between the machine parameter $\xi$ associated with
the B-H type cloning machine and the input state $\alpha^2$ is
obtained by putting $\lambda=1$ in equation (2.20).\\
Following the argument given in previous subsection 2.2.1, we find
that the hybrid quantum cloning machine (B-H type cloning
transformation + phase covariant quantum cloning transformation)
clones the same four states $\{|\psi^{\pm}\rangle_{1},
|\psi^{\pm}\rangle_{2}\}$ with minimum $D_{ab}$. Also it can be
easily verified that there is no improvement in the quality of
cloning of these four states. Therefore, this hybrid quantum
cloning machine does not give anything new because it neither
involves in cloning of new states nor it gives any improvement in
the fidelity of cloning.
\section{ \emph{State independent hybrid cloning transformation}}


In this section, we study one symmetric and two asymmetric
universal hybrid quantum cloning machines.

\subsection{\emph{ Hybridization of two BH type cloning transformations}}
In the preceding subsection 2.2.1, we have found that the quantum
cloning machine obtained by combining two BH type cloning
transformations is state dependent but in this section we will
show that a proper combination of two BH type cloning
transformations can serve as a state independent cloner also. A
hybrid quantum cloning machine (2.5) becomes state independent or
universal if the fidelity $F_{HCM}$ defined in equation (2.13) and
the deviation $D_{ab}$ defined in equation (2.16), are state
independent. From equation (2.15), it is clear that $F_{HCM}$ is
state independent. Therefore, the only remaining task is to show
the independence of the deviation $D_{ab}$.\\
$D_{ab}$ is input state independent if,
\begin{eqnarray}
\frac{\partial D_{ab}}{\partial \alpha^2}=0
\Longrightarrow[2(\lambda(1-2\xi)+(1-\lambda)(1-2\xi^{\prime}))-3]^2\nonumber\\
-[2(\eta\lambda-(1-\lambda){\eta^\prime})-2]^2+8[2\xi\lambda+2{\xi^\prime}(1-\lambda)]-5=0.
\end{eqnarray}
Using equation (2.14) in equation (2.27), we get
\begin{eqnarray}
\lambda=\frac{(6{\xi^{\prime}}-1)}{6({\xi^{\prime}}-\xi)},
\end{eqnarray}
provided $\xi\neq{\xi^{\prime}}$. \\
Using the value of $\lambda$ in (2.15), we get
\begin{eqnarray}
F_{HCM}=\frac{5}{6}.
\end{eqnarray}
If $\xi={\xi^{\prime}}=\frac{1}{6}$ , then there is nothing
special about the transformation (2.5) because the condition
$\xi={\xi^{\prime}}=\frac{1}{6}$, simply takes the transformation
(2.5) to B-H state independent quantum cloning transformation. If
$\xi\neq{\xi^{\prime}}~\textrm{and}~\xi'\neq\frac{1}{6}$, then the
hybrid quantum cloning machine (B-H type transformation + B-H type
transformation) will become state independent for all values of
$\xi$ and
${\xi^{\prime}}$~$\textrm{(provided}~\xi\neq{\xi^{\prime}}~\textrm{and}~\xi'\neq\frac{1}{6})$.
Therefore the newly defined hybrid cloning machine generates a
class of universal cloning machines for
$\lambda=\frac{(6{\xi^{\prime}}-1)}{6({\xi^{\prime}}-\xi)}$
$\textrm{(provided}~\xi\neq{\xi^{\prime}}~\textrm{and}~\xi'\neq\frac{1}{6})$.
The fidelity of the introduced universal hybrid cloning machine is
equal to $\frac{5}{6}$ which is the optimal fidelity one can
obtain. Although the machine is universal and optimal for an
unknown quantum state but it is different from B-H state
independent cloning machine. It is different in the sense that B-H
cloning machine is state independent for just only one value of
the machine parameter $\xi=\frac{1}{6}$ while the cloning machine
defined by (2.5) works as a universal cloner provided $\lambda$ is
given by equation (2.28) and for all values of $\xi$ and
${\xi^{\prime}}$
$\textrm{(provided}~\xi\neq{\xi^{\prime}}~\textrm{and}~\xi'\neq\frac{1}{6})$.
\subsection{ \emph{Hybridization of optimal universal symmetric B-H cloning transformation and optimal
universal asymmetric Pauli cloning transformation} }

An asymmetric quantum cloning machine can be constructed by
applying hybridization technique. Using the hybridization
procedure we can construct universal asymmetric hybrid quantum
cloning machine by combining universal symmetric B-H cloning
transformation and optimal universal asymmetric Pauli cloning transformation. \\
The Hybrid cloning transformation is given by
\begin{eqnarray}
&&|0\rangle |\Sigma\rangle |Q\rangle
|n\rangle\longrightarrow\sqrt{1-\lambda}[\sqrt{\frac{2}{3}}|0\rangle|0\rangle
|\uparrow\rangle
+\sqrt{\frac{1}{6}}(|0\rangle|1\rangle+|1\rangle|0\rangle)|\downarrow\rangle]
|i\rangle{}\nonumber\\&&
+\sqrt{\lambda}[(\frac{1}{\sqrt{1+p^2+q^2}})(|0\rangle|0\rangle|\uparrow\rangle
+(p|0\rangle|1\rangle+q|1\rangle|0\rangle)|\downarrow\rangle)]|j\rangle, \\
&&|1\rangle |\Sigma\rangle |Q\rangle
|n\rangle\longrightarrow\sqrt{1-\lambda}[\sqrt{\frac{2}{3}}|1\rangle|1\rangle
|\downarrow\rangle
+\sqrt{\frac{1}{6}}(|0\rangle|1\rangle+|1\rangle|0\rangle)|\uparrow\rangle]
|i\rangle{}\nonumber\\&&
+\sqrt{\lambda}[(\frac{1}{\sqrt{1+p^2+q^2}})(|1\rangle|1\rangle|\downarrow
\rangle+(p|1\rangle|0\rangle+q|0\rangle|1\rangle)|\uparrow\rangle)]|j\rangle,
\end{eqnarray}
\textrm where p + q =1.\\
The hybrid cloning machine (2.30-2.31) produces two asymmetric
copies of the input state $|\chi\rangle$. These asymmetric cloned
states are described by the reduced density operators $\rho_1$ and
$\rho_2$
\begin{eqnarray}
\rho_1=\lambda[(\frac{1}{1+p^2+q^2})((1-q^2+p^2)|\chi\rangle\langle\chi|+q^2I)]+(1-\lambda)[\frac{5}{6}|\chi\rangle\langle\chi|+\frac{1}{6}|\overline{\chi}\rangle\langle\overline{\chi}|], \\
\rho_2=\lambda[(\frac{1}{1+p^2+q^2})((1-p^2+q^2)|\chi\rangle\langle\chi|+p^2I)]+(1-\lambda)[\frac{5}{6}|\chi\rangle\langle\chi|+\frac{1}{6}|\overline{\chi}\rangle\langle\overline{\chi}|].
\end{eqnarray}
The quality of the asymmetric clones are given by
\begin{eqnarray}
(F_1)_{HCM}=\frac{5}{6}+(\frac{\lambda}{2})[\frac{(p^2+1)}{(p^2-p+1)}-\frac{5}{3} ~],\\
(F_2)_{HCM}=\frac{5}{6}+(\frac{\lambda}{2})[\frac{(p^2-2p+2)}{(p^2-p+1)}-\frac{5}{3}~].
\end{eqnarray}
Note: Equations (2.34) and (2.35) show that the hybrid quantum
cloning machine tends to B-H state independent quantum cloning
machine in the limiting sense when (i) $\lambda\rightarrow 0$ and
$0\leq p\leq1$ or (ii) $\lambda\rightarrow 1$ and $p=\frac{1}{2}$.\\
Next we show that the fidelities $(F_{1})_{HCM}$ and
$(F_{2})_{HCM}$ cannot be greater than the optimal value
$\frac{5}{6}$ simultaneously i.e. if $(F_1)_{HCM}$ greater than
$\frac{5}{6}$ then $(F_2)_{HCM}$ must be less than $\frac{5}{6}$
for all $\lambda's$ lying between 0 and 1 and
vice-versa.\\
Without any loss of generality, we assume
$(F_{1})_{HCM}>\frac{5}{6}$ for $0<\lambda<1$. Our task is to find
the values of p for which $(F_{1})_{HCM}>\frac{5}{6}$.
\begin{eqnarray}
(F_1)_{HCM}>\frac{5}{6} &\Longrightarrow& \frac{(p^2+1)}{(p^2-p+1)}>\frac{5}{3} {}\nonumber\\
& \Longrightarrow &   (2p-1)(p-2)<0 {}\nonumber\\
& \Longrightarrow&  (2p-1)>0, \textrm{ Since } p-2<0 {}\nonumber\\
& \Longrightarrow&  p>\frac{1}{2}. \nonumber
\end{eqnarray}
Now we have to show that if $p>\frac{1}{2}$ then
$(F_2)_{HCM}<\frac{5}{6}$. We prove this result by contradiction.\\
If possible, let $(F_2)_{HCM}>\frac{5}{6}$ for $p>\frac{1}{2}$.
\begin{eqnarray}
(F_2)_{HCM}>\frac{5}{6}&\Longrightarrow&
\frac{(p^2-2p+2)}{(p^2-p+1)}>\frac{5}{3}{}\nonumber\\
& \Longrightarrow & (2p-1)(p+1)<0{}\nonumber\\
& \Longrightarrow& (2p-1)<0 , \textrm{ Since } p+1>0  {}\nonumber\\
& \Longrightarrow&  p<\frac{1}{2}.\nonumber
\end{eqnarray}
which contradicts our assumption.\\
Hence $(F_2)_{HCM}<\frac{5}{6}$ for $p>\frac{1}{2}$. Therefore, we
can conclude that the fidelities $(F_{1})_{HCM}$ and
$(F_{2})_{HCM}$ given in (2.34) and (2.35) cannot cross the
optimal limit $\frac{5}{6}$ simultaneously.\\
Furthermore, we construct a table below in which we show the
tradeoff between the two fidelities.\\
\textsl{Table-2.3: Fidelity of the asymmetric copies produced from
asymmetric hybrid \\quantum cloning machine}\\
\begin{tabular}{| c| c| c| c| c|}
\hline
  p & $\lambda$ & $(F_1)_{HCM}=\frac{5}{6}+(\frac{\lambda}{2})\times$ &
  $(F_2)_{HCM}=\frac{5}{6}+(\frac{\lambda}{2})\times$&$(F_1)_{HCM}\sim(F_2)_{HCM}$
  \\ & &$[\frac{(p^2+1)}{(p^2-p+1)}-\frac{5}{3} ~]$ &$[\frac{(p^2-2p+2)}{(p^2-p+1)}-\frac{5}{3}~]$ & \\
  \hline
  [0.0,1.0] & 0.0  & 0.83 & 0.83 & 0.00 (symmetric copies)\\
  \hline
  0.0 & [0.1,0.9]&[0.80,0.53] & [0.85,0.98] & [0.05,0.45] \\
  \hline
  0.1 &[0.1,0.9] &[0.81,0.58] & [0.85,0.98] & [0.04,0.40] \\
  \hline
  0.2 & [0.1,0.9]&[0.81,0.64] & [0.85,0.96] & [0.04,0.32] \\
  \hline
  0.3 &[0.1,0.9] &[0.82,0.70] & [0.84,0.93] & [0.02,0.23] \\
  \hline
  0.4 &[0.1,0.9] & [0.83,0.77]& [0.84,0.89] & [0.01,0.12] \\
  \hline
  0.5 &[0.1,0.9] & 0.83 & 0.83 & 0.0 (Symmetric copies) \\
  \hline
  0.6 &[0.1,0.9] &[0.84,0.89] & [0.83,0.77] & [0.01,0.12] \\
  \hline
  0.7 & [0.1,0.9]&[0.84,0.93] & [0.82,0.70] & [0.02,0.23]\\
  \hline
  0.8 & [0.1,0.9]&[0.85,0.96] & [0.81,0.64] & [0.04,0.32] \\
  \hline
  0.9 &[0.1,0.9] &[0.85,0.98] & [0.81,0.58] & [0.04,0.40] \\
  \hline
  [0.0,1.0] & 1.0 & $(F_1)_{PCM}$ & $(F_2)_{PCM}$ &
  $(F_1)_{PCM}\sim(F_2)_{PCM}$\\
  \hline
\end{tabular}\\\\
\textbf{\textsl{Illustration of the table 2.3:\\
For some fixed value of p, we can construct different hybrid
quantum cloning machine by combining two independent quantum
cloners viz. optimal universal symmetric B-H cloner and optimal
universal asymmetric Pauli cloner with different probabilities
$(\lambda)$. These hybrid cloning machine produce two asymmetric
copies at the output. In particular, if $p=0.1~
\textrm{and}~\lambda=0.2$ then
$(F_{1})_{HCM}=0.78~and~F_{2})_{HCM}=0.78$. In general, for
$p=0.1$ we find that the quality described by the fidelity
$(F_{1})_{HCM}$ of one copy decreases from 0.81 to 0.58 while the
quality described by the fidelity $(F_{2})_{HCM}$ of other copy
increases from 0.85 to 0.98 as $\lambda$ varies from 0.1 to 0.9.
As a result the difference between the qualities of the two copies
increases from 0.04 to 0.40.}}\\\\
We note that the fidelity of the hybrid quantum cloning machine
(B-H cloner + Pauli cloner) depends on the parameter p and
$\lambda$ . From table we can observe that for $p=0.0$ to $p=0.4$
one of the output $(F_1)_{HCM}$ behave as a decreasing function
and another output of the asymmetric cloning machine $(F_2)_{HCM}$
behaves as an increasing function. The role of the fidelities
$(F_1)_{HCM}$ and $(F_2)_{HCM}$ are swapped for $p=0.6$ to
$p=0.9$. Here we observe that the asymmetric hybrid cloning
machine reduces to B-H symmetric cloning machine in two cases: (i)
when $\lambda$ = 0  and (ii) when $p=0.5$.\\
When $\lambda = 1.0$, our asymmetric hybrid cloner also reduces to
asymmetric Pauli cloner for all p $(0\leq p\leq1)$.

\subsection{\emph{Hybridization of universal B-H cloning
transformation and universal anti-cloning transformation}}

In this subsection, we introduce an interesting hybrid
quantum-cloning machine, which is a combination of universal B-H
cloning machine and a universal anti-cloning machine. The
introduced cloning machine is interesting in the sense that it
acts like anti-cloning machine which means that the spin direction
of the outputs of the cloner are antiparallel. We will show later
that the newly introduced Hybrid cloning machine (B-H cloner +
Anti-cloner) serves as a better anti-cloner than the existing
quantum anti-cloning machine \cite{song1}. Also we show that if
the values of the machine parameter $\lambda$ is in the
neighborhood of 1 then the values of
the two non-identical fidelities lie in the neighborhood of $\frac{5}{6}$ \\
The introduced hybrid anti-cloning transformation is defined by
\begin{eqnarray}
&&|0\rangle |\Sigma\rangle |Q\rangle
|n\rangle\longrightarrow\sqrt{\lambda}[\sqrt{\frac{2}{3}}|0\rangle|0\rangle|\uparrow\rangle
+\sqrt{\frac{1}{6}}(|0\rangle|1\rangle+|1\rangle|0\rangle)|\downarrow\rangle]|i\rangle
+(\sqrt{1-\lambda}){}\nonumber\\&&[\sqrt{\frac{1}{6}}|0\rangle|0\rangle|\uparrow\rangle
+(\frac{1}{\sqrt{2}}e^{icos^{-1}(\frac{1}{\sqrt{3}})}|0\rangle|1\rangle
-\frac{1}{\sqrt{6}}|1\rangle|0\rangle)|\rightarrow\rangle+\frac{1}{\sqrt{6}}|1\rangle|1\rangle|\leftarrow\rangle]|j\rangle,\\
&&|1\rangle |\Sigma\rangle |Q\rangle
|n\rangle\longrightarrow\sqrt{\lambda}[\sqrt{\frac{2}{3}}|1\rangle|1\rangle|\downarrow\rangle
+\sqrt{\frac{1}{6}}(|0\rangle|1\rangle+|1\rangle|0\rangle)|\uparrow\rangle]|i\rangle
+(\sqrt{1-\lambda}){}\nonumber\\&&[\sqrt{\frac{1}{6}}|1\rangle|1\rangle|\rightarrow\rangle
+(\frac{1}{\sqrt{2}}e^{icos^{-1}(\frac{1}{\sqrt{3}})}|1\rangle|0\rangle-\frac{1}{\sqrt{6}}|0\rangle|1\rangle)|\uparrow\rangle+\frac{1}{\sqrt{6}}|0\rangle|0\rangle|\downarrow\rangle]|j\rangle,
\end{eqnarray}
where $|\uparrow\rangle$, $|\downarrow\rangle$, $|\rightarrow\rangle$, $|\leftarrow\rangle$ are mutually orthogonal machine states.\\
The above defined cloning machine (2.36-2.37) produces two copies
which are described by the reduced density operator in mode `a'
and mode `b'
\begin{eqnarray}
\rho_a=|0\rangle\langle0|[\lambda(\frac{5\alpha^2}{6}+\frac{\beta^2}{6})+(1-\lambda)(\frac{2\alpha^2}{3}+\frac{\beta^2}{3})]+|0\rangle\langle1|[\lambda\frac{2\alpha\beta}{3}+(1-\lambda)\frac{\alpha\beta}{3}]
\nonumber\\+|1\rangle\langle0|[\lambda\frac{2\alpha\beta}{3}+(1-\lambda)\frac{\alpha\beta}{3}]+|1\rangle\langle1|[\lambda(\frac{5\beta^2}{6}+\frac{\alpha^2}{6})+(1-\lambda)(\frac{\alpha^2}{3}+\frac{2\beta^2}{3})], \\
\rho_b=|0\rangle\langle0|[\lambda(\frac{5\alpha^2}{6}+\frac{\beta^2}{6})+(1-\lambda)(\frac{\alpha^2}{3}+\frac{2\beta^2}{3})]+|0\rangle\langle1|[\lambda\frac{2\alpha\beta}{3}-(1-\lambda)\frac{\alpha\beta}{3}]
\nonumber\\+|1\rangle\langle0|[\lambda\frac{2\alpha\beta}{3}-(1-\lambda)\frac{\alpha\beta}{3}]+|1\rangle\langle1|[\lambda(\frac{5\beta^2}{6}+\frac{\alpha^2}{6})+(1-\lambda)(\frac{2\alpha^2}{3}+\frac{\beta^2}{3})].
\end{eqnarray}
Let $F_a$ and $F_b$ denote the fidelities of the two copies with
opposite spin direction. These fidelities are given by
\begin{eqnarray}
F_a=\frac{5\lambda}{6}+\frac{2(1-\lambda)}{3}, ~
F_b=\frac{5\lambda}{6}+\frac{(1-\lambda)}{3}.
\end{eqnarray}
It is clear from equation (2.40) that the introduced hybrid anti-
cloning machine is asymmetric in nature, i.e., the hybrid quantum
cloning machine resulting from Universal B-H cloning machine and
universal anti-cloning machine behaves as an asymmetric quantum
cloning machine for all values of the parameter $\lambda$ lying
between 0 and 1. The two different fidelities given in (2.40) of
the anti-cloning machine can approach the optimal value
$\frac{5}{6}$ when the parameter $\lambda$ approaches one. Here we
should note an important fact that both the fidelities tend to
$\frac{5}{6}$ but not equal to $\frac{5}{6}$ unless $\lambda=1$.
Hence the fidelities $F_a$ and $F_b$ takes different values in the
neighborhood of $\frac{5}{6}$ when the values of $\lambda$ are
lying in the neighborhood of 1. For further illustration we
construct a table
below:\\\\\\\\\\\\\\\\\\\\
\textsl{Table-2.4: Fidelity of the two asymmetric clone produced
from hybrid anti-cloning machine}\\
\begin{tabular}{| c| c| c| c|}
\hline
  parameter $(\lambda)$ & $F_a=\frac{5\lambda}{6}+\frac{2(1-\lambda)}{3}  $ & $F_b=\frac{5\lambda}{6}+\frac{(1-\lambda)}{3}$& Difference between qualities
  \\ & & & of the two copies \\ & & & $F_a\sim F_b$\\
  \hline
  0.0 & 0.67 & 0.33 & 0.34 \\
  \hline
  0.1 & 0.68 & 0.38 & 0.30 \\
  \hline
  0.2 & 0.70 & 0.43 & 0.27 \\
  \hline
  0.3 & 0.72 & 0.48 & 0.24 \\
  \hline
  0.4 & 0.73 & 0.53 & 0.20 \\
  \hline
  0.5 & 0.75 & 0.58 & 0.17 \\
  \hline
  0.6 & 0.77 & 0.63 & 0.14 \\
  \hline
  0.7 & 0.78 & 0.68 & 0.10 \\
  \hline
  0.8 & 0.80 & 0.73 & 0.07 \\
  \hline
  0.9 & 0.82 & 0.78 & 0.04 \\
  \hline
  1.0 & 0.83 & 0.83 & 0.00 (Symmetric copies) \\ \hline
\end{tabular}\\\\
\textbf{\textsl{Illustration of the Table-2.4:\\
When the two independent quantum cloner viz. universal B-H quantum
copier and universal anti-cloner occurs with probabilities 0.1 and
0.9 respectively in the hybrid cloning machine, it produces two
asymmetric copies with fidelity 0.68 and 0.38 respectively.}}\\\\
It is clear that both the fidelities of output copies with
opposite spins are increasing function of the parameter $\lambda$.
Therefore, as $\lambda$ increases, the values of the fidelities
$F_a$ and $F_b$ also increases and approaches towards the optimal
cloning fidelity 0.83. The above table shows that when $\lambda$ =
0, our hybrid anti-cloner reduces to anti-cloner introduced by
Song and Hardy \cite{song1}. Also when $\lambda=1$ , we observe
that the copies with opposite spin direction change into the
copies with same spin direction with optimal fidelity. Therefore,
we can conclude that the hybrid anti-cloner performs better than
the existing quantum anti-cloning machine.\\\\

\large \baselineskip .85cm

\chapter{Broadcasting of entanglement}
\setcounter{page}{73} \markright{\it CHAPTER~\ref{chap3}.
Broadcasting of entanglement}
\label{chap3}%
The true sign of intelligence is not knowledge but imagination -
Albert EINSTEIN\\\\
Bell's theorem is easy to understand but hard to believe - Nick
Herbert \\\\
God [could] vary the laws of Nature, and make worlds of several
sorts in several parts of the universe - Isaac Newton
\section{\emph{Prelude}}
Entanglement \cite{einstein1}, the heart of quantum information
theory plays a crucial role in computational and communicational
purposes. As a valuable resource in quantum information
processing, quantum entanglement has been widely used in quantum
cryptography \cite{ekert1,shor1}, quantum superdense coding
\cite{bennett2} and quantum teleportation \cite{bennett3}. An
astonishing feature of quantum information processing is that
information can be "encoded" in non-local correlations between two
separated particles. The more "pure" is the quantum entanglement,
the more "valuable" is the given two-particle state. Therefore, to
extract pure quantum entanglement from a partially entangled
state, researchers had done lot of works in the past years on
purification procedures \cite{bennett4,deutsch1}. In other words,
it is possible to compress locally an amount of quantum
information. Now generally a question arises: whether the opposite
is true or not i.e. can quantum correlations be "decompressed"?
This question was tackled by several researchers using the concept
of "Broadcasting of quantum inseparability"
\cite{bandyopadhyay1,buzek2,dobrzanski1,masiak1}. Broadcasting is
nothing but a local copying of non-local quantum correlations.
Among all the
problems regarding entanglement, broadcasting of entanglement is an important issue to consider.\\\\
\textbf{Definition 3.1:} Suppose two distant parties A and B share
two qubit-entangled state
$|s\rangle_{AB}=\alpha|00\rangle_{AB}+\beta|11\rangle_{AB}$, where
$\alpha$ is real and $\beta$ is complex with the condition $\alpha^{2}+ |\beta|^{2}=1$.\\
The first qubit belongs to A and the second belongs to B. Each of the two parties now
perform local copier on their own qubit and then the input entangled state $|s\rangle$
has been broadcast if for some values of the probability $\alpha^{2}$\\
(1) non-local output states are inseparable, and\\
(2) local output states are separable. \\\\
In classical theory one can always broadcast information but in
quantum theory, broadcasting is not always possible
\cite{barnum1,horo3}. H.Barnum et.al. showed that non-commuting
mixed states cannot be broadcasted \cite{barnum1}. However for
pure states broadcasting is equivalent to cloning. In the process
of broadcasting of entanglement, we generally use Peres-Horodecki
theorem for showing the inseparability of non-local outputs and
separability of local
outputs.\\
V.Buzek, V.Vedral, M.B.Plenio, P.L.Knight and M.Hillery
\cite{buzek2} were the first who showed that the decompression of
initial quantum entanglement is possible, i.e. that from a pair of
entangled particles, two less entangled pairs can be obtained by
local $1\rightarrow2$ optimal universal symmetric cloning machine.\\
When the universal B-H quantum cloners are applied locally on each
qubits of the entangled state
$|\phi\rangle_{AB}=\alpha|00\rangle_{AB}+\beta|11\rangle_{AB}$,
the local output described by the density operator is given by
\begin{eqnarray}
\rho_{AA'}=\rho_{BB'}=\frac{2\alpha^{2}}{3}|00\rangle\langle00|+\frac{1}{3}|+\rangle\langle+|
+\frac{2\beta^{2}}{3}|11\rangle\langle11|
\end{eqnarray}
where $|+\rangle=(\frac{1}{\sqrt{2}})(|01\rangle+\langle10|)$, $A'~\textrm{and}~B'$ denote the copies of the input A and
B respectively.\\
while the non-local output described by the density operator is given by
\begin{eqnarray}
\rho_{AB'}=\rho_{A'B}&=&\frac{(24\alpha^{2}+1)}{36}|00\rangle\langle00|+\frac{5}{36}(|01\rangle\langle01|+|10\rangle\langle10|)
+\frac{(24\beta^{2}+1)}{36}|11\rangle\langle11|+{}\nonumber\\&&\frac{4\alpha\beta}{9}(|00\rangle\langle11|+|11\rangle\langle00|)
\end{eqnarray}
From Peres-Horodecki criteria for separability, it follows that
$\rho_{AA'}(\rho_{BB'})$ is separable if
\begin{eqnarray}
\frac{1}{2}-\frac{\sqrt{48}}{16}\leq\alpha^{2}\leq\frac{1}{2}+\frac{\sqrt{48}}{16}
\end{eqnarray}
and $\rho_{AB'}(\rho_{A'B})$ is inseparable if
\begin{eqnarray}
\frac{1}{2}-\frac{\sqrt{39}}{16}\leq\alpha^{2}\leq\frac{1}{2}+\frac{\sqrt{39}}{16}
\end{eqnarray}
Therefore, the entanglement is broadcasted via local state independent quantum cloner
if the probability- amplitude-squared $\alpha^{2}$ is given by the range
\begin{eqnarray}
\frac{1}{2}-\frac{\sqrt{39}}{16}\leq\alpha^{2}\leq\frac{1}{2}+\frac{\sqrt{39}}{16}
\end{eqnarray}
The fidelity of broadcasting is given by
\begin{eqnarray}
F_{1}(\alpha^{2})= \langle\phi|\rho_{AB'}|\phi\rangle =
\frac{25}{36}-\frac{4\alpha^{2}(1-\alpha^{2})}{9}
\end{eqnarray}
It is observed from equation (3.6) that although the state
independent cloner is used as a local cloner for broadcasting
entanglement, the fidelity of copying an entanglement depends on
the input state. Thus, the actions of state independent cloners
locally on the entangled state does produce less entangled pairs
but its
quality depends on the input entangled state.\\
Hence, the average fidelity is given by
\begin{eqnarray}
\overline{F_{1}}=\int_{0}^{1}F_{1}(\alpha^{2})d\alpha^{2}=\frac{67}{108}\simeq0.62
\end{eqnarray}
Further S.Bandyopadhyay and G.Kar \cite{bandyopadhyay1} studied
the broadcasting of entanglement and showed that only those
universal quantum cloners whose fidelity is greater than
$\frac{1}{2}(1+\sqrt{\frac{1}{3}})$ are suitable for local copying
because only then the non-local output states becomes inseparable
for some values of the input parameter $\alpha$. They proved that
an entanglement is optimally broadcast only when optimal quantum
cloners are used for local copying and also showed that
broadcasting of entanglement into more than two entangled pairs is
not possible using only local operations. Later, I.Ghiu
\cite{ghiu1} investigated the broadcasting of entanglement by
using local $1\rightarrow2$ optimal universal asymmetric Pauli
machines and showed that the inseparability is optimally broadcast when symmetric cloners are applied.\\
This chapter is divided into two parts:\\
In the first part, we deal with the problem of how well one can
produce two entangled pairs starting from a given entangled pair
using state dependent cloner as a local copier. Here we construct
a state dependent cloner from B-H quantum cloning transformation
by relaxing one of the universality conditions viz.
$\frac{\partial D_{ab}}{\partial \alpha^{2}}= 0$, where $D_{ab}=
Tr[\rho_{ab}^{(out)}-\rho_{a}^{id}\otimes\rho_{b}^{id}]^{2}$.
$\rho_{ab}^{(out)}$ describes the entangled output state of the
cloner and $\rho_{a}^{id}$, $\rho_{b}^{id}$ describe the input
states in modes 'a' and 'b' respectively. Further we show that the
length of the interval for probability-amplitude-squared
$\alpha^{2}$ for broadcasting of entanglement using state
dependent cloner can be made larger than the length of the
interval for probability-amplitude-squared for broadcasting
entanglement using state independent cloner. Moreover, we show
that there exists local state dependent cloner which gives better
quality copy (in terms of average fidelity) of an entangled pair
than the local universal cloner.
This part is discussed in details in sections 3.2 and 3.3 of this chapter.\\
In the second part, we investigate the problem of secretly
broadcasting three-qubit entangled state between two distant
partners with universal quantum cloning machine and then the
result is generalized to generate secret entanglement among three
parties. Let us suppose that the two distant partners share an
entangled state
$|\psi\rangle_{13}=\alpha|00\rangle+\beta|11\rangle$. The two
parties then apply optimal universal quantum cloning machine on
their respective qubits to produce four qubit state
$|\chi\rangle_{1234}$. One party (say, Alice) then performs
measurement on her quantum cloning machine state vectors. After
that she informs Bob about her measurement result using Goldenberg
and Vaidman's quantum cryptographic scheme based on orthogonal
states \cite{goldenberg1}. Getting measurement result from Alice,
other partner (say, Bob) also performs measurement on his quantum
cloning machine state vectors and using the same cryptographic
scheme, he sends his measurement outcome to Alice. Since the
measurement results are interchanged secretly so Alice and Bob
share secretly a four qubit state. They again apply the cloning
machine on their respective qubits and generate six qubit state
$|\phi\rangle_{125346}$. Therefore, both parties have three qubits
each. Among six qubit state, we interestingly find that there
exists two three
qubit state shared by Alice and Bob which are entangled for some values of the input parameter $\alpha^{2}$.\\
Finally, we investigate the problem of secret entanglement
broadcasting among three distant parties. To solve this problem,
we start with the result of the first part i.e. we assume that the
two distant partners (say, Alice and Bob) share a three qubit
entangled state. Without any loss of generality, we assume that
among three qubits, two are with Alice and one with Bob. Then
Alice teleports one of the qubit to the third distant partner
(say, Carol). After the completion of the teleportation procedure,
we find that the three distant partners share a three qubit
entangled state for the same values of the input parameters
$\alpha^{2}$ as in the first part of the protocol. We discuss this
portion in sections 3.4 and 3.5 of this chapter.\\
This chapter is based on our works "Broadcasting of
Inseparability" \cite{adhikari3} and "Broadcasting of three-qubit
entanglement via local copying and entanglement swapping"
\cite{adhikari5}.
\section{\emph{State dependent B-H quantum cloning machine}}
In the literature, many state dependent quantum cloners were
known. In this section, we also introduce another state dependent
cloner. The introduced state dependent cloner is interesting in
the sense that it can be constructed from B-H quantum cloning
transformation by relaxing one universality condition viz.
$\frac{\partial D_{ab}}{\partial \alpha^{2}}= 0$, where $D_{ab}=
Tr[\rho_{ab}^{(out)}-\rho_{a}^{id}\otimes\rho_{b}^{id}]^{2}$.\\
The B-H cloning transformation is given by
\begin{eqnarray}
|0\rangle|\Sigma\rangle|Q\rangle\rightarrow|0\rangle|0\rangle|Q_{0}\rangle
+(|0\rangle|1\rangle+|1\rangle|0\rangle)|Y_{0}\rangle
\end{eqnarray}
\begin{eqnarray}
|1\rangle|\Sigma\rangle|Q\rangle\rightarrow|1\rangle|1\rangle|Q_{1}\rangle
+(|0\rangle|1\rangle+|1\rangle|0\rangle)|Y_{1}\rangle
\end{eqnarray}
The unitarity of the transformation gives
\begin{eqnarray}
\langle Q_{i}|Q_{i}\rangle+2\langle Y_{i}|Y_{i}\rangle=1,~~~i=0,1
\end{eqnarray}
\begin{eqnarray}
\langle Y_{0}|Y_{1}\rangle=\langle Y_{1}|Y_{0}\rangle=0
\end{eqnarray}
We assume \begin{eqnarray} \langle Q_{0}|Y_{0}\rangle=\langle
Q_{1}|Y_{1}\rangle=\langle Q_{1}|Q_{0}\rangle=0
\end{eqnarray}
Let \begin{eqnarray} |\psi\rangle=\alpha|0\rangle+\beta|1\rangle
\end{eqnarray}
with $\alpha^{2}+|\beta|^{2}=1$, be the input state. Here we assume $\alpha$ is real and $\beta$ is complex.\\
The cloning transformation (3.8-3.9) copies the information of the
input state (3.13) partially into two identical states described
by the density operators $\rho_{a}^{(out)}$ and $\rho_{b}^{(out)}$ respectively.\\
The reduced density operator $\rho_{a}^{(out)}$ is given by
\begin{eqnarray}
\rho_{a}^{(out)}&=&|0\rangle\langle0|[\alpha^{2}+(|\beta|^{2}\langle
Y_{1}|Y_{1}\rangle-\alpha^{2}\langle
Y_{0}|Y_{0}\rangle)]+|0\rangle\langle1|\alpha\beta^{*}[\langle
Q_{1}|Y_{0}\rangle+\langle
Y_{1}|Q_{0}\rangle]+{}\nonumber\\&&|1\rangle\langle0|\alpha\beta[\langle
Q_{1}|Y_{0}\rangle+\langle
Y_{1}|Q_{0}\rangle]+|1\rangle\langle1|[|\beta|^{2}-(|\beta|^{2}\langle
Y_{1}|Y_{1}\rangle-\alpha^{2}\langle
Y_{0}|Y_{0}\rangle)]{}\nonumber\\&=&|0\rangle\langle0|[\alpha^{2}+\lambda(|\beta|^{2}-\alpha^{2})]
+|0\rangle\langle1|\alpha\beta^{*}\mu+|1\rangle\langle0|\alpha\beta\mu+
{}\nonumber\\&&|1\rangle\langle1|[|\beta|^{2}-\lambda(|\beta|^{2}-\alpha^{2})]
\end{eqnarray}
where \begin{eqnarray} \langle Y_{0}|Y_{0}\rangle=\langle Y_{1}|Y_{1}\rangle=\lambda
\end{eqnarray}
\begin{eqnarray}
\langle Q_{0}|Y_{1}\rangle=\langle Q_{1}|Y_{0}\rangle=\langle
Y_{1}|Q_{0}\rangle=\langle Y_{0}|Q_{1}\rangle=\frac{\mu}{2}
\end{eqnarray}
The output state described by the density operator
$\rho_{b}^{(out)}$ looks the same as $\rho_{a}^{(out)}$.\\
The distortion of the qubit in mode 'a' is
\begin{eqnarray}
D_{a}=2\lambda^{2}(4\alpha^{4}-4\alpha^{2}+1)+2\alpha^{2}(1-\alpha^{2})(\mu-1)^{2}
\end{eqnarray}
The distortion $D_{ab}$ is defined by
\begin{eqnarray}
D_{ab}&=&Tr[\rho_{ab}^{(out)}-\rho_{a}^{id}\otimes\rho_{b}^{id}]^{2} {}\nonumber\\&=&
U_{11}^{2}+ 2|U_{12}|^{2}+ 2|U_{12}|^{2}+ U_{22}^{2}+ 2|U_{23}|^{2}+ U_{33}^{2}
\end{eqnarray}
where \begin{eqnarray}
U_{11}= \alpha^{4}-\alpha^{2}(1-2\lambda)\\
U_{12}= \sqrt{2}\alpha^{3}\beta^{*}-\frac{\mu}{\sqrt{2}}\alpha\beta^{*},
U_{21}= (U_{12})^{*}\\
U_{13}= \alpha^{2}(\beta^{*})^{2}, U_{31}= (U_{13})^{*}\\
U_{22}= 2\alpha^{2}|\beta|^{2}-2\lambda\\
U_{23}=\sqrt{2}\alpha\beta^{*}|\beta|^{2}-\frac{\mu}{\sqrt{2}}\alpha\beta^{*},
U_{32}= (U_{23})^{*}\\
U_{33}= |\beta|^{4}-|\beta|^{2}(1-2\lambda)
\end{eqnarray}
The cloning transformation (3.8-3.9) is input state independent if
$D_{a}$ and $D_{ab}$ are input state independent. To make the
cloning transformation (3.8-3.9) input state dependent, we assume
$D_{ab}$ is input state dependent i.e.
\begin{eqnarray}
\frac{\partial D_{ab}}{\partial \alpha^{2}}\neq 0
\end{eqnarray}
The relation between the machine parameters $\lambda$ and $\mu$ is established by
solving the equation $\frac{\partial D_{a}}{\partial \alpha^{2}}= 0 $.\\
Therefore,
\begin{eqnarray}
\frac{\partial D_{a}}{\partial \alpha^{2}}= 0 \Longrightarrow
\mu=1-2\lambda
\end{eqnarray}
The value of the machine parameter $\lambda$ is restricted from
the condition $\frac{\partial D_{ab}}{\partial \alpha^{2}}\neq 0$.
The above condition (3.25) implies that $\lambda$ can take any
value between 0 and $\frac{1}{2}$, except $\frac{1}{6}$. However,
if $\lambda=\frac{1}{6}$, then $\frac{\partial D_{a}}{\partial
\alpha^{2}}= 0$ and $\frac{\partial D_{ab}}{\partial \alpha^{2}}=
0$, therefore the machine becomes
universal in the sense that $D_{a}$ and $D_{ab}$ does not depend on the input state.\\
Putting $\mu= 1-2\lambda$ in equation (3.19-3.24) and using
equation (3.18), we get
\begin{eqnarray}
D_{ab}&=&[\alpha^{4}-\alpha^{2}(1-2\lambda)]^{2}+
4\alpha^{2}(1-\alpha^{2})(\alpha^{2}-\frac{(1-2\lambda)}{2})^{2}+
2\alpha^{4}(1-\alpha^{2})^{2}+{}\nonumber\\&&(2\alpha^{2}(1-\alpha^{2})-2\lambda)^{2}
+ 4\alpha^{2}(1-\alpha^{2})(1-\alpha^{2}-\frac{(1-2\lambda)}{2})^{2} +{}\nonumber\\&&
(1-\alpha^{2})^{2}(2\lambda-\alpha^{2})^{2}
\end{eqnarray}
Now, we are in search of the machine parameter $'\lambda'$ for
which $D_{ab}$ attains its minimum value.\\
For maximum or minimum value of $D_{ab}$, we have
\begin{eqnarray}
\frac{\partial D_{ab}}{\partial \lambda}= 0 \Longrightarrow
\lambda=\frac{3\alpha^{2}(1-\alpha^{2})}{4}
\end{eqnarray}
Again, \begin{eqnarray} \frac{\partial^{2} D_{ab}}{\partial \lambda^{2}}=16 > 0
\end{eqnarray}
Thus for the value of $\lambda$ given in equation (3.28), $D_{ab}$
is minimum.\\
\textsl{Table 3.1: Comparison between B-H state dependent and
independent cloner.}\\
\begin{tabular}{|c|c c|c c|}
\hline &  For B-H state- & dependent cloner & For B-H state-&
independent cloner \\
\hline Input state&
Machine & Distance & Machine & Distance \\
parameter $(\alpha)$ & parameter& between & parameter& between
\\&$\lambda=\frac{3\alpha^{2}(1-\alpha^{2})}{4}$
& input and & $\lambda=\frac{1}{6}$ &
input and \\ &  & output states, & & output states, \\
& & $D_{a}=2\lambda^{2}$ & & $D_{a}$\\
\hline
0.1 & 0.007  & 0.000098 & 0.167 & 0.055556 \\

0.2 & 0.029 & 0.001682 & 0.167 & 0.055556\\

0.3 & 0.061 & 0.007442 & 0.167 & 0.055556\\

0.4 & 0.101 & 0.020402 & 0.167 & 0.055556\\

0.5 & 0.141 & 0.039762 &  0.167 & 0.055556\\

0.6 & 0.173 & 0.059858 & 0.167 & 0.055556\\

0.7 & 0.187 & 0.069938 & 0.167 & 0.055556\\

0.8 & 0.173 & 0.059858 & 0.167 & 0.055556\\

0.9 & 0.115 & 0.026450 & 0.167 & 0.055556\\
\hline
\end{tabular}\\\\
\textbf{\textsl{Illustration of the table 3.1:\\
For the input state $(0.1)|0\rangle+\sqrt{0.99}|1\rangle$, we can
construct a quantum copying machine with parameter $\lambda=0.007$
which produces noisy outputs. As a result of copying procedure,
the identical copies at the output are distorted by the amount
0.000098. The quality of the copy of B-H state independent cloner
remains same for all input states.}}\\\\
Equation (3.29) shows that $D_{ab}$ has minimum value when the
machine parameter $\lambda$ takes the form given in equation
(3.28). Thus we are able to construct a quantum-cloning machine
where machine state vectors depend on input state and therefore
the quality of the copy depends on the input state i.e. for
different input states, machine state vectors take different values and hence the quality of the copy changes.\\
Putting $\mu= 1-2\lambda$ in (3.17), we get
$D_{a}(\alpha^{2})=2\lambda^{2}$, Since $\lambda$ depends on
$\alpha^{2}$.
\section{\emph{Broadcasting of entanglement using state dependent B-H quantum
cloning machine}} In this section we show that to broadcast an
entanglement, state dependent quantum cloning machine is more
effective than state independent B-H quantum cloning machine.\\
Let us consider a general pure entangled state
\begin{eqnarray}
|\chi\rangle_{AB}=\alpha_{1}|00\rangle+
\beta_{1}|11\rangle+\gamma_{1}|10\rangle + \delta_{1}|01\rangle
\end{eqnarray}
where we assume that $\alpha_{1}, \beta_{1}, \gamma_{1},
\delta_{1}$  are real and satisfy the condition $\alpha_{1}^{2}+
\beta_{1}^{2}+ \gamma_{1}^{2}+ \delta_{1}^{2}=1$. The first qubit
(A) belongs to Alice and the second qubit (B)
belongs to Bob.\\
The two distant partners Alice and Bob apply their respective
state dependent quantum cloner on their qubits to produce two
output systems $A'$ and $B'$ respectively. Now to investigate the
existence of non-local correlations in two systems described by
the non-local density operators $\rho_{AB'}~\textrm{and}~
\rho_{A'B}$, we use Peres-Horodecki criteria. The same criteria is
used to test the separability of the local
outputs described by the density operators $\rho_{AA'}~ \textrm{and}~ \rho_{BB'}$.\\
The two non-local output states of a copier are described by the density operator
$\rho_{AB'}$ and $\rho_{A'B}$,
\begin{eqnarray}
\rho_{AB'}&=&\rho_{A'B}= C_{11}|00\rangle\langle00|+
C_{44}|11\rangle\langle11|+C_{22}|01\rangle\langle01|+C_{33}|10\rangle\langle10|+
{}\nonumber\\&& C_{23}(|00\rangle\langle11|+|11\rangle\langle00|)+
C_{12}(|01\rangle\langle00|+ |00\rangle\langle01|)+
C_{13}(|00\rangle\langle10|+|10\rangle\langle00|)+ {}\nonumber\\&&
C_{14}(|01\rangle\langle10|+|10\rangle\langle01|)+
C_{24}(|01\rangle\langle11|+|11\rangle\langle01|)+
C_{34}(|11\rangle\langle10|+{}\nonumber\\&&|10\rangle\langle11|)
\end{eqnarray}
where
\begin{eqnarray}
C_{11}=\alpha_{1}^{2}(1-\lambda)^{2}+\beta_{1}^{2}\lambda^{2}+\lambda(1-\lambda)(\delta_{1}^{2}+\gamma_{1}^{2})\\
C_{12}=\beta_{1}\gamma_{1}\lambda\mu+\delta_{1}\alpha_{1}\mu(1-\lambda)\\
C_{13}=\beta_{1}\delta_{1}\lambda\mu+\alpha_{1}\gamma_{1}\mu(1-\lambda)\\
C_{14}=\mu^{2}\gamma_{1}\delta_{1}\\
C_{22}=\delta_{1}^{2}(1-\lambda)^{2}+\gamma_{1}^{2}\lambda^{2}+\lambda(1-\lambda)(\alpha_{1}^{2}+\beta_{1}^{2})\\
C_{23}=\mu^{2}\alpha_{1}\beta_{1}\\
C_{24}=\alpha_{1}\gamma_{1}\lambda\mu+\beta_{1}\delta_{1}\mu(1-\lambda)\\
C_{33}=\gamma_{1}^{2}(1-\lambda)^{2}+\delta_{1}^{2}\lambda^{2}+\lambda(1-\lambda)(\alpha_{1}^{2}+\beta_{1}^{2})\\
C_{34}=\alpha_{1}\delta_{1}\lambda\mu+\beta_{1}\gamma_{1}\mu(1-\lambda)\\
C_{44}=\alpha_{1}^{2}\lambda^{2}+\beta_{1}^{2}(1-\lambda)^{2}+\lambda(1-\lambda)(\delta_{1}^{2}+\gamma_{1}^{2})
\end{eqnarray}
The two local output states of a copier are described by the
density operators $\rho_{AA'}$ and $\rho_{BB'}$,
\begin{eqnarray}
{}\nonumber\\&&\rho_{AA'}= K_{11}|00\rangle\langle00|+ K_{44}|11\rangle\langle11|+
K_{22}|01\rangle\langle01|+ K_{33}|10\rangle\langle10|+ {}\nonumber\\&&
K_{12}(|01\rangle\langle00|+ |00\rangle\langle01|)+
K_{13}(|00\rangle\langle10|+|10\rangle\langle00|)+ {}\nonumber\\&&
K_{14}(|01\rangle\langle10|+|10\rangle\langle01|)+
K_{24}(|01\rangle\langle11|+|11\rangle\langle01|)+
{}\nonumber\\&&K_{34}(|11\rangle\langle10|+|10\rangle\langle11|)
\end{eqnarray}
where
\begin{eqnarray}
K_{11} = (1-2\lambda)(\alpha_{1}+\delta_{1})^{2} \\
K_{12} = K_{13}= K_{24} = K_{34} =
(\frac{\mu}{2})(\alpha_{1}+\delta_{1})(\beta_{1}+\gamma_{1})\\
K_{14}= K_{22} = K_{33} = \lambda + 2\lambda
(\alpha_{1}\delta_{1} + \beta_{1}\gamma_{1})\\
K_{32} = K_{23} = 0, K_{44} = (1-2\lambda)(\beta_{1}+ \gamma_{1})^{2}
\end{eqnarray}
\begin{eqnarray}
{}\nonumber\\&&\rho_{BB'}= K'_{11}|00\rangle\langle00|+ K'_{44}|11\rangle\langle11|+
K'_{22}|01\rangle\langle01|+ K'_{33}|10\rangle\langle10|+ {}\nonumber\\&&
K'_{12}(|01\rangle\langle00|+ |00\rangle\langle01|)+
K'_{13}(|00\rangle\langle10|+|10\rangle\langle00|)+ {}\nonumber\\&&
K'_{14}(|01\rangle\langle10|+|10\rangle\langle01|)+
K'_{24}(|01\rangle\langle11|+|11\rangle\langle01|)+
{}\nonumber\\&&K'_{34}(|11\rangle\langle10|+|10\rangle\langle11|)
\end{eqnarray}
where
\begin{eqnarray}
K'_{11} = (1-2\lambda)(\alpha_{1}+\gamma_{1})^{2} \\
K'_{12} = K'_{13}= K'_{24} = K'_{34} =
(\frac{\mu}{2})(\alpha_{1}+\gamma_{1})(\beta_{1}+\delta_{1})\\
K'_{14} = K'_{22} = K'_{33} = \lambda + 2\lambda
(\alpha_{1}\gamma_{1} + \beta_{1}\delta_{1})\\
K'_{32} = K'_{23} = 0, K'_{44} = (1-2\lambda)(\beta_{1}+ \delta_{1})^{2}
\end{eqnarray}
The composite systems described by the density operators $\rho_{AB'}$ and $\rho_{A'B}$
are inseparable if at least one of the determinants $W_{3}$ and $W_{4}$ is negative
and $W_{2}$ is non-negative, where
\begin{eqnarray}
 {}\nonumber\\&& W_3= \begin{tabular}{|c c c|}
 $C_{11}$ & $C_{12}$ & $C_{13}$ \\
 $C_{12}$ & $C_{22}$ & $C_{23}$ \\
 $C_{13}$ & $C_{23}$ & $C_{33}$ \\
\end{tabular}~,~~
W_4=\begin{tabular}{|c c c c|}
  $C_{11}$ & $C_{12}$ & $C_{13}$ & $C_{14}$ \\
  $C_{12}$ & $C_{22}$ & $C_{23}$ & $C_{24}$ \\
  $C_{13}$ & $C_{23}$ & $C_{33}$ & $C_{34}$ \\
  $C_{14}$ & $C_{24}$ & $C_{34}$ & $C_{44}$ \\
\end{tabular}~,~~{}\nonumber\\&&
W_2= \begin{tabular}{|c c|}
$C_{11}$ & $C_{12}$ \\
$C_{12}$ & $C_{22}$\\
\end{tabular}
\end{eqnarray}
The entries in the determinants are given by the equations (3.32-3.41).\\
The local output state in Alice's Hilbert space described by the density operator
$\rho_{AA'}$ is separable if
\begin{eqnarray}
 {}\nonumber\\&&W_3= \begin{tabular}{|c c c|}
 $K_{11}$ & $K_{12}$ & $K_{13}$ \\
 $K_{12}$ & $K_{22}$ & $K_{23}$ \\
 $K_{13}$ & $K_{23}$ & $K_{33}$ \\
\end{tabular}~ \geq 0~,~~
W_4=\begin{tabular}{|c c c c|}
  $K_{11}$ & $K_{12}$ & $K_{13}$ & $K_{14}$ \\
  $K_{12}$ & $K_{22}$ & $K_{23}$ & $K_{24}$ \\
  $K_{13}$ & $K_{23}$ & $K_{33}$ & $K_{34}$ \\
  $K_{14}$ & $K_{24}$ & $K_{34}$ & $K_{44}$ \\
\end{tabular}~ \geq 0~,{}\nonumber\\&&
W_2= \begin{tabular}{|c c|}
$K_{11}$ & $K_{12}$ \\
$K_{12}$ & $K_{22}$ \\
\end{tabular}~ \geq 0
\end{eqnarray}
The entries in the determinants are given by the equations
(3.43-3.46).\\
The local output state in Bob's Hilbert space described by the density operator
$\rho_{BB'}$ is separable if
\begin{eqnarray}
 {}\nonumber\\&&W_3= \begin{tabular}{|c c c|}
 $K'_{11}$ & $K'_{12}$ & $K'_{13}$ \\
 $K'_{12}$ & $K'_{22}$ & $K'_{23}$ \\
 $K'_{13}$ & $K'_{23}$ & $K'_{33}$ \\
\end{tabular}~ \geq 0~,~~
W_4=\begin{tabular}{|c c c c|}
  $K'_{11}$ & $K'_{12}$ & $K'_{13}$ & $K'_{14}$ \\
  $K'_{12}$ & $K'_{22}$ & $K'_{23}$ & $K'_{24}$ \\
  $K'_{13}$ & $K'_{23}$ & $K'_{33}$ & $K'_{34}$ \\
  $K'_{14}$ & $K'_{24}$ & $K'_{34}$ & $K'_{44}$ \\
\end{tabular}~ \geq 0~,{}\nonumber\\&&
W_2= \begin{tabular}{|c c|}
$K'_{11}$ & $K'_{12}$ \\
$K'_{12}$ & $K'_{22}$ \\
\end{tabular}~ \geq 0
\end{eqnarray}
The entries in the determinants are given by the equations (3.48-3.51).\\
The general pure entangled state (3.30) can be broadcast if the equations (3.52)-(3.54) are satisfied.\\
For simplicity and without any loss of generality, we assume that the two distant
parties Alice and Bob share a pair of particles prepared in the pure entangled state
\begin{eqnarray}
|\chi\rangle=\alpha_{1}|00\rangle_{AB}+\beta_{1}|11\rangle_{AB}
\end{eqnarray}
where $\alpha_{1}$ is real and $\beta_{1}$ is a complex number
such that $\alpha_{1}^{2}+|\beta_{1}|^{2}=1$.\\
Alice and Bob then apply the state dependent quantum cloner as a
local copier on their qubits. As a result, the two non-local
output states described by the density operators $\rho_{AB'}$ and
$\rho_{A'B}$ are given by
\begin{eqnarray}
\rho_{AB'} = \rho_{A'B}&=&
|00\rangle\langle00|[\alpha_{1}^{2}(1-2\lambda)+\lambda^{2}]+\lambda(1-\lambda)(|01\rangle\langle01|
+|10\rangle\langle10|)+{}\nonumber\\&&|11\rangle\langle11|[|\beta_{1}|^{2}(1-2\lambda)+\lambda^{2}]+
\alpha_{1}\beta_{1}^{*}\mu^{2}|00\rangle\langle11|+{}\nonumber\\&&\alpha_{1}\beta_{1}\mu^{2}|11\rangle\langle00|
\end{eqnarray}
It follows from the Peres-Horodecki theorem that $\rho_{AB'}$ and
$\rho_{A'B}$ are inseparable if\\\\
$W_4=\begin{tabular}{|c c c c|}
  $(1-2\lambda)\alpha_{1}^{2}$ & $0$ & $0$ & $0$ \\
  $0$ & $\lambda(1-\lambda)$ & $\alpha_{1}\beta_{1}^{*}\mu^{2}$ & $0$ \\
  $0$ & $\alpha_{1}\beta_{1}\mu^{2}$ & $\lambda(1-\lambda)$ & $0$ \\
  $0$ & $0$ & $0$ & $|\beta_{1}|^{2}(1-2\lambda)+\lambda^{2}$ \\
\end{tabular}~ < 0$\\\\
$~~~~~\Rightarrow
\alpha_{1}^{4}\mu^{4}-\alpha_{1}^{2}\mu^{4}+\lambda^{2}(1-\lambda)^{2}<0\\\\
~~~~~\Rightarrow
\frac{1}{2}-(\frac{\sqrt{\mu^{4}-4\lambda^{2}(1-\lambda)^{2}}}{2\mu^{2}})<\alpha_{1}^{2}<
\frac{1}{2}+(\frac{\sqrt{\mu^{4}-4\lambda^{2}(1-\lambda)^{2}}}{2\mu^{2}})\\\\
~~~~~\Rightarrow
\frac{1}{2}-(\frac{\sqrt{(1-2\lambda)^{4}-4\lambda^{2}(1-\lambda)^{2}}}{2(1-2\lambda)^{2}})<\alpha_{1}^{2}<
\frac{1}{2}+(\frac{\sqrt{(1-2\lambda)^{4}-4\lambda^{2}(1-\lambda)^{2}}}{2(1-2\lambda)^{2}})$\\\\
Also we note that $W_{3}<0$ and $W_{2}\geq0$.\\
The local density operators $\rho_{AA'}$ and $\rho_{BB'}$ are
given by
\begin{eqnarray}
\rho_{AA'} = \rho_{BB'}&=&
|00\rangle\langle00|\alpha_{1}^{2}(1-2\lambda)+\lambda(|01\rangle\langle01|+
|10\rangle\langle10|+|01\rangle\langle10|+|10\rangle\langle01|)+
{}\nonumber\\&&|11\rangle\langle11||\beta_{1}|^{2}(1-2\lambda)
\end{eqnarray}
Now $\rho_{AA'}$ and $\rho_{BB'}$ are separable if $W_{2}\geq0$,
$W_{3}\geq0$ and $W_{4}\geq0$.\\\\
$W_4=\begin{tabular}{|c c c c|}
  $(1-2\lambda)\alpha_{1}^{2}$ & $0$ & $0$ & $\lambda$ \\
  $0$ & $\lambda$ & $0$ & $0$ \\
  $0$ & $0$ & $\lambda$ & $0$ \\
  $\lambda$ & $0$ & $0$ & $|\beta_{1}|^{2}(1-2\lambda)$ \\
\end{tabular}~ \geq 0 \\\\\\
~~~~~\Rightarrow
\alpha_{1}^{4}(1-2\lambda)^{2}-\alpha_{1}^{2}(1-2\lambda)^{2}+\lambda^{2}\leq0\\\\
~~~~~\Rightarrow \frac{1}{2}-\frac{\sqrt{1-4\lambda}}{2(1-2\lambda)}\leq
\alpha_{1}^{2}\leq
\frac{1}{2}+\frac{\sqrt{1-4\lambda}}{2(1-2\lambda)}$\\\\\\\\\\\\
\textsl{Table 3.2: Intervals representing the separability and
inseparability between two systems.}
\begin{tabular}{| c| c| c| c|}
\hline Machine& Interval$(I_{1})$ & Interval$(I_{2})$ &Common
interval\\ parameter,$\lambda$ & for inseparability & for
separability &  between $(I_{1})$\\& between systems $(A-B')$ &
between systems $(A-A')$ & and $(I_{2})$\\ & and $(A'-B)$ & and
$(B-B')$ &  \\
\hline
0.007 & (0.00005, 0.99994)  & (0.00005, 0.99994) & (0.00005, 0.99994) \\

0.029 & (0.00101, 0.99899) & (0.00094, 0.99905) & (0.00101, 0.99899) \\

0.061 & (0.00555, 0.99444) & (0.00485, 0.99514) & (0.00555, 0.99444) \\

0.101 & (0.02076, 0.97923)& (0.01628, 0.98371) & (0.02076, 0.97923) \\

0.115 & (0.03038, 0.96961) & (0.02282, 0.97717) &  (0.03038, 0.96961)\\

0.141 & (0.05863, 0.94136) & (0.04017, 0.95982) & (0.05863, 0.94136) \\

0.159 & (0.09091, 0.90908) & (0.05768, 0.94231) &  (0.09091, 0.90908) \\

0.173 & (0.12836, 0.87163) & (0.07570, 0.92429) & (0.12836, 0.87163) \\

0.187 & (0.18458, 0.81541)& (0.09904, 0.90095) & (0.18458, 0.81541) \\
\hline
\end{tabular}\\\\
\textbf{\textsl{Illustration of the table 3.2:\\
If locally the quantum cloning machine with parameter
$\lambda=0.029$ is used for broadcasting an entanglement, then the
2-qubit systems $(A-B')$ and $(A'-B)$ possess non-local properties
for $0.00101<\alpha_{1}^{2}< 0.99899$ and the systems $(A-A')$ and
$(B-B')$ possess local properties for $0.00094<\alpha_{1}^{2}<
0.99905$. Therefore, the class of input entangled states with
parameter $\alpha_{1}^{2}$ lying in the interval $(0.00101,
0.99899)$ has been broadcast when the local
copying machine with parameter $\lambda=0.029$ is used.}}\\\\
We can observe from the above table that in the last two cases,
the length of the intervals for broadcasting via state dependent
cloner are smaller than the length of the interval for
broadcasting discussed by Buzek et.al. while the situation is opposite in the remaining cases.\\
Now to see how well the local state dependent quantum cloners produce two entangled
pairs from a single pair, we have to calculate the amount of overlapping between the
input entangled state and the output entangled state described by the density
operator $\rho_{AB'}(\rho_{A'B})$.\\
The fidelity of broadcasting of inseparability is given by
\begin{eqnarray}
F(\alpha_{1}^{2})=\langle\chi|\rho_{AB'}|\chi\rangle=(1-\lambda)^{2}
-4\alpha_{1}^{2}(1-\alpha_{1}^{2})\lambda(1-2\lambda)
\end{eqnarray}
\textsl{Table 3.3: Quality of the copies of an entangled state
produced from state dependent B-H quantum cloning machine}\\
\begin{tabular}{|c|c|c|}
\hline Amplitude&
Machine & $F(\alpha_{1}^{2})=(1-\lambda)^{2}+4\lambda(1-2\lambda)\alpha_{1}^{2}(1-\alpha_{1}^{2})$ \\
$(\alpha)$ & parameter&
\\&$\lambda=\frac{3\alpha^{2}(1-\alpha^{2})}{4}$ &\\
\hline
0.1 & 0.007  & 0.99  \\

0.2 & 0.029 & 0.94 \\

0.3 & 0.061 & 0.86 \\

0.4 & 0.101 & 0.76 \\

0.5 & 0.141 & 0.66 \\

0.6 & 0.173 & 0.58 \\

0.7 & 0.187 & 0.54 \\

0.8 & 0.173 & 0.58 \\

0.9 & 0.115 & 0.72 \\
\hline
\end{tabular}\\\\
\textbf{\textsl{Illustration of the table 3.3:\\
The state dependent B-H quantum cloning machine with parameter
$\lambda=0.007$ produce two less entangled copies of the input
entangled state $(0.1)|0\rangle+\sqrt{0.99}|1\rangle$ with
fidelity 0.99.}}
\section{\emph{Secret broadcasting of 3-qubit entangled state between two distant
partners}} All the previous works on the broadcasting of
entanglement dealt with the generation of two 2-qubit entangled
state starting from a 2-qubit entangled state using either optimal
universal symmetric cloner or asymmetric cloner. The generated two
qubit entangled state can be used as a quantum channel in quantum
cryptography, quantum teleportation etc. The advantage of our
protocol over other protocols of broadcasting is that we are able
to provide a protocol which generates secret quantum channel
between distant partners. The introduced protocol generates two
3-qubit entangled states between two distant partners starting
from a 2-qubit entangled state and also provides the security of
the generated quantum channel. Not only that we also generalize
our protocol from two parties to three parties and show that the
generated 3-qubit entangled states can serve as a secured quantum
channel between three distant parties.\\\\
\textbf{Few definitions}\\
Let the shared entangled state described by the two qubit density operator be
$\rho_{13}$. Using B-H quantum cloning machine twice by the distant partners (Alice
and Bob) on their respective qubits, they generate total six-qubit state
$\rho_{125346}$ between them. Therefore, Alice has three qubits
'1','2' and '5' and Bob possesses three qubits '3', '4' and '6'.\\
\textbf{Definition 3.2:} The three-qubit entanglement is said to
be broadcast if (i) Any of the two local outputs (say
($\rho_{12}$,$\rho_{15}$) in Alice's side and
($\rho_{34}$,$\rho_{36})$ in Bob's side) are separable (ii) One
local output (say $\rho_{25}$ in Alice's side and $\rho_{46}$ in
Bob's side) is inseparable and associated with these local
inseparable output, two non-local outputs (say
($\rho_{23}$,$\rho_{35}$) and ($\rho_{14}$,$\rho_{16})$) are inseparable.\\
\textbf{Definition 3.3:} An entanglement is said to be closed if
each party has non-local correlation with other parties. For
instance, any three particle entangled state described by the
density operator $\rho_{325}$ is closed if $\rho_{32}$,$\rho_{25}$
and $\rho_{35}$ are entangled state. Otherwise, it is said to be
an open entanglement (See figure 3.7).
\section{\emph{Discussion of quantum cryptographic scheme based on orthogonal states}}


Since non-orthogonal states cannot be cloned so many quantum
cryptographic protocols were designed on the basis of No-cloning
principle which uses non-orthogonal states as the carriers of
information. In 1995, L.Goldenberg and L.Vaidman introduced a
quantum cryptographic scheme which was based on orthogonal states
\cite{goldenberg1}. The two distant partners Alice and Bob uses
Goldenberg and Vaidman's quantum cryptographic scheme to send
their measurement results. Now we describe Goldenberg and
Vaidman's quantum cryptographic scheme adopted by Alice to send
her measurement result to Bob.\\
Without any loss of generality, let Alice consider the same
experimental setup as used in \cite{goldenberg1} to send her
measurement result to Bob. The set up consists of a Mach-Zehnder
interferometer with two storage rings, $SR_1$ and $SR_2$, of equal
time delays. The set up is described in figure-3.1. Alice can
transmit a bit by sending a single particle either from the source
$S_{|\uparrow\rangle^A}$ (sending 0)
or from the source $S_{|\downarrow\rangle^A}$ (sending 1), where\\
\begin{eqnarray}
|\uparrow\rangle^A= \frac{1}{\sqrt{2}}(|a\rangle+|b\rangle)
\end{eqnarray}
\begin{eqnarray}
|\downarrow\rangle^A= \frac{1}{\sqrt{2}}(|a\rangle-|b\rangle)
\end{eqnarray}
Alice then registered the sending time $t_s$ for later use. The
particle passes through the first beam-splitter $BS_1$ and evolves
into a superposition of two localized wavepackets:$|a\rangle$,
moving in the upper channel and $|b\rangle$, moving in the bottom
channel. The wavepacket $|b\rangle$ is delayed for some time
$\tau$ in the storage ring $SR_1$ while $|a\rangle$ is moving in
the upper channel. When $|a\rangle$ arrives to the storage ring
$SR_2$ at Bob's site, wavepacket $|b\rangle$ starts moving on the
bottom channel towards Bob. Let us assume for simplicity that the
travelling time of the particles from Alice to Bob be $\epsilon$
smaller than the delayed time $\tau$. During the flight-time of
$|b\rangle$, wavepacket $|a\rangle$ is delayed in $SR_2$. Finally,
the two wavepackets arrive simultaneously to the second
beam-slitter $BS_2$ and interfere. A particle started in the state
$|\uparrow\rangle^A$ emerges at the detector
$D_{|\uparrow\rangle}$, and a particle started in the state
$|\downarrow\rangle^A$ emerges at the detector
$D_{|\downarrow\rangle}$. Bob, detecting the arriving particle,
receives the bit sent by Alice: $D_{|\uparrow\rangle}$ activated
means $'0'$ and $D_{|\downarrow\rangle}$ activated means $'1'$. As
soon as Bob detects the arriving particle, he registers the
receiving time of the particle $t_r$. Now the only task remaining
for Alice and Bob is to detect the third party $'Eve'$. To do the
same, Alice and Bob perform two tests (using a classical channel).
First, they compare the sending time $t_s$ with the receiving time
$t_r$ for each particle, where $t_r=t_s+\tau+\epsilon$. Second,
they look for changes in the data by comparing a portion of the
transmitted bits with the same portion of the received bits. If,
for any checked bit, the timing is not respected or
anti-correlated bits are found, the users learn about the
intervention of Eve. In this way, Alice sends her measurement
result secretly (either $|\uparrow\rangle^A$ or $|\downarrow\rangle^A$) to Bob. \\
If Eve wants to learn the message in a mid-way sending from Alice
to Bob and at the same time if she wants to omit her presence in
the running protocol then she has to obey the two basic
requirements: 1) since the data is encoded in the relative phase
between the two wavepackets $|a\rangle$ and $|b\rangle$ so she has
to do something which makes the phase same at $t_s$ and $t_r$. 2)
She have to keep in mind that the two wavepackets must arrive
together to $BS_2$ at the correct time, otherwise a timing problem occurs.\\\\
\textbf{Description of protocol}\\
Let us consider that the two qubit entangled state
$|\psi\rangle_{13}$ is shared between two distant partners
popularly known as Alice and Bob. The particles '1' possessed by
Alice and the paticle '3' possessed by Bob respectively. Alice and
Bob then operate quantum cloning machines on their respective
qubits. After cloning procedure, Alice performs measurement on the
quantum cloning machine state vector and sends the measurement
result to Bob using the Goldenberg and Vaidman's quantum
cryptographic scheme based \cite{goldenberg1} on orthogonal
states. After getting measurement result from Alice; Bob performs
measurement on his quantum cloning machine state vector and sends
the measurement result to Alice using the same cryptographic
scheme adopted by Alice. Consequently, the two distant partners
share a four qubit state $|\zeta\rangle_{1234}$. Now Alice has two
qubits '1' and '2' and Bob '3' and '4' respectively. Both of them
again operates quantum cloning machine on one of the qubits, they
holds. As a result, the distant parties now share six qubit state
$|\phi\rangle_{125346}$ in which three qubits '1','2' and
'5'possessed by Alice and remaining three qubits '3','4' and '6'
possessed by Bob. Here we show that there exist two 3-qubit
entangled states between two distant partners for some values of
the input parameter $\alpha^{2}$ and therefore it is possible to
show that secret broadcasting of 3-qubit entangled state using
only universal quantum cloning machine. The word 'secret' is
justified by observing an important fact that the transmission of
measurement results from Alice to Bob and Bob to Alice have been
done by using Goldenberg and Vaidman's quantum cryptographic
scheme. Therefore, message regarding measurement results have been
transmitted secretly between two distant partners. Hence, the
broadcasted three-qubit entangled state is only known to Alice and
Bob and not to the third party 'Eve'. As a result, these newly
generated three-qubit entangled states can be used as secret
quantum channels in various quantum cryptographic scheme. Also, it
is very difficult to hack the quantum information flowing between
two distant partners via our proposed quantum channel because to
hack the quantum information the hacker (say, Eve) has to do two
things: First, she has to gather knowledge about the initially
shared entangled state and secondly, she has to collect
information about the measurement result performed by two distant
partners. Therefore, the quantum channel generated by our protocol
is more secured and hence can be used in various protocols viz. quantum key distribution protocols \cite{cabello1,li1} etc.\\
Now to understand our protocol more clearly, we again discuss the whole protocol below
by considering a specific example.\\
\underline{\textbf{Step -1}}\\
Let the two particle entangled state shared by two distant
partners Alice and Bob be given by
\begin{eqnarray}
|\psi\rangle_{13}=\alpha|00\rangle+\beta|11\rangle
\end{eqnarray}
where $\alpha$ is real and $\beta$ is complex with $\alpha^{2}+|\beta|^{2}=1$.
(See figure-3.2).\\
\underline{\textbf{Step-2}}\\
The B-H quantum copier is given by the transformation\\
\begin{eqnarray}
|0\rangle|\Sigma\rangle|Q\rangle\rightarrow
\sqrt{\frac{2}{3}}|00\rangle|Q_{0}\rangle+\frac{1}{\sqrt{3}}|\psi^+\rangle|Q_{1}\rangle
\end{eqnarray}
\begin{eqnarray}
|1\rangle|\Sigma\rangle|Q\rangle\rightarrow
\sqrt{\frac{2}{3}}|11\rangle|Q_{1}\rangle+\frac{1}{\sqrt{3}}|\psi^+\rangle|Q_{0}\rangle
\end{eqnarray}
where $|\psi^+\rangle=\frac{1}{\sqrt{2}}(|01\rangle+|10\rangle$
and $|Q_{0}\rangle,|Q_{1}\rangle$ are orthogonal quantum cloning
machine state vectors. Alice and Bob then operates B-H quantum
cloning machine locally to copy the state of their respective
particles. Therefore, the combined system of four qubits is given
by
\begin{eqnarray}
|\chi\rangle_{1234}&=&[(\frac{2\alpha}{3}|0000\rangle+\frac{\beta}{3}|\psi^+\rangle|\psi^+\rangle)|Q_{0}\rangle^B+
(\frac{\sqrt{2}\alpha}{3}|00\rangle|\psi^+\rangle+\frac{\sqrt{2}\beta}{3}|\psi^+\rangle|11\rangle)
|Q_{1}\rangle^B]{}\nonumber\\&&|Q_{0}\rangle^A+[(\frac{\sqrt{2}\alpha}{3}|\psi^+\rangle|00\rangle+
\frac{\sqrt{2}\beta}{3}|11\rangle|\psi^+\rangle)|Q_{0}\rangle^B+(\frac{\alpha}{3}|\psi^+\rangle|\psi^+\rangle
+{}\nonumber\\&&\frac{2\beta}{3}|1111\rangle)|Q_{1}\rangle^B]|Q_{1}\rangle^A
\end{eqnarray}
The subscripts 1,2 and 3,4 refer to two approximate copy qubits in
the Alice's and Bob's Hilbert space respectively. Also
$|\rangle^A$ and $|\rangle^B$ denote quantum cloning machine state
vectors in Alice's and Bob's Hilbert space respectively.
(See figure-3.3).\\
Alice then performs measurement on the quantum cloning machine
state vectors in the basis $\{|Q_{0}\rangle^A,|Q_{1}\rangle^A\}$.
Thereafter Alice informs Bob about her measurement result using
Goldenberg and Vaidman's \cite{goldenberg1} quantum cryptographic
scheme based on orthogonal states. After getting measurement
result from Alice, Bob also performs measurement on the quantum
cloning machine state vectors in the basis
$\{|Q_{0}\rangle^B,|Q_{1}\rangle^B\}$ and then using the same
cryptographic scheme, he sends his measurement outcome to Alice.
In this way Alice and Bob interchange their measurement
results secretly.\\
\underline{\textbf{Step-3}}\\
After measurement, let the state shared by Alice and Bob be given
by
\begin{eqnarray}
|\zeta_a\rangle_{1234}= \frac{1}{\sqrt{N}}
[\frac{2\alpha}{3}|0000\rangle+\frac{\beta}{3}|\psi^+\rangle|\psi^+\rangle]
\end{eqnarray}
Where $N=\frac{3\alpha^{2}+1}{9}$ represents the normalization factor. \\
Afterward, Alice and Bob again operate their respective cloners on
the qubits '2' and '4' respectively and therefore, the total state
of six qubits is given by
\begin{eqnarray}
|\phi\rangle_{125346}&=& \frac{1}{\sqrt{N}}
[\frac{2\alpha}{3}[|0\rangle_1\otimes(\sqrt{\frac{2}{3}}|00\rangle|Q_{0}\rangle+\frac{1}{\sqrt{3}}|\psi^+\rangle|Q_{1}\rangle)_{25}
\otimes|0\rangle_3\otimes(\sqrt{\frac{2}{3}}|00\rangle|Q_{0}\rangle+{}\nonumber\\&&\frac{1}{\sqrt{3}}|\psi^+\rangle|Q_{1}\rangle)_{46}
+\frac{\beta}{6}[|0\rangle_1\otimes(\sqrt{\frac{2}{3}}|11\rangle|Q_{1}\rangle+\frac{1}{\sqrt{3}}|\psi^+\rangle|Q_{0}\rangle)_{25}
\otimes|0\rangle_3\otimes{}\nonumber\\&&(\sqrt{\frac{2}{3}}|11\rangle|Q_{1}\rangle+\frac{1}{\sqrt{3}}|\psi^+\rangle|Q_{0}\rangle)_{46}
+|0\rangle_1\otimes(\sqrt{\frac{2}{3}}|11\rangle|Q_{1}\rangle+\frac{1}{\sqrt{3}}|\psi^+\rangle|Q_{0}\rangle)_{25}
{}\nonumber\\&&\otimes|1\rangle_3\otimes(\sqrt{\frac{2}{3}}|00\rangle|Q_{0}\rangle+\frac{1}{\sqrt{3}}|\psi^+\rangle|Q_{1}\rangle)_{46}
+|1\rangle_1\otimes(\sqrt{\frac{2}{3}}|00\rangle|Q_{0}\rangle+{}\nonumber\\&&\frac{1}{\sqrt{3}}|\psi^+\rangle|Q_{1}\rangle)_{25}
\otimes|0\rangle_3\otimes(\sqrt{\frac{2}{3}}|11\rangle|Q_{1}\rangle+{\frac{1}{\sqrt{3}}|\psi^+\rangle|Q_{0}\rangle)_{46}
+|1\rangle_1\otimes}\nonumber\\&&(\sqrt{\frac{2}{3}}|00\rangle|Q_{0}\rangle+\frac{1}{\sqrt{3}}|\psi^+\rangle|Q_{1}\rangle)_{25}
\otimes|1\rangle_3\otimes(\sqrt{\frac{2}{3}}|00\rangle|Q_{0}\rangle+{}\nonumber\\&&\frac{1}{\sqrt{3}}|\psi^+\rangle|Q_{1}\rangle)_{46}]
\end{eqnarray}
Now our task is to see whether we can generate two 3-qubit
entangled state from above six qubit state or not. To examine the
above fact, we have to consider two 3-qubit states described by
the density operators $\rho_{146}$ and $\rho_{325}$ (See figure-3.4).\\
The density operator $\rho_{146}$ is given by
\begin{eqnarray}
\rho_{146}&=&\frac{1}{N}[
\frac{4\alpha^2}{9}(\frac{2}{3}|000\rangle\langle000|+\frac{1}{3}|0\psi^+\rangle\langle0\psi^+|)+\frac{\alpha\beta^*}{9}(\frac{\sqrt{2}}{3}|000\rangle\langle1\psi^+|
+\frac{\sqrt{2}}{3}|0\psi^+\rangle\langle111|)+{}\nonumber\\&&\frac{\alpha\beta}{9}(\frac{\sqrt{2}}{3}|111\rangle\langle0\psi^+|
+\frac{\sqrt{2}}{3}|1\psi^+\rangle\langle000|)+\frac{|\beta|^2}{36}(\frac{2}{3}|011\rangle\langle011|+\frac{2}{3}|0\psi^+\rangle\langle0\psi^+|
+{}\nonumber\\&&\frac{2}{3}|000\rangle\langle000|+\frac{2}{3}|111\rangle\langle111|+\frac{2}{3}|1\psi^+\rangle\langle1\psi^+|
+\frac{2}{3}|100\rangle\langle100|)]
\end{eqnarray}
The density operator $\rho_{325}$ describes the other 3-qubit state looks exactly the same as $\rho_{146}$.\\
Now to show the state described by the density operator $\rho_{146}$ is entangled, we
have to show that the two qubit states described by the density operators
$\rho_{14}$,$\rho_{16}$ and $\rho_{46}$ are entangled i.e. we have to show that there
exist some values of the input state parameter $\alpha^{2}$ for
which the three-qubit state is a closed entangled state.\\
The reduced density operators $\rho_{14}$,$\rho_{16}$ and $\rho_{46}$ are given by
\begin{eqnarray}
\rho_{16}=\rho_{14}&=& \frac{1}{N}
[\frac{4\alpha^2}{9}(\frac{5}{6}|00\rangle\langle00|+\frac{1}{6}|01\rangle\langle01|)+\frac{2\alpha\beta^*}{27}|00\rangle\langle11|
+\frac{2\alpha\beta}{27}|11\rangle\langle00|+{}\nonumber\\&&\frac{|\beta|^2}{36}(|00\rangle\langle00|+|01\rangle\langle01|+|10\rangle\langle10|+|11\rangle\langle11|)]
\end{eqnarray}
\begin{eqnarray}
\rho_{46}&=& \frac{1}{N}
[\frac{4\alpha^2}{9}(\frac{2}{3}|00\rangle\langle00|+\frac{1}{6}(|01\rangle\langle01|+|01\rangle\langle10|+|10\rangle\langle01|+|10\rangle\langle10|))+
{}\nonumber\\&&\frac{|\beta|^2}{36}(\frac{4}{3}|00\rangle\langle00|+
\frac{4}{3}|11\rangle\langle11|+
\frac{2}{3}(|01\rangle\langle01|+|01\rangle\langle10|+|10\rangle\langle01|+{}\nonumber\\&&|10\rangle\langle10|))]
\end{eqnarray}
Using Peres-Horodecki theorem, we find that the states described
by the density operators $\rho_{16}$ and $\rho_{14}$ are entangled
if $0.18<\alpha^2<1$ and the state described by the density
operator $\rho_{46}$ is entangled if $0.61<\alpha^2<1$. Therefore,
we can say that the state described by the density operator
$\rho_{146}$ is a closed three qubit entangled state if
$0.61<\alpha^2<1$. Similarly, the other reduced density operator
$\rho_{325}$ describe a closed entangled state if $0.61<\alpha^2<1$.\\
Also the other two-qubit states described by the density operators
$\rho_{12}$,$\rho_{15}$,$\rho_{34}$ and $\rho_{36}$ are given by
\begin{eqnarray}
\rho_{12}=\rho_{15}=\rho_{34}=\rho_{36}&=& \frac{1}{N}
[\frac{4\alpha^2}{9}(\frac{5}{6}|00\rangle\langle00|+\frac{1}{6}|01\rangle\langle01|)
+\frac{|\beta|^2}{36}(\frac{1}{3}|00\rangle\langle00|+\frac{5}{3}|01\rangle\langle01|+{}\nonumber\\&&\frac{4}{3}|01\rangle\langle10|+
\frac{4}{3}|10\rangle\langle01|+\frac{5}{3}|10\rangle\langle10|+\frac{1}{3}|11\rangle\langle11|)]
\end{eqnarray}
These density operators are separable only when $0.27<\alpha^2<1$.
Hence, broadcasting of three-qubit entangled state is possible when $0.61<\alpha^2<1$.\\
Now, our task is to find out how is the entanglement distributed
over the state i.e. how much are the two qubit density operators
$\rho_{16}$, $\rho_{14}$ and $\rho_{46}$ are entangled. To
evaluate the amount of entanglement, we have to calculate the
concurrence defined by Wootters \cite{wootters2} and hence entanglement of formation.\\
After some tedious calculations, we find that the concurrence and
hence the entanglement of formation depends on the probability
$\alpha^{2}$. Therefore, we have to calculate the amount of
entanglement in the 2-qubit states described by the reduced
density operators $\rho_{16},\rho_{14}$ and $\rho_{46}$ in the
range $0.61<\alpha^2<1$ because the two qubit reduced density
operators are entangled in this range of the input state parameter
$\alpha^2$. Since concurrence depends on $\alpha^{2}$ so it varies
as $\alpha^{2}$ varies. Therefore, when $0.61<\alpha^2<1$, the
concurrences for the mixed states described by density operators
$\rho_{16}, \rho_{14}$ varies from 0.17 to 0.29 while the
concurrence for the mixed states described by density operators
$\rho_{46}$ varies from 0.08 to 0.15 respectively. Using equations
(1.9) and (1.10), we find that the amount of entanglement between
the states described by the density oprators $\rho_{16},
\rho_{14}$ varies from 0.06 to 0.15 while the amount of
entanglement between the states described by the density operator
$\rho_{46}$ varies from 0.01 to 0.03 respectively. Therefore, the
generated three-qubit entangled state is a weak closed entangled
state in the sense that the amount of entanglement in the
two-qubit density operators are very low. Further, the above
results show that the entanglement between the qubits 1 and 6 (1
and 4) is higher than that between the the qubits 4 and 6.\\
Furthermore, if the measurement results are
$\frac{\sqrt{2}\alpha}{3}|00\rangle|\psi^{+}\rangle
+\frac{\sqrt{2}\beta}{3}|\psi^{+}\rangle|11\rangle$ or
$\frac{\sqrt{2}\alpha}{3}|\psi^{+}\rangle|00\rangle
+\frac{\sqrt{2}\beta}{3}|11\rangle|\psi^{+}\rangle$, then the two
3-qubit entangled state described by the density operators
$\rho_{146}$ and $\rho_{325}$ are different and the broadcasting
is possible for $0.6<\alpha^{2}<1$ or $0.14<\alpha^{2}<0.4$
according to the measurement results. Also if the outcome of the
measurement is
$\frac{\alpha}{3}|\psi^{+}\rangle|\psi^{+}\rangle+\frac{2\beta}{3}|1111\rangle$,
then the state described by the density operators $\rho_{146}$ and
$\rho_{325}$ are identical and the broadcasting is possible for
$0.38<\alpha^{2}<0.73$.
\section{\emph{Secret generation of two 3-qubit entangled state between three
distant partners}} In this section, we investigate a question, can
we secretly generate two 3-qubit entangled state shared between
three distant partners using LOCC? The answer is in affirmative.
To generate three-qubit entangled states between three distant
partners, we require only two well-known concept: (i) quantum
cloning and (ii) entanglement swapping
\cite{bose1,zukowski1}.\\\\
Let us suppose for the implementation of any particular
cryptographic scheme, three distant partners Alice, Bob and Carol
want to generate two three qubit entangled states between them. To
do the same task, let us assume that initially Alice-Bob and
Carol-Alice share two qubit entangled states described by the
density operators $\rho_{13}$, $\rho_{78}$, where Alice has qubits
'1' and '8', Bob and Carol possess qubits '3' and '7'
respectively. Then Alice and Bob adopting the broadcasting process
described in the previous section to generate two three-qubit
entangled state in between them. Therefore, Alice and Bob now have
two 3-qubit entangled states described by the density operators
$\rho_{146}$ and $\rho_{325}$ where Alice has qubits '1','2'and
'5' and Bob possesses '3','4' and '6'. Now we are in a position
for the illustration of the generation of 3-qubit entangled state
between three parties at distant places by using the concept of entanglement swapping.\\
Without any loss of generality, we take a three-qubit entangled
state between two distant parties described by the density
operator $\rho_{325}$.\\
The density operator $\rho_{325}$ can be rewritten as
\begin{eqnarray}
\rho_{325}&=&\frac{1}{N}
[\frac{4\alpha^2}{9}(\frac{2}{3}|000\rangle\langle000|+\frac{1}{3}|0\psi^+\rangle\langle0\psi^+|)+\frac{\alpha\beta^*}{9}(\frac{\sqrt{2}}{3}|000\rangle\langle1\psi^+|
+\frac{\sqrt{2}}{3}|0\psi^+\rangle\langle111|)+{}\nonumber\\&&\frac{\alpha\beta}{9}(\frac{\sqrt{2}}{3}|111\rangle\langle0\psi^+|
+\frac{\sqrt{2}}{3}|1\psi^+\rangle\langle000|)+\frac{|\beta|^2}{36}(\frac{2}{3}|011\rangle\langle011|+\frac{2}{3}|0\psi^+\rangle\langle0\psi^+|
+{}\nonumber\\&&\frac{2}{3}|000\rangle\langle000|+\frac{2}{3}|111\rangle\langle111|+\frac{2}{3}|1\psi^+\rangle\langle1\psi^+|
+\frac{2}{3}|100\rangle\langle100|)]
\end{eqnarray}
where qubits 2 and 5 are possessed by Alice and qubit 3 is possessed by Bob respectively.\\
To achieve the goal of the generation of three qubit entangled state between three
distant partners, we proceed in the following way:\\
Let Alice and Carol shared a singlet state
\begin{eqnarray}
|\psi^-\rangle_{87}=(\frac{1}{\sqrt{2}})(|01\rangle-|10\rangle)
\end{eqnarray}
where particles 8 and 7 are possessed by Alice and Carol respectively.\\
The combined state between Alice,Bob and Carol is given by the
\begin{eqnarray}
\rho_{32587}=\rho_{325}\otimes|\psi^-\rangle_{78}\langle\psi^-|
\end{eqnarray}
Alice then performs Bell state measurement on the particles 2 and
8 in the basis $\{|B^{\pm}_1\rangle,|B^{\pm}_2\rangle\}$, where
$|B^{\pm}_1\rangle=(\frac{1}{\sqrt{2}})(|00\rangle\pm|11\rangle)$,
$|B^{\pm}_2\rangle=(\frac{1}{\sqrt{2}})(|01\rangle\pm|10\rangle)$\\
If the measurement result is $|B^{+}_1\rangle$, then the 3-qubit density operator is
given by
\begin{eqnarray}
\rho_{357}&=&\frac{1}{N}
[\frac{4\alpha^2}{9}[\frac{2}{3}|001\rangle\langle001|+\frac{1}{6}(|011\rangle\langle011|-|011\rangle\langle000|-|000\rangle\langle011|+|000\rangle\langle000|)]
+{}\nonumber\\&&\frac{\alpha\beta^*}{27}(|001\rangle\langle111|-|001\rangle\langle100|+|000\rangle\langle110|-|011\rangle\langle110|)
+\frac{\alpha\beta}{27}(-|110\rangle\langle011|+{}\nonumber\\&&|110\rangle\langle000|+|111\rangle\langle001|-|100\rangle\langle001|)
+\frac{|\beta|^2}{36}[\frac{2}{3}(|010\rangle\langle010|+|001\rangle\langle001|+{}\nonumber\\&&|110\rangle\langle110|+|101\rangle\langle101|)
+\frac{1}{3}(|011\rangle\langle011|-|011\rangle\langle000|-|000\rangle\langle011|+|000\rangle\langle000|
{}\nonumber\\&&+|111\rangle\langle111|-|111\rangle\langle100|-|100\rangle\langle111|+|100\rangle\langle100|)]
\end{eqnarray}
After Bell-state measurement, Alice announces publicly the measurement result.
Thereafter, Alice,Bob and Carol operate an unitary operator
$U_1=I_3\otimes(\sigma_z)_5\otimes(\sigma_x)_7$ on their respective particles to
retrieve the state described by the density operator $\rho_{325}$.\\
If the measurement result is $|B^{-}_1\rangle$ or $|B^{+}_2\rangle $ or
$|B^{-}_2\rangle$ then accordingly they operate an unitary operator
$U_2=I_3\otimes(I_5)\otimes(\sigma_x)_7$ or $U_3=I_3\otimes(I_5)\otimes(\sigma_z)_7$
or $U_4=I_3\otimes(I_3)\otimes(I_7)$  on their respective particles to retrieve the
state described by the density operator $\rho_{325}$.\\
For the remaining cases given in step-3 of the previous section
3.5, we can proceed in a similar manner as above. Hence, in every
cases we find that after getting the measurement result, each
party (Alice, Bob and Carol) apply suitable unitary operators on
their respective particles to share the 3-qubit entangled state
in between them, which is previously shared between only two distant partners Alice and Bob.
The above protocol is described pictorially in figures 3.5 and 3.6.\\
Therefore, we describe here the secret generation of 3-qubit
entangled state between three distant partners starting from
3-qubit entangled state shared between two distant partners using
quantum cloning and entanglement swapping. This quantum channel
generated by the above procedures can be regarded as a secret
quantum channel because the result of the measurement on the
machine state vectors are transmitted secretly by quantum cryptographic scheme.\\

\large \baselineskip .85cm

\chapter{On universal quantum deletion transformation}
\setcounter{page}{102} \markright{\it CHAPTER~\ref{chap4}. On
universal quantum deletion transformation}
\label{chap4}%
The most remarkable discovery ever made by scientists was science
itself - Jacob Bronowski\\\\
The whole of science is nothing more than a refinement of everyday
thinking - Albert Einstein\\\\
Science cannot solve the ultimate mystery of Nature. And it is
because in the last analysis we ourselves are part of the mystery
we are trying to solve - Max Planck

\section{\emph{Prelude}}
The complementary theory of "quantum no-cloning theorem" is the
"quantum no-deleting" principle \cite{pati5}. It states that
linearity of quantum theory forbids deleting one unknown quantum
state against a copy in either a reversible or an irreversible
manner. We can understand the principle behind quantum deletion
more clearly, if we compare quantum deletion with the "Landauer
erasure principle" \cite{land1}. It tells us that a single copy of
some classical information can be erased with some energy cost. It
is an irreversible operation. In quantum theory, erasure of a
single unknown state may be thought of as swapping it with some
standard state and then dumping it into the environment. Unlike
quantum erasure, quantum deletion is a different concept. Quantum
deletion \cite{horo1,pati6} is more like reversible 'uncopying' of
an unknown quantum state. Although there is not a perfect deleting
machine, the corresponding no-deleting principle does not prohibit
us from constructing the approximate deleting machine
\cite{adhikari1,adhikari2,adhikari4,pati6}. If quantum deleting
could be done, then one would create a standard blank state onto
which an unknown quantum state is copied approximately by
deterministic cloning or exactly by probabilistic cloning process
\cite{duan1,pati2,zhang3}. We can apply the quantum
deleting machine in a situation when scarcity of memory in a quantum computer occurs.\\
In this chapter, we discuss the problem of constructing an
efficient universal deletion machine in the sense of high fidelity
of deletion. We construct two types of "universal quantum deletion
machine" which approximately deletes a copy such that the fidelity
of deletion does not depend on the input state. Also we classify
two types of universal quantum deletion machines: (1) a
conventional deletion machine described by one unitary operator
viz. Deleter and (2) a modified deletion machine described by two
unitary operators viz. Deleter and Transformer. Here it is shown
that the modified deletion machine deletes a qubit with fidelity
$\frac{3}{4}$, which is the maximum limit for deleting an unknown
quantum state. In addition to this we also show that the modified
deletion machine retains the qubit in the first mode with average
fidelity 0.77 (approx.) which is slightly greater than the
fidelity of measurement for two given identical states
\cite{massar1}. We also show that the deletion machine itself is
input state independent, i.e., the information is not hidden in
the deleting machine, and hence we can delete the information completely from the deletion machine.\\
Next, we study the quantum deletion machine with two transformers,
and show that the deletion machine with a single transformer
performs better than the deletion machine with more than two
transformers. We also observe that the fidelity of deletion
depends on the blank state used in the deleter, and so for
different blank states the fidelity is different. Furthermore, we
study the Pati-Braunsein deleter with transformer.\\
This chapter is based on our works "Quantum deletion: Beyond
no-deletion principle"\cite{adhikari2} and "Improving the fidelity
of deletion"\cite{adhikari4}
\section{\emph{Quantum deletion machines}}
In this section, we discuss two types of deletion machines. The
first type of deletion machine is conventional i.e. it just
deletes a qubit while the second type of deletion machine not only
deletes a qubit but also transforms the state after deletion
operation. The newly defined deletion machine i.e. the modified
deletion machine, consists of two parts, the deleter and the transformer.\\
1. Deleter: This is nothing but a unitary transformation U used to
delete one copy from among two given copies of an unknown quantum state.\\
A unitary transformation U which describes a deleter is given below:
\begin{eqnarray}
U |00\rangle_{ab}|A\rangle_{c}\rightarrow
|0\rangle_{a}|\Sigma\rangle_{b}|A_{0}\rangle_{c}+
[|0\rangle_{a}|1\rangle_{b}+|1\rangle_{a}|0\rangle_{b}]|B_{0}\rangle_{c}\\
U |01\rangle_{ab}|A\rangle_{c}\rightarrow
|0\rangle_{a}|\Sigma_{\perp}\rangle_{b}|D_{0}\rangle_{c}+
|1\rangle_{a}|0\rangle_{b}|C_{0}\rangle_{c}\\
U |10\rangle_{ab}|A\rangle_{c}\rightarrow
|1\rangle_{a}|\Sigma\rangle_{b}|D_{0}\rangle_{c}+
|0\rangle_{a}|1\rangle_{b}|C_{0}\rangle_{c}\\
U |11\rangle_{ab}|A\rangle_{c}\rightarrow
|1\rangle_{a}|\Sigma_{\perp}\rangle_{b}|A_{1}\rangle_{c}+
[|0\rangle_{a}|1\rangle_{b}+|1\rangle_{a}|0\rangle_{b}]|B_{1}\rangle_{c}
\end{eqnarray}
where $|A\rangle$ is the initial and
$|A_{i}\rangle,|B_{i}\rangle,|C_{j}\rangle,|D_{j}\rangle$ (i=0,1;j=0) are the final
machine state vectors. $|\Sigma\rangle$ is some standard state and
$|\Sigma_{\perp}\rangle$ denotes a state orthogonal to $|\Sigma\rangle$.\\
We assume
\begin{eqnarray}
\langle A_{0}|B_{0}\rangle=\langle A_{0}|D_{0}\rangle=\langle
A_{1}|D_{0}\rangle=\langle A_{1}|B_{1}\rangle=\langle A_{0}|A_{1}\rangle=\langle
B_{0}|C_{0}\rangle=\langle
B_{0}|D_{0}\rangle=0,\\
\langle A_{0}|B_{1}\rangle=\langle A_{1}|B_{0}\rangle=\langle
B_{0}|B_{1}\rangle=\langle B_{1}|D_{0}\rangle=\langle
C_{0}|A_{1}\rangle=\langle B_{1}|C_{0}\rangle=0,{}\nonumber\\
\langle A|A_{0}\rangle=\langle A|D_{0}\rangle=\langle A|A_{1}\rangle=Y,\langle
A|B_{0}\rangle=\langle A|C_{0}\rangle=\langle A|B_{1}\rangle=0.
\end{eqnarray}
The normalization condition of the transformation (4.1-4.4) gives
\begin{eqnarray}
\langle A_{i}|A_{i}\rangle + 2\langle B_{i}|B_{i}\rangle=1,~~i=0,1
{}\nonumber\\\langle C_{0}|C_{0}\rangle + \langle D_{0}|D_{0}\rangle=1
\end{eqnarray}
The orthogonality condition to be satisfied for the transformation
(4.1-4.4) is
\begin{eqnarray}
\langle A_{0}|C_{0}\rangle=\langle D_{0}|C_{0}\rangle=0
\end{eqnarray}
2. Transformer: It is described by a unitary transformation T. It
is used in the deletion machine to increase the
fidelity of deletion and minimize the distortion of the undeleted qubit.\\
The unitary operator T \cite{gorbachev1} is defined by
\begin{eqnarray}
T=|\psi^{+}\rangle\langle00|+|11\rangle\langle01|+
|\psi^{-}\rangle\langle10|+|00\rangle\langle11|
\end{eqnarray}
where
$|\psi^{\pm}\rangle=(\frac{1}{\sqrt{2}})(|01\rangle\pm|10\rangle)$\\\\
\textbf{A few Definitions: }\\
Let $|\psi\rangle=\alpha|0\rangle+\beta|1\rangle$ with
$\alpha^{2}+|\beta|^{2}=1$ be any unknown quantum state.\\
Without any loss of generality we can take $\alpha$ real and $\beta$ complex.\\
Let $\rho_{a}$ and $\rho_{b}$ be the reduced density operators describing the state of
the undeleted qubit in mode 'a' and the state of the deleted qubit in mode 'b'
respectively and $\rho_{c}$ denote the density operator of the machine state after the
deletion operation.\\
Let $F_{a}=\langle\psi|\rho_{a}|\psi\rangle$,
$F_{b}=\langle\Sigma'|\rho_{b}|\Sigma'\rangle$, where
$|\Sigma'\rangle=(\frac{1}{\sqrt{2}})(|\Sigma\rangle+|\Sigma_{\perp}\rangle)$ denotes
the fidelity of the qubit in the modes a and b, respectively, after deletion operation
and $F_{c}=\langle A|\rho_{c}|A\rangle$ denotes the overlapping between
the initial and final machine state vectors.\\
\textbf{Definition 4.1: \textit{(State dependent deletion
machine)}} A deletion machine is said to be state dependent if
$F_{a}$, $F_{b}$ and $F_{c}$ depend on the input state.\\
\textbf{Definition 4.2: \textit{(Universal deletion machine)}} A
deletion machine is said to be universal if $F_{b}$ and $F_{c}$
are independent of the input state. The machine is optimal if it maximizes $F_{a}$ and $F_{b}$.\\
\textbf{Definition 4.3: \textit{(Ideal deletion machine)}} A
deletion machine is said to be ideal if $F_{a}$, $F_{b}$ and
$F_{c}$ are input state independent and the machine is optimal if it maximizes $F_{a}$
and $F_{b}$.\\
Note: From definition 4.2 and 4.3, we can say that every ideal
deletion machine is universal but converse is not true.
\section{\emph{Conventional deletion machine (deletion machine without
transformer)}} In this section, we show that the conventional
deletion machine (4.1-4.4), i.e., a deletion machine without
transformer, becomes either a universal deletion machine or an
ideal deletion machine in some restricted cases. The conventional
deletion machine is described in figure-4.1.\\
The deletion machine can be shown to be universal or ideal following three steps:\\
\textbf{Step 1:} The reduced density operator in the mode '1' is
given by
\begin{eqnarray}
\rho_{1}&=&Tr_{23}(\rho_{123})=|0\rangle\langle0|[\alpha^{4}(\langle
A_{0}|A_{0}\rangle+\langle
B_{0}|B_{0}\rangle)+\alpha^{2}|\beta|^{2}+|\beta|^{4}\langle
B_{1}|B_{1}\rangle]+{}\nonumber\\&&|1\rangle\langle1|[\alpha^{4}\langle
B_{0}|B_{0}\rangle+\alpha^{2}|\beta|^{2}+|\beta|^{4}(\langle
A_{1}|A_{1}\rangle+\langle B_{1}|B_{1}\rangle)]
\end{eqnarray}
Let us assume
\begin{eqnarray}
\langle A_{0}|A_{0}\rangle=\langle A_{1}|A_{1}\rangle=\langle
D_{0}|D_{0}\rangle=1-2\lambda\\
\langle B_{0}|B_{0}\rangle=\langle B_{1}|B_{1}\rangle=\frac{\langle
C_{0}|C_{0}\rangle}{2}=\lambda
\end{eqnarray}
with $0\leq\lambda\leq\frac{1}{2}$ which follows from the Schwarz inequality.
\begin{eqnarray}
\rho_{1}=|0\rangle\langle0|[\alpha^{4}(1-\lambda)+\alpha^{2}|\beta|^{2}+|\beta|^{4}\lambda]
+|1\rangle\langle1|[\alpha^{4}\lambda+\alpha^{2}|\beta|^{2}+|\beta|^{4}(1-\lambda)]
\end{eqnarray}
The overlapping between the input state $|\psi\rangle$ and the density operator
$\rho_{1}$ is given by
\begin{eqnarray}
F_{1}=\langle\psi|\rho_{1}|\psi\rangle=(1-\lambda)+2\alpha^{2}(1-\alpha^{2})(2\lambda-1)
\end{eqnarray}
Therefore $F_{1}$ depends on $\alpha^{2}$ and the parameter $\lambda$.\\
Now it is interesting to discuss results for two different values
of $\lambda$ in two different cases. In the first case $F_{1}$  is
found to be input
state independent and in the second case it depends on the input state.\\
\textbf{Case I:} If $\lambda\rightarrow\frac{1}{2}$ then
$F_{1}\rightarrow\frac{1}{2}$. Although we are able to make $F_{1}$ input state
independent, the performance of the deletion machine is not very satisfactory since it
fails to retain the qubit in the first mode faithfully after the deletion operation.\\
\textbf{Case II:} If $\lambda\rightarrow0$, then $F_{1}\rightarrow
1-2\alpha^{2}(1-\alpha^{2})$, which is input state dependent and
therefore we have to calculate the average value.\\
The average fidelity is given by
\begin{eqnarray}
\overline{F_{1}}=\int_{0}^{1}F_{1}(\alpha^{2})d\alpha^{2}\rightarrow\frac{2}{3}
\end{eqnarray}
This value is equal to the fidelity of measurement for a given single unknown state.
Although $F_{1}$ is input state dependent the average value $\overline{F_{1}}$ exceeds
the fidelity discussed in case I.\\
\textbf{Step 2:} The reduced density operator in the mode 2 is given by
\begin{eqnarray}
\rho_{2}&=&Tr_{13}(\rho_{123})=|0\rangle\langle0|[\alpha^{4}\langle
B_{0}|B_{0}\rangle+\alpha^{2}|\beta|^{2}\langle C_{0}|C_{0}\rangle+|\beta|^{4}\langle
B_{1}|B_{1}\rangle]+{}\nonumber\\&&|1\rangle\langle1|[\alpha^{4}\langle
B_{0}|B_{0}\rangle+\alpha^{2}|\beta|^{2}\langle C_{0}|C_{0}\rangle+|\beta|^{4}\langle
B_{1}|B_{1}\rangle]+|\Sigma\rangle\langle\Sigma|[\alpha^{4}\langle
A_{0}|A_{0}\rangle+{}\nonumber\\&&\alpha^{2}|\beta|^{2}\langle
D_{0}|D_{0}\rangle]+|\Sigma_{\perp}\rangle\langle\Sigma_{\perp}|[|\beta|^{4}\langle
A_{1}|A_{1}\rangle+\alpha^{2}|\beta|^{2}\langle D_{0}|D_{0}\rangle]
\end{eqnarray}
Using equations (4.11) and (4.12), equation (4.16) reduces to
\begin{eqnarray}
\rho_{2}&=&|0\rangle\langle0|[\alpha^{4}\lambda+2\alpha^{2}|\beta|^{2}\lambda+|\beta|^{4}\lambda]+
|1\rangle\langle1|[\alpha^{4}\lambda+2\alpha^{2}|\beta|^{2}\lambda+|\beta|^{4}\lambda]
+{}\nonumber\\&&|\Sigma\rangle\langle\Sigma|[\alpha^{2}(1-2\lambda)]
+|\Sigma_{\perp}\rangle\langle\Sigma_{\perp}|[|\beta|^{2}(1-2\lambda)]
\end{eqnarray}
The fidelity of deletion is defined by
\begin{eqnarray}
F_{2}=\langle\Sigma'|\rho_{2}|\Sigma'\rangle=(\frac{1}{2})[(1-2\lambda)+(K_{1}+K_{2})\lambda]
\end{eqnarray}
where
\begin{eqnarray}
K_{1}=\langle\Sigma|0\rangle^{2}+|\langle\Sigma|1\rangle|^{2}+\langle\Sigma|0\rangle\langle0|\Sigma_{\perp}\rangle
+\langle\Sigma|1\rangle\langle1|\Sigma_{\perp}\rangle\\
K_{2}=|\langle\Sigma_{\perp}|0\rangle|^{2}+\langle\Sigma_{\perp}|1\rangle^{2}+\langle\Sigma|0\rangle\langle0|\Sigma_{\perp}\rangle
+\langle\Sigma|1\rangle\langle1|\Sigma_{\perp}\rangle
\end{eqnarray}
The standard state $|\Sigma\rangle$ can be written as
$|\Sigma\rangle=m_{1}|0\rangle+m_{2}|1\rangle$, where without any loss of generality
we can take $m_{1}$ real and $m_{2}$ complex satisfying the relation
\begin{eqnarray}
m_{1}^{2}+|m_{2}|^{2}=1.
\end{eqnarray}
A state orthogonal to $|\Sigma\rangle$ is given by
\begin{eqnarray}
|\Sigma_{\perp}\rangle=-m_{2}^{*}|0\rangle+m_{1}|1\rangle
\end{eqnarray}
Therefore,
\begin{eqnarray}
\langle\Sigma|0\rangle=\langle\Sigma_{\perp}|1\rangle=m_{1}~~,
\langle\Sigma|1\rangle=m_{2}^{*},\langle\Sigma_{\perp}|0\rangle=-m_{2}
\end{eqnarray}
Using equations (4.19), (4.20), (4.21) and (4.23), we get
\begin{eqnarray}
K_{1}+ K_{2}=2
\end{eqnarray}
Putting the value of $(K_{1}+ K_{2})$ in equation (4.18), we get
\begin{eqnarray}
F_{2}=\frac{1}{2}
\end{eqnarray}
Here we note that the fidelity of deletion neither depends on
input state nor on machine state. The value of the fidelity of
deletion is calculated to be $\frac{1}{2}$, which is not a very
satisfactory result at all. The same value of the fidelity is also
obtained by Qiu \cite{qiu1} for his deletion machine and it
emphasizes the difficulty of improving its fidelity. We also find
here that the fidelity of deletion for our prescribed deletion
machine cannot be improved further if the machine is kept in its
present form but the fidelity may be improved if we define
a deletion machine in another way, which we discuss in details in the next section.\\
Step 3: The reduced density operator in the mode '3' is given by
\begin{eqnarray}
\rho_{3}=Tr_{12}(\rho_{123})&=&\alpha^{4}(|A_{0}\rangle\langle
A_{0}|+m_{2}^{*}|A_{0}\rangle\langle B_{0}|+m_{2}|B_{0}\rangle\langle
A_{0}|+2|B_{0}\rangle\langle
B_{0}|)+{}\nonumber\\&&\alpha^{3}\beta^{*}(m_{2}|A_{0}\rangle\langle
C_{0}|+2m_{1}|B_{0}\rangle\langle D_{0}|+2|B_{0}\rangle\langle
C_{0}|)+\alpha^{3}\beta(m_{2}^{*}|C_{0}\rangle\langle
A_{0}|{}\nonumber\\&&+2m_{1}|D_{0}\rangle\langle B_{0}|+2|C_{0}\rangle\langle
B_{0}|)+\alpha^{2}|\beta|^{2}[m_{2} |A_{0}\rangle\langle
B_{1}|+m_{2}^{*}|B_{1}\rangle\langle
A_{0}|-{}\nonumber\\&&m_{2}^{*}|A_{1}\rangle\langle B_{0}|-m_{2} |B_{0}\rangle\langle
A_{1}|+2(|B_{0}\rangle\langle B_{1}|+ |B_{1}\rangle\langle
B_{0}|)+2(|C_{0}\rangle\langle C_{0}|+{}\nonumber\\&&|D_{0}\rangle\langle
D_{0}|)+2m_{1}(|C_{0}\rangle\langle D_{0}|+|D_{0}\rangle\langle
C_{0}|)]+\alpha|\beta|^{2}\beta[2m_{1}(|B_{1}\rangle\langle
D_{0}|+{}\nonumber\\&&|D_{0}\rangle\langle B_{1}|)-m_{2}^{*}|A_{1}\rangle\langle
C_{0}|-m_{2}|C_{0}\rangle\langle A_{1}|+2(|B_{1}\rangle\langle
C_{0}|+|C_{0}\rangle\langle B_{1}|)]+{}\nonumber\\&&|\beta|^{4}(|A_{1}\rangle\langle
A_{1}|-m_{2}^{*}|A_{1}\rangle\langle B_{1}|-m_{2}|B_{1}\rangle\langle
A_{1}|+2|B_{1}\rangle\langle B_{1}|)
\end{eqnarray}
Using equation (4.6), (4.26) and the relation
$\alpha^{2}+|\beta|^{2}=1$, we get
\begin{eqnarray}
\langle A|\rho_{3}|A\rangle=Y^{2}
\end{eqnarray}
which is independent of $\alpha^{2}$. This means that the information is not hidden in
the deletion machine and hence it deletes the state completely because we cannot
retrieve the state by applying a unitary transformation from the deletion machine.\\
Note: (1) If $\lambda\rightarrow\frac{1}{2}$, then $F_{1}$,
$F_{2}$ and $\langle A|\rho_{3}|A\rangle$ are independent of
$\alpha^{2}$. Also $F_{1}\rightarrow\frac{1}{2}$,
$F_{2}=\frac{1}{2}$. Therefore, for
$\lambda\rightarrow\frac{1}{2}$, the conventional deletion machine
becomes ideal deletion machine in the limiting sense but the machine is not optimal.\\
(2) If $\lambda\neq\frac{1}{2}$ then also $F_{2}$ and $\langle
A|\rho_{3}|A\rangle$ are independent of $\alpha^{2}$ because they
do not depend on $\lambda$ so the conventional deletion machine
becomes universal deletion machine for all values of
$\lambda$ $(0\leq\lambda<\frac{1}{2})$.\\
Now case-2  is interesting in the sense that if
$\lambda\rightarrow0$, then the average value of $F_{1}$ tends to
the maximum limit $\frac{2}{3}$ that is also obtained by state
dependent Pati-Braunstein deletion machine. Moreover, the fidelity
of a qubit in mode 1, i.e., $F_{1}$ is found to be greater than
the fidelity of deletion $F_{2}$.
\section{\emph{Modified deletion machine (deletion machine with single
transformer)}} In the preceding section 4.3, we discussed the
deletion machine without considering a vital part of it. In this
section we take into account that important part of the deletion
machine without which we cannot improve the fidelity of deletion.
In addition to a unitary transformation U (named the deleter) that
deletes a qubit, a unitary operator T (named the transformer) must
be used in the deletion machine. The role of the transformer is to
transform the resultant state immediately obtained after the
deletion operation, thereby improving the fidelity of deletion of
the qubit in the second mode and increasing the fidelity of the
retained qubit in the first mode. The modified deletion machine is
described in figure-4.2.\\
In the first chamber the deletion process is completed. Thereafter
the deleted state described by the density operator $\rho_{123}$
enters into the second chamber where another unitary operator
called transformer transforms it into the state $\rho'_{123}$,
\begin{eqnarray}
\rho'_{123}=(I\otimes T)\rho_{123}(I\otimes T)^{\dagger}
\end{eqnarray}
The reduced density operator describing the state $\rho'_{1}$ is given by
\begin{eqnarray}
\rho'_{1}&=&Tr_{23}(\rho'_{123})=|0\rangle\langle0|(\frac{1}{2})(\alpha^{4}[m_{1}^{2}(1-2\lambda)
+\lambda]+\alpha^{2}|\beta|^{2}\{[3|m_{2}|^{2}-m_{1}(m_{2}+m_{2}^{*}){}\nonumber\\&&+m_{1}^{2}](1-2\lambda)
+2\lambda\}+|\beta|^{4}[(|m_{2}|^{2}+2m_{1}^{2})(1-2\lambda)+\lambda])+|0\rangle\langle1|
(\frac{1}{\sqrt{2}})\times{}\nonumber\\&&(\alpha^{4}[m_{1}m_{2}^{*}(1-2\lambda)+\lambda]
+\alpha^{2}|\beta|^{2}\{2\lambda+[m_{1}^{2}-
m_{2}^{2}-m_{1}(m_{2}+m_{2}^{*})](1-2\lambda)\}+{}\nonumber\\&&|\beta|^{4}[\lambda+m_{1}m_{2}(1-2\lambda)])+
|1\rangle\langle0|(\frac{1}{\sqrt{2}})
(\alpha^{4}[m_{1}m_{2}(1-2\lambda)+\lambda]+\alpha^{2}|\beta|^{2}\{2\lambda+{}\nonumber\\&&[m_{1}^{2}-
(m_{2}^{*})^{2}-m_{1}(m_{2}+m_{2}^{*})](1-2\lambda)\}+|\beta|^{4}[\lambda+
m_{1}m_{2}^{*}(1-2\lambda)])
+{}\nonumber\\&&|1\rangle\langle1|(\frac{1}{2})(\alpha^{4}[(m_{1}^{2}+2|m_{2}|^{2})
(1-2\lambda)+3\lambda]+\alpha^{2}|\beta|^{2}\{[|m_{2}|^{2}+m_{1}(m_{2}+m_{2}^{*})+{}\nonumber\\&&3m_{1}^{2}](1-2\lambda)+6\lambda\}
+|\beta|^{4}[|m_{2}|^{2}(1-2\lambda)+3\lambda]).
\end{eqnarray}
The fidelity of the qubit in mode 1 is given by
\begin{eqnarray}
F_{3}=\langle\psi|\rho'_{1}|\psi\rangle\rightarrow\frac{3}{4}-\frac{\alpha^{2}}{2}+
\frac{\alpha(\beta+\beta^{*})}{2\sqrt{2}}~~~~~for~~\lambda\rightarrow\frac{1}{2}
\end{eqnarray}
If $\beta$ is real, then the average fidelity of this mode is
\begin{eqnarray}
\overline{F_{3}}=\int_{0}^{1}F_{3}(\alpha^{2})d\alpha^{2}\rightarrow\frac{1}{2}+\frac{\pi}{8\sqrt{2}}
=0.77~~(approx.)
\end{eqnarray}
The state described by the reduced density operator $\rho'_{2}$ is
given by
\begin{eqnarray}
\rho'_{2}&=&Tr_{13}(\rho'_{123})=
|0\rangle\langle0|(\frac{1}{2})(\alpha^{4}[m_{1}^{2}(1-2\lambda)
+\lambda]+\alpha^{2}|\beta|^{2}\{[3|m_{2}|^{2}+m_{1}(m_{2}+m_{2}^{*}){}\nonumber\\&&+m_{1}^{2}](1-2\lambda)
+2\lambda\}+|\beta|^{4}[(|m_{2}|^{2}+2m_{1}^{2})(1-2\lambda)+\lambda])+|0\rangle\langle1|
(\frac{1}{\sqrt{2}})\times{}\nonumber\\&&(\alpha^{4}[m_{1}m_{2}^{*}(1-2\lambda)-\lambda]
-\alpha^{2}|\beta|^{2}\{2\lambda+[m_{1}^{2}+
m_{2}^{2}+m_{1}(m_{2}^{*}-m_{2})](1-2\lambda)\}{}\nonumber\\&&-|\beta|^{4}[\lambda+m_{1}m_{2}(1-2\lambda)])+
|1\rangle\langle0|(\frac{1}{\sqrt{2}})
(\alpha^{4}[m_{1}m_{2}(1-2\lambda)-\lambda]-\alpha^{2}|\beta|^{2}\{2\lambda+{}\nonumber\\&&[m_{1}^{2}+
(m_{2}^{*})^{2}+m_{1}(m_{2}-m_{2}^{*})](1-2\lambda)\}-|\beta|^{4}[\lambda+
m_{1}m_{2}^{*}(1-2\lambda)])
+{}\nonumber\\&&|1\rangle\langle1|(\frac{1}{2})(\alpha^{4}[(m_{1}^{2}+2|m_{2}|^{2})
(1-2\lambda)+3\lambda]+\alpha^{2}|\beta|^{2}\{[|m_{2}|^{2}-m_{1}(m_{2}+m_{2}^{*})+{}\nonumber\\&&3m_{1}^{2}](1-2\lambda)+6\lambda\}
+|\beta|^{4}[|m_{2}|^{2}(1-2\lambda)+3\lambda]).
\end{eqnarray}
 The fidelity of the qubit in mode 2 is given by
\begin{eqnarray}
F_{4}=\langle\Sigma'|\rho'_{2}|\Sigma'\rangle&=&\frac{1}{2}[R_{1}(m_{1}-m_{2})(m_{1}-m_{2}^{*})
+R_{2}(m_{1}+m_{2})(m_{1}+m_{2}^{*})+{}\nonumber\\&&R_{3}(m_{1}-m_{2})(m_{1}+m_{2})+
R_{4}(m_{1}-m_{2}^{*})(m_{1}+m_{2}^{*})],
\end{eqnarray}
where
\begin{eqnarray}
R_{1}&=&(\frac{1}{2})\{\alpha^{4}[m_{1}^{2}(1-2\lambda)+\lambda]+\alpha^{2}|\beta|^{2}
\{[3|m_{2}|^{2}-m_{1}(m_{2}+m_{2}^{*})+m_{1}^{2}](1-2\lambda){}\nonumber\\&&+2\lambda\}+
|\beta|^{4}[(|m_{2}|^{2}+2m_{1}^{2})(1-2\lambda)+\lambda]\}\\
R_{2}&=&(\frac{1}{2})\{\alpha^{4}[m_{1}^{2}+2|m_{2}|^{2}(1-2\lambda)+3\lambda]+
\alpha^{2}|\beta|^{2}\{[|m_{2}|^{2}+m_{1}(m_{2}+m_{2}^{*}){}\nonumber\\&&+3m_{1}^{2}](1-2\lambda)+6\lambda\}+
|\beta|^{4}[|m_{2}|^{2}(1-2\lambda)+3\lambda]\}\\
R_{3}&=&(\frac{1}{\sqrt{2}})\{\alpha^{4}[m_{1}m_{2}^{*}(1-2\lambda)-\lambda]-\alpha^{2}|\beta|^{2}
\{[m_{1}^{2}+m_{2}^{2}+m_{1}(m_{2}^{*}-m_{2})]\times{}\nonumber\\&&(1-2\lambda)+2\lambda\}-
|\beta|^{4}[m_{1}m_{2}(1-2\lambda)+\lambda]\}\\
R_{4}&=&(\frac{1}{\sqrt{2}})\{\alpha^{4}[m_{1}m_{2}(1-2\lambda)-\lambda]-\alpha^{2}|\beta|^{2}
\{[m_{1}^{2}+(m_{2}^{*})^{2}+m_{1}(m_{2}-m_{2}^{*})]\times{}\nonumber\\&&(1-2\lambda)+2\lambda\}-
|\beta|^{4}[m_{1}m_{2}^{*}(1-2\lambda)+\lambda]\}
\end{eqnarray}
If $m_{1}=m_{2}=\frac{1}{\sqrt{2}}$, then the expression for
$F_{4}$ given in equation (4.33) reduces to
\begin{eqnarray}
F_{4}=R_{2}\rightarrow\frac{3}{4}=0.75 ~~for~~\lambda\rightarrow\frac{1}{2}
\end{eqnarray}
Since the machine states are invariant under the unitary transformation T, so $\langle
A|\rho'_{3}|A\rangle=\langle A|\rho_{3}|A\rangle=Y^{2}$, which is independent of
$\alpha^{2}$.\\
Hence the deletion machine with transformer becomes a universal
deletion machine when the machine parameter
$\lambda\rightarrow\frac{1}{2}$ and
$m_{1}=m_{2}=\frac{1}{\sqrt{2}}$. This universal deletion machine
deletes a qubit with fidelity $\frac{3}{4}$ (in the limiting
sense), which is the maximum limit for deleting an unknown qubit.
In addition, the average fidelity of the qubit in the first mode
is found to be 0.77, which is greater than the average fidelity
$(\overline{F_{a}} = 0.66)$ obtained by Pati-Braunstein deletion
machine.\\
Furthermore, if the machine parameter $\lambda$ tends to
$\frac{1}{2}$ then for all real blank state parameters $m_{1}$ and
$m_{2}$ the limiting fidelity of deletion $F_{4}$ given in
equation (4.33) goes towards $F_{4}'$ i.e.
\begin{eqnarray}
F_{4}\rightarrow F_{4}'=\frac{1}{2}[1+ m_{1}m_{2}-\frac{m_{1}^{2}-
m_{2}^{2}}{\sqrt{2}}],~~as~~
\lambda\rightarrow\frac{1}{2}~~(\textrm{for real}~~m_{1}~
\textrm{and}~~m_{2})
\end{eqnarray}
Equation (4.39) shows that the fidelity of deletion remains the
same for all input states $\alpha^{2}$ and it depends only on the
blank state. Since the limiting fidelity of deletion of the
deletion machine with one transformer depends on the parameters
$m_{1}$ and $m_{2}$ so the variation of the limiting fidelity of
deletion with $m_{1}$ and $m_{2}$ is studied and given in the table 4.1.\\
\textsl{Table-4.1: Limiting fidelities for deletion machine with one transformer}\\
\begin{tabular}{| c| c| c| c|}
\hline
  $m_{1}^{2}$ & $m_{2}^{2}$ & $m_{1}^{2}-m_{2}^{2}$ & $Limiting fidelity (F_{4}')
  =\frac{1}{2}[1\pm\frac{\sqrt{1-(m_{1}^{2}-m_{2}^{2}})}{2}-\frac{m_{1}^{2}-m_{2}^{2}}{\sqrt{2}}]$\\
  &  &  & $\textrm{according as}~~ m_{1}m_{2}>0~~\textrm{or}~~
  m_{1}m_{2}<0$ \\ & & & (up to two significant figures)\\
  \hline
  0.0 & 1.0 & -1.0 & 0.85 \\
  \hline
  0.1 & 0.9 & -0.8 & 0.93~~\textrm{or}~~0.63 \\
  \hline
  0.2 & 0.8 & -0.6 & 0.91~~\textrm{or}~~0.51 \\
  \hline
  0.3 & 0.7 & -0.4 & 0.87~~\textrm{or}~~0.41\\
  \hline
  0.4 & 0.6 & -0.2 & 0.81~~\textrm{or}~~0.32 \\
  \hline
  0.5 & 0.5 & 0.0 & 0.75 \\
  \hline
  0.6 & 0.4 & 0.2 & 0.67~~\textrm{or}~~0.18 \\
  \hline
  0.7 & 0.3 & 0.4 & 0.58~~\textrm{or}~~0.12\\
  \hline
  0.8 & 0.2 & 0.6 & 0.48~~\textrm{or}~~0.08 \\
  \hline
  0.9 & 0.1 & 0.8 & 0.36~~\textrm{or}~~0.06 \\
  \hline
  1.0 & 0.0 & 1.0 & 0.14 \\ \hline
\end{tabular}\\\\
\textbf{\textsl{Illustration of the table 4.1:\\
If the blank state is of the form
$|\Sigma\rangle=\sqrt{0.1}|0\rangle+\sqrt{0.9}|1\rangle$, then the
deletion machine with one transformer deletes a qubit with
fidelity 0.93 and if the blank state either take the form
$|\Sigma\rangle=-\sqrt{0.1}|0\rangle+\sqrt{0.9}|1\rangle$ or
$|\Sigma\rangle=\sqrt{0.1}|0\rangle-\sqrt{0.9}|1\rangle$, then the
fidelity of deletion is found to be 0.63.}}\\\\
Therefore, we can observe from the above table that if the product
of the parameter of the blank state is negative (i.e. when either
$m_{1}$ or $m_{2}$ is negative), then the deletion machine deletes
the state with lower fidelity of deletion but if the product of
the parameter of the blank state is positive (i.e. when both
$m_{1}$ and $m_{2}$ are negative or positive), then the deletion
machine performs well in the sense of high fidelity of deletion.
In this work we have discussed the quantum deletion machine with
one transformer for various values of the parameters $m_{1}$ and
$m_{2}$ and find that the quantum deletion machine really works
well for some blank states and, with the help of those blank
states, quantum deletion machine deletes a quantum state with
fidelity of deletion higher than 0.75.
\section{\emph{Quantum deletion machine with two transformers}}
In this section, we study the quantum deletion machine (4.1-4.4)
with two transformers. Since the introduction of the transformer
increases the fidelity of deletion so one may expect that the
application of a transformer more than one time increases the
fidelity of deletion further and therefore there may exist a
threshold number of the transformers (i.e. maximum number of
transformers) whose application on the deleted state increases the
fidelity of deletion to its optimal value. But, we
will show that this is not necessarily true.\\
Let $|\psi\rangle=\alpha|0\rangle+\beta|1\rangle$ with $\alpha^{2}+|\beta|^{2}=1$ be
any unknown quantum state, where $\alpha$ is real and $\beta$ is complex.\\
The modified deletion machine with two transformers deletes one of
the copies of an input state $|\psi\rangle|\psi\rangle$ and then,
after transforming the deleted qubit, the final output state of
the deletion machine is described by the density operator
$\rho_{12c}^{out}=(T)^{2}|\chi_{d}^{out}\rangle_{12c}\langle\chi_{d}^{out}|(T^{\dagger})^{2}$
where $|\chi_{d}^{out}\rangle$ represents a state after passing
through the deleter (4.1-4.4). Since we are interested to see the
performance of the deletion machine with two transformers in the
sense of how well it deletes a qubit, we only consider the state
of the qubit in mode 2. Therefore, the reduced density operator
describes the state of the qubit in mode 2 is given by
$\rho_{2}^{out}$. But as, $\lambda\rightarrow\frac{1}{2}$,
$\rho_{2}^{out}\rightarrow\rho_{2}'^{out}$ where
\begin{eqnarray}
\rho_{2}'^{out}=\frac{5}{8}|0\rangle\langle0|+\frac{1}{4}(\frac{1}{\sqrt{2}}-1)|0\rangle\langle1|
+\frac{1}{4}(\frac{1}{\sqrt{2}}-1)|1\rangle\langle0|+\frac{3}{8}|1\rangle\langle1|
\end{eqnarray}
Equation (4.40) shows that the state described by the density
operator $\rho_{2}'^{out}$ is input state independent i.e.
whatever be the input quantum state, after passing through the
deleter and two transformers, the resulting output state remains the same.\\
The limiting fidelity of deletion is given by
\begin{eqnarray}
F=\langle\Sigma'|\rho_{2}'^{out}|\Sigma'\rangle, \textrm{where
}|\Sigma'\rangle=(\frac{1}{\sqrt{2}})(|\Sigma\rangle+|\Sigma_{\perp}\rangle)
\end{eqnarray}
Hence,
\begin{eqnarray}
F&=&\frac{1}{2}[\frac{5}{8}(1+m_{1}m_{2}^{*}-m_{1}m_{2})+\frac{1}{4}(\frac{1}{\sqrt{2}}-1)
(2m_{1}^{2}-(m_{2}^{*})^{2}-m_{2}^{2})+{}\nonumber\\&&\frac{3}{8}(1+m_{1}m_{2}^{*}+m_{1}m_{2})]
\end{eqnarray}
Equation (4.42) shows that the fidelity of deletion varies as
$m_{1}$ and $m_{2}$, i.e., the limiting fidelity depends on the
blank state used in the deleter but not on the arbitrary input
state.\\
In particular, if we assume $m_{1}$ and $m_{2}$ to be real, then the variation of
fidelity with the amplitudes of the blank state $m_{1}$ and $m_{2}$ is given
in the table-4.2:\\
\textsl{Table-4.2: Limiting fidelities for deletion machine with two transformers}\\
\begin{tabular}{| c| c| c| c|}
\hline
  $m_{1}^{2}$ & $m_{2}^{2}$ & $m_{1}^{2}-m_{2}^{2}$ & $Limiting fidelity (F)
  =\frac{1}{2}[1\mp\frac{\sqrt{1-(m_{1}^{2}-m_{2}^{2})^{2}}}{4}
  +(\frac{1}{2})(\frac{1}{\sqrt{2}}-1)(m_{1}^{2}-m_{2}^{2})]$\\
  &  &  & $\textrm{according as}~~ m_{1}m_{2}>0~~\textrm{or}~~
  m_{1}m_{2}<0$ \\ & & & (up to two significant figures)\\
  \hline
  0.0 & 1.0 & -1.0 & 0.57 \\
  \hline
  0.1 & 0.9 & -0.8 & 0.48~~\textrm{or}~~0.63 \\
  \hline
  0.2 & 0.8 & -0.6 & 0.44~~\textrm{or}~~0.64 \\
  \hline
  0.3 & 0.7 & -0.4 & 0.41~~\textrm{or}~~0.64\\
  \hline
  0.4 & 0.6 & -0.2 & 0.39~~\textrm{or}~~0.63 \\
  \hline
  0.5 & 0.5 & 0.0 & 0.37 \\
  \hline
  0.6 & 0.4 & 0.2 & 0.36~~\textrm{or}~~0.60 \\
  \hline
  0.7 & 0.3 & 0.4 & 0.35~~\textrm{or}~~0.58\\
  \hline
  0.8 & 0.2 & 0.6 & 0.35~~\textrm{or}~~0.55 \\
  \hline
  0.9 & 0.1 & 0.8 & 0.36~~\textrm{or}~~0.51 \\
  \hline
  1.0 & 0.0 & 1.0 & 0.42 \\ \hline
\end{tabular}\\\\
\textbf{\textsl{Illustration of the table 4.2:\\
If the blank state is of the form
$|\Sigma\rangle=\sqrt{0.1}|0\rangle+\sqrt{0.9}|1\rangle$, then the
deletion machine with two transformers delete a qubit with
fidelity 0.48 and if the blank state either takes the form
$|\Sigma\rangle=-\sqrt{0.1}|0\rangle+\sqrt{0.9}|1\rangle$ or
$|\Sigma\rangle=\sqrt{0.1}|0\rangle-\sqrt{0.9}|1\rangle$, then the
fidelity of deletion is found to be 0.63.}}\\\\
On the contrary, we observe here that if the product of the
parameter of the blank state is negative (i.e. when either $m_{1}$
or $m_{2}$ is negative), then the deletion machine with two
transformers delete the state with fidelity of deletion higher
than the case when the product of the parameter of the blank state
is positive (i.e. when both $m_{1}$ and $m_{2}$ are negative or
positive). Also we note that the deletion machine with two
transformers deletes a state with fidelity of deletion 0.37 when
we use the blank state with parameter $m_{1}=m_{2}=0.5$. In
addition to this, If we compare the quantum deletion machine with
two transformer with the quantum deletion machine with a single
transformer, then we find that the deletion machine with a single
transformer works better when the product of the amplitudes of the
blank state $m_{1}$ and $m_{2}$ is positive, while the deletion
machine with two transformers works better when the product of the
parameters $m_{1}$ and $m_{2}$ is negative.
\section{\emph{PB deleting machine with transformer}}
In this section, we study the conditional PB deleting machine with
the addition of a unitary operator called transformer T. We will
show in this section that the addition of a transformer in the
quantum deletion machine not only increases the fidelity of
deletion but also makes the fidelity of deletion input state independent.\\
The conditional PB deleting transformation (PB deleter) is defined
by equations (1.160-1.163) in chapter-1.\\
Now if we consider the deletion machine (PB deleter + Transformer)
to delete a copy from two copies of the input state
$|\psi\rangle=\alpha|0\rangle+\beta|1\rangle$ with
$\alpha^{2}+|\beta|^{2}=1$ then we find that the final output
state from the deletion machine is given by
$\rho_{12c}^{out}=T|\psi_{d}^{out}\rangle_{12c}\langle\psi_{d}^{out}|T^{\dagger}$,
where $|\psi_{d}^{out}\rangle_{12c}$ denotes the state after
passing through the PB deleter.\\
The reduced density operator in mode 2 is given by
\begin{eqnarray}
\rho_{2}&=&Tr_{1c}(T|\psi_{d}^{out}\rangle_{12c}\langle\psi_{d}^{out}|T^{\dagger})=
|0\rangle\langle0|(\frac{\alpha^{4}m_{1}^{2}}{2}+\frac{\alpha^{2}|\beta|^{2}}{2}
+\frac{|\beta|^{4}m_{1}^{2}}{2}+|\beta|^{4}m_{2}^{2}){}\nonumber\\&&+|0\rangle\langle1|
(\frac{\alpha^{4}m_{1}m_{2}^{*}}{\sqrt{2}}-\frac{\alpha^{2}|\beta|^{2}}{\sqrt{2}}
+\frac{|\beta|^{4}m_{1}m_{2}}{\sqrt{2}})+|1\rangle\langle0|
(\frac{\alpha^{4}m_{1}m_{2}}{\sqrt{2}}-\frac{\alpha^{2}|\beta|^{2}}{\sqrt{2}}
{}\nonumber\\&&+\frac{|\beta|^{4}m_{1}m_{2}^{*}}{\sqrt{2}})+|1\rangle\langle1|
(\frac{\alpha^{4}m_{1}^{2}}{2}+\frac{3\alpha^{2}|\beta|^{2}}{2}+\frac{|\beta|^{4}m_{1}^{2}}{2}
+\alpha^{4}|m_{2}|^{2})
\end{eqnarray}
Now to see how well our deleting system deletes a qubit, we have to calculate the
fidelity of deletion defined by
$F_{2}=\langle\Sigma|\rho_{2}|\Sigma\rangle$.\\
If we assume $m_{1}$ and $m_{2}$ to be real, then
\begin{eqnarray}
F_{2}&=&m_{1}^{2}[\frac{m_{1}^{2}}{2}+\alpha^{2}|\beta|^{2}
(\frac{1-2m_{1}^{2}}{2})+|\beta|^{4}m_{2}^{2}]+m_{1}m_{2}[\frac{m_{1}m_{2}}{\sqrt{2}}
-\alpha^{2}|\beta|^{2}(\frac{1+2m_{1}m_{2}}{\sqrt{2}})
]+{}\nonumber\\&&m_{1}m_{2}[\frac{m_{1}m_{2}}{\sqrt{2}}
-\alpha^{2}|\beta|^{2}(\frac{1+2m_{1}m_{2}}{\sqrt{2}}]+m_{2}^{2}[\frac{m_{1}^{2}}{2}+\alpha^{2}|\beta|^{2}
(\frac{3-2m_{1}^{2}}{2})+\alpha^{4}m_{2}^{2}]{}\nonumber\\
\end{eqnarray}
If $m_{1}=\frac{1}{\sqrt{2}}$ and $m_{2}=-\frac{1}{\sqrt{2}}$, then
$F_{2}=\frac{1}{2}+\frac{1}{2\sqrt{2}}=0.85$
(approximately).\\
Equation (4.44) shows that there exists a blank state for which
the fidelity of deletion is input state independent and also its
value approaches the optimal cloning fidelity, which we expect
from our universal deletion machine. Therefore, the advantage of
using the transformer in the quantum deletion machine with PB
deleter is that the machine deletes a qubit with fidelity 0.85,
which remains the same for all input states. In addition to this,
we can observe that the average fidelity of deletion (0.85) for a
deletion machine with a PB deleter and transformer is greater than
the average fidelity (0.83) for simply a PB deleter.\\\\

\large \baselineskip .85cm

\chapter{Concatenation of quantum cloning and deletion machines}
\setcounter{page}{118} \markright{\it CHAPTER~\ref{chap5}.
Concatenation of quantum cloning and deletion machines }
\label{chap5}%
The fundamental concept in social science is Power, in the same
sense in which Energy is the fundamental concept in PHYSICS -
Bertrand Russell \\\\
The idea that time may vary from place to place is a difficult
one, but it is the idea Einstein used, and it is correct - believe
it or not - Richard Feynman\\\\
It is wrong to think that the task of physics is to find out how
nature is. Physics concerns what we can say about nature - Niels
Bohr
\section{\emph{Prelude}}
The emerging field of quantum computation and Information
technology investigated the possibility of exploiting greater
information processing ability using qubits \cite{galvao1} .
Therefore, manipulation and extraction of quantum information are
important tasks in building quantum computer. The copying and
deleting of information in a classical computer are inevitable
operations whereas similar operations cannot be realized perfectly
in quantum computers. Linear evolution makes these quantum
operations impossible on arbitrary superpositions of quantum
states. Quantum cloning and deleting can both be regarded as
devices which perform a unitary operation to distill classical
information from quantum information. This doesn't mean that
quantum deleting is just the reverse of quantum cloning. The
difference between quantum cloning and quantum deleting can be
explained by the following two points: (i) A quantum cloning
process can be thought as a swapping operation between the blank
qubit and the cloning machine state, but a swapping operation
between the cloning state and the deleting machine state can not
be thought as a successful deleting process. (ii) The fidelity of
each mode is different for the deleting operations whereas it has
the same value in the cloning process in symmetric case.\\
The purpose of this chapter is to find the effect on an arbitrary
qubit as a result of the concatenation of quantum cloning and
deleting machines. It is a known fact that quantum deleting
machine can be applied in a situation when scarcity of memory in a
quantum computer occurs. Naturally, a situation may arise in which
an arbitrary quantum state is needed to be copied by the imperfect
quantum cloner. After performing the given task with cloned qubit
if one finds that scarcity of memory occurs then in this situation
one has to delete one copy among two cloned copies to store new
information in a quantum computer. Consequently in this chapter we
will study the concatenation of two quantum operations viz.
unitary quantum cloning and deleting transformations. At first, we
construct a state dependent quantum deleting machine and show that
the minimum average distortion of the input qubit and maximum
fidelity of deletion approach to $\frac{1}{3}$ and $\frac{5}{6}$
respectively. Thereafter, we have studied the concatenation of
cloning and deletion machines. This chapter is based on our work
entitled "Deletion of Imperfect cloned copies"\cite{adhikari1}.
\section{\emph{State dependent quantum deletion machine}}
A state dependent quantum deleting transformation can be defined by
\begin{eqnarray}
U|0\rangle|0\rangle|Q\rangle\rightarrow |0\rangle|\Sigma\rangle|A_{0}\rangle\\
U|1\rangle|1\rangle|Q\rangle\rightarrow |1\rangle|\Sigma\rangle|A_{1}\rangle\\
U|0\rangle|1\rangle|Q\rangle\rightarrow(a_{0}|0\rangle|1\rangle+b_{0}|1\rangle|0\rangle)|Q\rangle\\
U|1\rangle|0\rangle|Q\rangle\rightarrow(a_{1}|0\rangle|1\rangle+b_{1}|1\rangle|0\rangle)|Q\rangle
\end{eqnarray}
where $|Q\rangle,|A_{0}\rangle,|A_{1}\rangle$ and $|\Sigma\rangle$
have their usual meanings and $a_{i}, b_{i}$ (i =0,1) are the complex numbers.\\
Due to the unitarity of the transformation (5.1-5.4) the following
relations hold:
\begin{eqnarray}
\langle A_{i}|A_{i}\rangle=1~~~~~~~~~~~~~~~~~~(i=0,1)\\
|a_{i}|^{2}+|b_{i}|^{2}=1~~~~~~~~~~~~~~~~~~~~~(i=0,1)\\
a_{i}a_{1-i}^{*}+b_{i}b_{1-i}^{*}=0~~~~~~~~~~~(i=0,1)\\
\langle A_{1}|Q\rangle=\langle A_{0}|Q\rangle=0.
\end{eqnarray}
Further we assume that
\begin{eqnarray}
\langle A_{1}|A_{0}\rangle=\langle A_{0}|A_{1}\rangle=0
\end{eqnarray}
A general pure state is given by
\begin{eqnarray}
|\psi\rangle=\alpha|0\rangle+\beta|1\rangle,~~~~~~~\alpha^{2}+|\beta|^{2}=1
\end{eqnarray}
Without any loss of generality we can assume that $\alpha$ and
$\beta$ are real numbers.\\
Using the transformation relation (5.1-5.4) and exploiting
linearity of U, we have
\begin{eqnarray}
U|\psi\rangle|\psi\rangle|Q\rangle&=&\alpha^{2}U|0\rangle|0\rangle|Q\rangle+\alpha\beta
U|0\rangle|1\rangle|Q\rangle+\alpha\beta
U|1\rangle|0\rangle|Q\rangle+\beta^{2}U|1\rangle|1\rangle|Q\rangle{}\nonumber\\&=&
\alpha^{2}|0\rangle|\Sigma\rangle|A_{0}\rangle+\alpha\beta[g|0\rangle|1\rangle+h|1\rangle|0\rangle]|Q\rangle
+\beta^{2}|1\rangle|\Sigma\rangle|A_{1}\rangle{}\nonumber\\&\equiv&|\psi\rangle_{12}^{(out)}
\end{eqnarray}
where $g=a_{0}+a_{1}$, $h=b_{0}+b_{1}$.\\
The reduced density operators of the output state in mode'1' and
'2' are given by
\begin{eqnarray}
\rho_{1}^{(out)}&=&tr_{2}[\rho_{12}^{(out)}]=tr_{2}[|\psi\rangle_{12}^{(out)}\langle\psi|]
{}\nonumber\\&=&[\alpha^{4}+\alpha^{2}\beta^{2}gg^{*}]|0\rangle\langle0|+[\beta^{4}+
\alpha^{2}\beta^{2}hh^{*}]|1\rangle\langle1|\\
\rho_{2}^{(out)}&=&tr_{1}[\rho_{12}^{(out)}]=tr_{1}[|\psi\rangle_{12}^{(out)}\langle\psi|]
{}\nonumber\\&=&\alpha^{4}|\Sigma\rangle\langle\Sigma|+\alpha^{2}\beta^{2}[gg^{*}]|1\rangle\langle1|
+hh^{*}|0\rangle\langle0|]+\beta^{4}|\Sigma\rangle\langle\Sigma|
\end{eqnarray}
Now to see the performance of the machine, we must calculate the distortion of the
input state and the fidelity of deletion.\\
Therefore, the H-S distance between the density operators
$\rho_{a}^{id}=|\psi\rangle\langle\psi|$ and $\rho_{1}^{(out)}$
given in equation (5.12) is
\begin{eqnarray}
D_{1}(\alpha^{2})&=&tr[\rho_{1}^{(out)}-\rho_{a}^{id}]^{2}{}\nonumber\\&=&k\alpha^{4}\beta^{4}
+2\alpha^{2}\beta^{2}
\end{eqnarray}
where $k = (gg^{*}-1)^{2}+ (hh^{*}-1)^{2}$.\\
Since $D_{1}$ depends on $\alpha^{2}$, so average distortion of input qubit in mode 1
is given by
\begin{eqnarray}
\overline{D_{1}}=\int_{0}^{1}D_{1}(\alpha^{2})d\alpha^{2}=
\frac{1}{3}(1+\frac{(gg^{*}-1)^{2}+ (hh^{*}-1)^{2}}{10})
\end{eqnarray}
The reduced density matrix of the qubit in the mode 2 i.e.
$\rho_{2}^{(out)}$ contains error due to imperfect deleting and
the error can be measured by calculating the fidelity. Thus the
fidelity is given by
$F_{1}=\langle\Sigma|\rho_{2}^{(out)}|\Sigma\rangle=
1-k_{1}\alpha^{2}\beta^{2}$, where
$k_{1}=2-gg^{*}M^{2}-hh^{*}(1-M^{2})$,$M=\langle\Sigma|1\rangle$.\\
Since fidelity of deletion depends on the input state, so the average fidelity over
all input state is given by
\begin{eqnarray}
\overline{F_{1}}=\int_{0}^{1}F_{1}(\alpha^{2})d\alpha^{2}=1-\frac{k_{1}}{6}=
\frac{2}{3}+\frac{(gg^{*}-hh^{*})M^{2}+ hh^{*}}{6}
\end{eqnarray}
From equation (5.15) and (5.16), we observe that the minimum
average distortion of the state in mode '1' from the input state
is $\frac{1}{3}$  and the minimum average fidelity of deletion is
$\frac{2}{3}$. So our prime task is to construct a deleting
machine or in other words, to find the value of the machine
parameters $a_{0}, a_{1}, b_{0}, b_{1}$ which will maximize the
fidelity of deletion but keep the average distortion at its minimum value.\\
To solve the above discussed problem, we take
$gg^{*}-hh^{*}=\epsilon$ and $hh^{*} =1+\epsilon_{1}$, where
$\epsilon$ and $\epsilon_{1}$ are very small quantities. Then
equations (5.15) and (5.16) reduce to
\begin{eqnarray}
\overline{D_{1}}=\frac{1}{3}+\frac{(\epsilon_{1})^{2}+(\epsilon+\epsilon_{1})^{2}}{30}\\
\overline{F_{1}}=\frac{5}{6}+\frac{\epsilon M^{2}+\epsilon_{1}}{6}
\end{eqnarray}
Therefore, $\overline{D_{1}}\rightarrow\frac{1}{3}$,
$\overline{F_{1}}\rightarrow\frac{5}{6}$ as $\epsilon,~~\epsilon_{1}\rightarrow0$.\\
The above equation shows that if we choose machine parameters $a_{0}, a_{1}, b_{0},
b_{1}$ in such a way that $gg^{*}$ and $hh^{*}$ both are very close to unity then only
we are able to keep the distortion at its minimum level and increase the average
fidelity to $\frac{5}{6}$.
\section{\emph{Concatenation of cloning and deletion machines}}
In this section, we study the effect of concatenation of cloning
and deleting machine. We investigate how well one can delete one
copy from the two imperfectly cloned copies of an unknown quantum
state. We consider here only the imperfect cloned copies obtained
from WZ cloning machine and BH cloning machine.
\subsection{\emph{Concatenation of WZ cloning machine and PB deleting machine}}
Let an unknown quantum state (5.10) be cloned by WZ cloning
machine. Using cloning transformation (1.17-1.18), an unknown
quantum state (5.10) cloned to
\begin{eqnarray}
\alpha|0\rangle|0\rangle|Q_{0}\rangle+\beta|1\rangle|1\rangle|Q_{1}\rangle
\end{eqnarray}
Now, operating deleting machine (4.1-4.4) on the cloned state
(5.19), we get the final output state as
\begin{eqnarray}
|\phi\rangle_{xy}^{(out)}=\alpha|0\rangle|\Sigma\rangle|A_{0}\rangle+\beta|1\rangle|\Sigma\rangle|A_{1}\rangle
\end{eqnarray}
The reduced density operator describing the output state in modes
x and y are given by
\begin{eqnarray}
\rho_{x}^{(out)}=tr_{y}(\rho_{xy})=\alpha^{2}|0\rangle\langle0|+\beta^{2}|1\rangle\langle1|\\
\rho_{y}^{(out)}=tr_{x}(\rho_{xy})=|\Sigma\rangle\langle\Sigma|
\end{eqnarray}
The H-S distance between the density operators
$\rho_{a}^{id}=|\psi\rangle\langle\psi|$ and $\rho_{x}^{(out)}$
given in equation (5.21) is
\begin{eqnarray}
D_{3}(\alpha^{2})=tr[\rho_{x}^{(out)}-\rho_{a}^{id}]^{2}=2\alpha^{2}(1-\alpha^{2})
\end{eqnarray}
The average distortion of input qubit after cloning and deleting operation is given by
\begin{eqnarray}
\overline{D_{3}}=\int_{0}^{1}D_{3}(\alpha^{2})d\alpha^{2}=0.33
\end{eqnarray}
The fidelity of deletion is given by
\begin{eqnarray}
F_{3}=\langle\Sigma|\rho_{y}|\Sigma\rangle=1
\end{eqnarray}
The above equations shows that if we clone an unknown quantum
state by WZ cloning machine, and delete a copy qubit by Pati and
Braunstein's deleting machine then the fidelity of deletion is
found to be 1 for arbitrary input state but the concatenation of
the cloning and deleting machine cannot retain the input qubit in
its original state.
\subsection{\emph{Concatenation of BH cloning machine and PB deleting machine}}
Let an unknown quantum state (5.10) be cloned by B-H cloning
machine. Using cloning transformation (1.48-1.49), quantum state
(5.10) is cloned to
\begin{eqnarray}
\alpha[|0\rangle|0\rangle|Q_{0}\rangle+(|0\rangle|1\rangle+|1\rangle|0\rangle)|Y_{0}\rangle]
+\beta[|1\rangle|1\rangle|Q_{1}\rangle+(|0\rangle|1\rangle+|1\rangle|0\rangle)|Y_{1}\rangle]
\end{eqnarray}
After operating deleting machine (4.1-4.4) to the cloned state
(5.26), the output state is given by
\begin{eqnarray}
|\phi\rangle_{xy}^{(out)}&=&\frac{1}{\sqrt{1+2\xi}}\{\alpha[|0\rangle|\Sigma\rangle|A_{0}\rangle
+(|0\rangle|1\rangle+|1\rangle|0\rangle)|Y_{0}\rangle]+\beta[|1\rangle|\Sigma\rangle|A_{1}\rangle
+(|0\rangle|1\rangle+{}\nonumber\\&&|1\rangle|0\rangle)|Y_{1}\rangle]
\end{eqnarray}
The reduced density operators describing the output state in mode
x and y are given by
\begin{eqnarray}
\rho_{x}^{(out)}=tr_{y}(\rho_{xy})=tr_{y}(|\phi\rangle_{xy}^{(out)}\langle\phi|)
=\frac{1}{1+2\xi}\{(\alpha^{2}+\xi)|0\rangle\langle0|+(\beta^{2}+\xi)|1\rangle\langle1|\}\\
\rho_{y}^{(out)}=tr_{x}(\rho_{xy})=tr_{y}(|\phi\rangle_{xy}^{(out)}\langle\phi|)=
\frac{1}{1+2\xi}\{|\Sigma\rangle\langle\Sigma|+I\xi\}
\end{eqnarray}
where I is the identity matrix in two dimensional Hilbert space.\\
The distance between the density operators
$\rho_{a}^{id}=|\psi\rangle\langle\psi|$ and $\rho_{x}^{(out)}$
given in equation (5.28) is
\begin{eqnarray}
D_{4}(\alpha^{2})=tr[\rho_{x}^{(out)}-\rho_{a}^{id}]^{2}=
\frac{2\xi^{2}+2\alpha^{2}\beta^{2}(1+4\xi)}{(1+2\xi)^{2}}
\end{eqnarray}
The average distortion of input state is given by
\begin{eqnarray}
\overline{D_{4}}=\int_{0}^{1}D_{4}(\alpha^{2})d\alpha^{2}=\frac{6\xi^{2}+4\xi+1}{3(1+2\xi)^{2}}
=\frac{11}{32},~\textrm{for B-H cloning machine}~ \xi=\frac{1}{6}
\end{eqnarray}
Also the fidelity of deletion of one qubit from two identical
cloned qubits is given by
\begin{eqnarray}
F_{4}=\langle\Sigma|\rho_{y}|\Sigma\rangle=\frac{1+\xi}{1+2\xi}=\frac{7}{8},~\textrm{for
B-H cloning machine}~ \xi=\frac{1}{6}
\end{eqnarray}
\subsection{\emph{Concatenation of WZ cloning machine and deleting machine(5.1-5.4)}}
After operating deleting machine (5.1-5.4) on the cloned state
(5.19), we get the output state as
\begin{eqnarray}
|\phi\rangle_{xy}^{(out)}=\alpha|0\rangle|\Sigma\rangle|A_{0}\rangle
+\beta[|1\rangle|\Sigma\rangle|A_{1}\rangle
\end{eqnarray}
The reduced density operators describing the output state in mode
x and y are given by
\begin{eqnarray}
\rho_{x}^{(out)}=tr_{y}(\rho_{xy})=tr_{y}(|\phi\rangle_{xy}^{(out)}\langle\phi|)
=\alpha^{2}|0\rangle\langle0|+\beta^{2}|1\rangle\langle1|\\
\rho_{y}^{(out)}=tr_{x}(\rho_{xy})=tr_{y}(|\phi\rangle_{xy}^{(out)}\langle\phi|)=
|\Sigma\rangle\langle\Sigma|
\end{eqnarray}
The H-S distance between the density operators
$\rho_{a}^{id}=|\psi\rangle\langle\psi|$ and $\rho_{x}^{(out)}$
given in equation (5.34) is
\begin{eqnarray}
D_{5}(\alpha^{2})=tr[\rho_{x}^{(out)}-\rho_{a}^{id}]^{2}= 2\alpha^{2}(1-\alpha^{2})
\end{eqnarray}
Since $D_{5}$ depends on $\alpha^{2}$, so average distortion of deletion is given by
\begin{eqnarray}
\overline{D_{5}}=\int_{0}^{1}D_{5}(\alpha^{2})d\alpha^{2}=0.33
\end{eqnarray}
The fidelity of the second qubit is given by
\begin{eqnarray}
F_{5}=\langle\Sigma|\rho_{y}|\Sigma\rangle=1
\end{eqnarray}
\subsection{ \emph{Concatenation of BH cloning machine and deleting machine(5.1-5.4)}}
After operating deleting machine (5.1-5.4) on the cloned state
(5.26), we get
\begin{eqnarray}
|\phi\rangle_{xy}^{(out)}&=&\{\alpha[|0\rangle|\Sigma\rangle|A_{0}\rangle
+(g|0\rangle|1\rangle+h|1\rangle|0\rangle)|Y_{0}\rangle]+\beta[|1\rangle|\Sigma\rangle|A_{1}\rangle
+{}\nonumber\\&&(g|0\rangle|1\rangle+h|1\rangle|0\rangle)|Y_{1}\rangle]\}
\end{eqnarray}
We assume
\begin{eqnarray}
\langle A_{0}|Y_{0}\rangle=\langle A_{1}|Y_{1}\rangle=0
\end{eqnarray}
The reduced density operators describing the output state in two
different modes are given by
\begin{eqnarray}
\rho_{x}^{(out)}&=&tr_{y}(\rho_{xy})=tr_{y}(|\phi\rangle_{xy}^{(out)}\langle\phi|)
{}\nonumber\\&=&\frac{1}{1+(gg^{*}+hh^{*})\xi}\{(\alpha^{2}+\xi
gg^{*})|0\rangle\langle0|+ (\beta^{2}+\xi hh^{*})|1\rangle\langle1|\}
\end{eqnarray}
\begin{eqnarray}
\rho_{y}^{(out)}&=&tr_{x}(\rho_{xy})=tr_{y}(|\phi\rangle_{xy}^{(out)}\langle\phi|)
{}\nonumber\\&=&\frac{1}{1+(gg^{*}+hh^{*})\xi}\{|\Sigma\rangle\langle\Sigma|+(\xi
hh^{*})|0\rangle\langle0|+ (\xi gg^{*})|1\rangle\langle1|\}
\end{eqnarray}
Now in order to measure the degree of distortion, we evaluate the
distance between the density operators
$\rho_{a}^{id}=|\psi\rangle\langle\psi|$ and $\rho_{x}^{(out)}$
given in equation (5.41) is given by
\begin{eqnarray}
D_{6}(\alpha^{2})=tr[\rho_{x}^{(out)}-\rho_{a}^{id}]^{2} =
\frac{2\xi^{2}(gg^{*}\beta^{2}-hh^{*}\alpha^{2})^{2}}{[1+(gg^{*}+hh^{*})\xi]^{2}}
\end{eqnarray}
The average distortion of input qubit is given by
\begin{eqnarray}
\overline{D_{6}}&=&\int_{0}^{1}D_{6}(\alpha^{2})d\alpha^{2}=\frac{1}{3}+
\frac{2\xi^{2}[(gg^{*})^{2}+(hh^{*})^{2}-(gg^{*})(hh^{*})]}{3[1+(gg^{*}+hh^{*})\xi]^{2}}
{}\nonumber\\&=&\frac{1}{3}+\frac{2}{3}
(\frac{(gg^{*})^{2}+(hh^{*})^{2}-(gg^{*})(hh^{*})}{[6+gg^{*}+hh^{*}]^{2}}),~\textrm{for}~
\xi=\frac{1}{6}
\end{eqnarray}
The fidelity of deletion is given by
\begin{eqnarray}
F_{6}&=&\langle\Sigma|\rho_{y}|\Sigma\rangle=\frac{1+\xi
M^{2}(gg^{*}-hh^{*})+\xi(hh^{*})}{1+(gg^{*}+hh^{*})\xi}{}\nonumber\\&=&\frac{6+
M^{2}(gg^{*}-hh^{*})+(hh^{*})}{6+gg^{*}+hh^{*}},~~\textrm{for B-H cloning machine}
~\xi=\frac{1}{6}
\end{eqnarray}
In particular, For $a_{0}=\frac{\sqrt{3}}{2}, a_{1}=\frac{i}{2}, b_{0}=\frac{i}{2},
b_{1}=\frac{\sqrt{3}}{2}$, we get $gg^{*}=hh^{*}=1$. In this case, we find that the
fidelity of deletion and the average distortion is same as in the case of  B-H cloning
machine and P-B deleting machine.\\

 \large \baselineskip .85cm

\setcounter{page}{126}\markright{\Large \textbf{List of
Publications}}

1. Title: Deletion of Imperfect cloned copies\\
   Authors: Satyabrata Adhikari, B.S.Choudhury\\
   Journal Ref.: Journal of Physics A: Mathematical and General \textbf{37}, 1
   (2004).\\\\
2. Title: Quantum Deletion: Beyond the no-deletion principle\\
   Authors: Satyabrata Adhikari\\
   Journal Ref.: Phys. Rev. A \textbf{72}, 052321 (2005).\\\\
3. Title: Broadcasting of Inseparability\\
   Authors: Satyabrata Adhikari, B.S.Choudhury, I.Chakrabarty\\
   Journal Ref.: J. Phys. A: Math. Gen. \textbf{39}, 8439 (2006).\\\\
4. Title: Improving the fidelity of deletion\\
   Authors: Satyabrata Adhikari, B.S.Choudhury\\
   Journal Ref.: Phys. Rev. A \textbf{73}, 054303 (2006).\\\\
5. Title: Probabilistic exact deletion and Probabilistic
   no-signalling\\
   Authors: I.Chakrabarty, Satyabrata Adhikari, B.S.Choudhury\\
   Journal Ref.: Physica Scripta \textbf{74}, 555 (2006).\\\\
6. Title: Broadcasting of three-qubit entanglement via local copying and entanglement swapping\\
   Authors: Satyabrata Adhikari, B.S.Choudhury\\
   Journal Ref.: Phys. Rev. A \textbf{74}, 032323 (2006).\\\\
7. Title: Inseparability of quantum parameters\\
   Authors: I.Chakrabarty, Satyabrata Adhikari, Prashant, B.S.Choudhury\\
   Comment: Accepted in International Journal of Theoretical
   Physics.\\\\
8. Title: Hybrid quantum cloning machine\\
   Authors: Satyabrata Adhikari, A.K.Pati, B.S.Choudhury, I.Chakrabarty \\
   Comment: Submitted to Quantum Information and Computation.\\\\

\end{document}